%% file: hobbs.tex
\documentclass{mn2e}
\usepackage{graphicx}
\usepackage{amssymb}
\footnotesize
\newdimen\minuswidth    %define @ width of minus sign for tables
\setbox0=\hbox{$-$}
\minuswidth=\wd0
\catcode`@=\active
\def@{\kern\minuswidth}
\newdimen\digitwidth    %define ! a one digit width for tables
\setbox0=\hbox{\rm0}
\digitwidth=\wd0
\catcode`!=\active
\def!{\kern\digitwidth}
\normalsize
\title[An analysis of the timing irregularities for 366 pulsars]
{An analysis of the timing irregularities for 366 pulsars}
\author[G. Hobbs et al.]
{G. Hobbs,$^1$
A. G. Lyne$^2$
\& M. Kramer$^2$
\\
$^1$ Australia Telescope National Facility, CSIRO, PO~Box~76, Epping
NSW~1710, Australia \\
$^2$ University of Manchester, Jodrell Bank Observatory, Macclesfield,
Cheshire SK11~9DL \\}

\let\leqslant=\leq %Authors with AMS fonts and mssymb.tex can comment
\date{}
\begin{document}
\maketitle
\newcommand{\setthebls}{
%                 de-comment this line for double spacing:
%\baselineskip=20pt
}
\setthebls

\begin{abstract}

We provide an analysis of timing irregularities observed for 366 pulsars. Observations were obtained using the 76-m Lovell radio telescope at the Jodrell Bank Observatory over the past 36 years.  These data sets have allowed us to carry out the first large-scale analysis of pulsar timing noise over time scales of $> 10$\,yr, with multiple observing frequencies and for a large sample of pulsars. Our sample includes both normal and recycled pulsars.   The timing residuals for the pulsars with the smallest characteristic ages are shown to be dominated by the recovery from glitch events, whereas the timing irregularities seen for older pulsars are quasi-periodic.  We emphasise that previous models that explained timing residuals as a low-frequency noise process are not consistent with observation.
\end{abstract}


\section{Introduction}

The Jodrell Bank data archive of pulsar observations contains over 6000 years of pulsar rotational history. The pulsar timing method (for a general overview see Lorimer \& Kramer 2005\nocite{lk05}, Lyne \& Graham-Smith 2004\nocite{ls04} or Manchester \& Taylor 1977\nocite{mt77}. Details are provided in Edwards et al. 2006)\nocite{ehm06} allows the observed pulse times of arrival (TOAs) to be compared with a model of the pulsar's astrometric, orbital and rotational parameters. The differences between the predicted arrival times and the actual arrival times are known as the pulsar ``timing residuals''.  For a perfect model the timing residuals would be dominated by measurement errors and have a ``white'' spectrum. Any features observed in the timing residuals indicate the presence of unmodelled effects which may include calibration errors, orbital companions or spin-down irregularities. There are two main types of irregularity, namely ``glitches'' which are sudden increases in rotation rate followed by a period of relaxation, and ``timing noise'', which consists of low-frequency structures.

A more complete understanding of pulsar timing noise will lead to many important results.  For instance, explaining the cause of timing noise and glitches may also allow us to relate these phenomena and hence provide an insight into the interior structure of neutron stars.
Pulsar timing array projects are being developed around the world with the aim of detecting gravitational waves by looking for irregularities in the timing of millisecond pulsars (see e.g. Hobbs 2005\nocite{hob05} and references therein). A stochastic background formed by coalescing supermassive binary black hole systems will induce signatures in the timing residuals with amplitudes of approximately 100\,ns.  If the intrinsic timing noise for millisecond pulsars is at a higher level then it becomes more difficult to extract the gravitational wave signal from the observations.  

In this paper we study the timing noise in the residuals of 366 pulsars that have been regularly observed over the past 10 to 36 years. The basic observational parameters for the pulsars in our sample were provided by Hobbs et al. (2004)\nocite{hlk+04}; hereafter H04.  These timing ephemerides included the pulsar positions, rotational frequencies and their first two derivatives, dispersion measures and their derivatives and proper motions.  The proper motions were subsequently updated with more recent observations, combined with other measurements of pulsar proper motions available from the literature and analysed by Hobbs et al. (2005)\nocite{hllk05}. This work allowed us to present a new velocity distribution for the pulsar population.

In order to obtain the precise spin and astrometric parameters presented in H04, it was necessary to  remove the low-frequency timing noise. This pre-whitening procedure was undertaken using a simple high-pass filter where harmonically related sinusoids were fitted to the residuals and the lowest frequency waves subtracted. The low-frequency noise was not subsequently studied in the previous papers. In this paper, we address the properties of this timing noise.

Low-frequency structures previously observed in pulsar data sets have been explained by random processes (e.g. Cordes \& Helfand 1980\nocite{ch80}, Lyne 1999\nocite{lyn99}), unmodelled planetary companions (e.g. Cordes 1993\nocite{cor93}) or free-precession (e.g. Stairs, Lyne \& Shemar 2000\nocite{sls00}). However, the physical phenomenon underlying most of the timing noise still has not been explained.  Much of the basic theoretical work was described by Boynton et al. (1972)\nocite{bgh+72} who analysed the arrival times for the Crab pulsar over a two-year period.  In their paper, an attempt was made to describe the timing noise as either phase, frequency or slowing-down noise corresponding to random walks in these parameters.  Later, Cordes \& Helfand (1980) found that out of a sample of 11 pulsars, seven showed timing noise consistent with frequency noise, two from slowing-down noise and two from phase-noise.  They concluded that 1) timing noise is widespread in pulsars, 2) it is correlated with period derivative and weakly with period, 3) it is not correlated with height above the Galactic plane, luminosity nor with pulse shape changes.  However, the data sets used were small in number, short and small-scale pulse shape variations would have been undetectable. The assumption that timing noise is a red-noise process has continued to date. However, Cordes \& Downs (1985)\nocite{cd85} showed that the idealised, random walk model was too simple.  They developed a more detailed model where discrete `micro-jumps' in one or more of the timing parameters were superimposed on the random walk process.  D'Alessandro et al. (1995)\nocite{dmh+95} analysed the timing residuals for 45 pulsars with data spanning up to seven years. They observed very weak timing noise for 19 of their pulsars, for seven the activity was attributed to random walk processes comprising  a large number of events in one of the rotation variables, a further seven were explained as resolved jumps in $\nu$ the pulse frequency and $\dot{\nu}$, its derivative, seven more as resolved jumps on a low-level background and for the remaining five the timing noise could not be explained as a pure random walk process nor resolved jumps.

Timing noise analyses with large samples of pulsars have been limited by the relatively short data spans studied.  Long data-spans have been analysed in a few papers, but for only a small number of pulsars.  For instance, Baykal et al. (1999)\nocite{babd99} analysed four pulsars timed for 14\,yr. Shabanova (1995)\nocite{sha95} observed PSR~B0329$+$54 for 16\,yr,  Stairs, Lyne \& Shemar (2000)\nocite{sls00} reported on 13\,yr of PSR~B1828$-$11 observations and Shabanova, Lyne \& Urama (2001)\nocite{slu01} analysed PSR~B1642$-$03 over a 30\,yr data span. More recently, the millisecond pulsar PSR~J1713$+$0747 was observed for 12\,yr (Splaver et al. 2005)\nocite{sns+05} and the young pulsar PSR~B1509$-$58 for 21\,yr (Livingstone et al. 2005)\nocite{lkgm05}.  Most of the analysis has concentrated on obtaining high quality spectral estimates of the timing residuals or fitting a simple model to the timing residuals of a single pulsar.  We note that our sample is $\sim$20 times larger than the previous large study of pulsar timing noise (D'Alessandro et al. 1995) in terms of the number of years of rotational history studied.

Glitches are thought to represent a sudden unpinning of superfluid vortices in the interior of the neutron star (Lyne, Shemar \& Graham-Smith 2000)\nocite{lsg00}.  The relationship between glitches and timing noise is not understood although Janssen \& Stappers (2006)\nocite{js06} showed that it is possible to model the timing noise in PSR~B1951+32 as multiple small glitches.  Glitches are discrete events that occur more commonly for young pulsars\footnote{We use the terms ``young'' and ``old'' throughout referring to a pulsar's characteristic age which is only an approximation of the pulsar's true age.} (although for two pulsars with similar rotational parameters, one may glitch frequently while the other may never have been observed to glitch) and have a wide range of sizes with fractional frequency increases between $10^{-9}$ and $10^{-5}$ (e.g. Lyne, Shemar \& Graham-Smith 2000\nocite{lsg00}, Hobbs et al. 2002)\nocite{hlj+02}.  No model currently predicts the time between glitches or the size of any given event although pulsars with large glitches tend to show larger intervals between glitches (Lyne et al. 2000)\nocite{lsg00}.  Melatos, Peralta \& Wyithe (2008)\nocite{mpw08} recently showed that, for most pulsars in their sample, the waiting time between glitch events followed an exponential distribution.  This distribution was subsequently modelled by Warszawski \&  Melatos (2008)\nocite{wm08} using a cellular automaton model of pulsar glitches. However, it is still not clear whether the glitch and timing noise phenomena are related.

In this paper we describe the observing system and present the measured timing residuals (\S\ref{sec:results}), rule out some models of timing noise (\S\ref{sec:ruleOut}) and highlight various properties of the timing noise (\S\ref{sec:properties}).  
  
\section{Results}\label{sec:results}

\begin{figure}
\includegraphics[angle=-90,width=8cm]{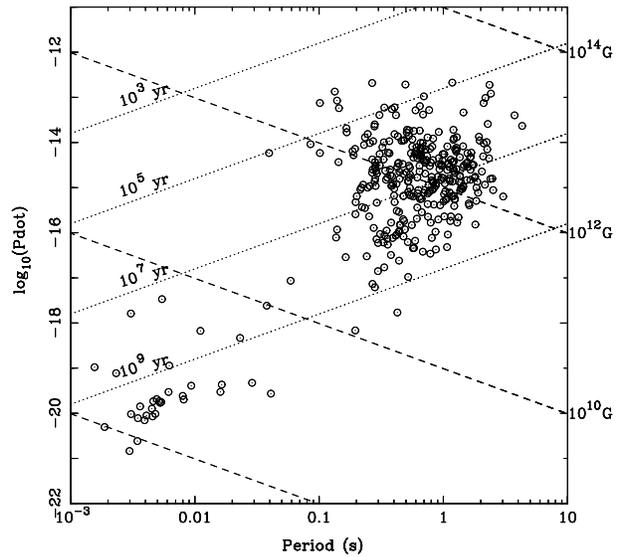}
\caption{Period--period-derivative diagram for the pulsars in our sample.}\label{fg:ppdot}
\end{figure}

\begin{figure}
\includegraphics[angle=-90,width=8cm]{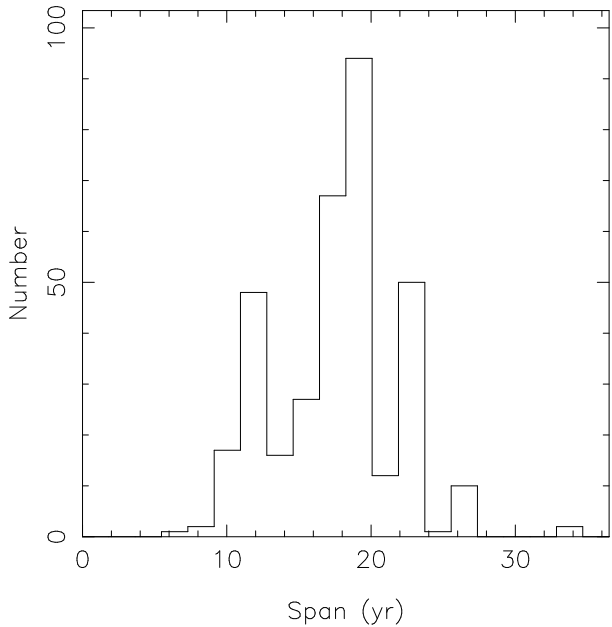}
\caption{Histogram of the time-span of our observations.}\label{fg:tspan}
\end{figure}

The timing solutions for the pulsars in our sample have been updated since H04 with more recent data from the Jodrell Bank Observatory.  The timing solutions were obtained using standard pulsar timing techniques as described in Paper~1.  In brief, the majority of the observations were obtained from the 76-m Lovell radio telescope.  The earliest times-of-arrival (TOAs) for 18 pulsars were obtained from observations using the NASA Deep Space Network (Downs \& Reichley 1983; Downs \& Krause-Polstorff 1986\nocite{dk86})\nocite{dr83}.  Observations using the Lovell telescope were carried out predominately at frequencies close to 408, 610, 910, 1410 and 1630\,MHz with a few early observations at 235 and 325\,MHz.  The last observations used in our data set were obtained around MJD~53500. The signals were combined to produce, for every observation, a total intensity profile.  TOAs were subsequently determined by convolving, in the time domain, the profile with a template corresponding to the observing frequency. The pulsar timing residuals described in this paper were obtained by fitting a timing model to the TOAs using the \textsc{tempo2} pulsar timing software (Hobbs, Edwards \& Manchester 2006)\nocite{hem06}. 

For those pulsars in which the residuals are dominated by large glitches it is difficult to obtain phase coherent timing solutions across many years of observation.  For  example, for PSR~B1800$-$21 we have data spanning from the year 1985 to the present.  However, for this analysis we can only use data between the years 1991 and 1997.  We have a similarly short data-span for PSR~B1727$-$33.  Although PSRs~B0531$+$21, B1758$-$23 and B1757$-$24 have been regularly observed we cannot obtain a useful data-set for the analysis described in this paper.  These pulsars are therefore not included in our sample.

In Table~\ref{tb:params} we present the basic parameters for our sample of pulsars.  In column order, we first provide the pulsar's J2000 and B1950 names, spin-frequency, $\nu$, frequency derivative, $\dot{\nu}$ and frequency second derivative, $\ddot{\nu}$.   The following three columns give the epoch of the centre of the data span, the number of observations and the time span of the observations.  We then provide various measures characterising the amount of timing noise for each data set. We present $\sigma_1$, the unweighted rms of the residuals after fitting for $\nu$ and $\dot{\nu}$, $\sigma_2$ the unweighted rms of the residuals after removal of $\nu$ and its first two derivatives and $\sigma_3$, the rms after whitening the data set by fitting, and removing, harmonically related sinusoids (as described in H04). The last two columns contain two stability measures ($\Delta_8$ and $\sigma_z$) that are discussed in \S\ref{sec:stab}.  

A period--period-derivative diagram\footnote{Note: we have attempted to be consistent in our use of pulse frequency and its derivatives. However, in order to make a direct comparison with earlier publications we sometimes use the pulse period and its derivatives where $P=1/\nu$, $\dot{P} = -\dot{\nu}/\nu^2$.} for the pulsars in our sample is shown in Figure~\ref{fg:ppdot} to indicate the range of pulsar parameters included in our sample.  The diagram includes dotted lines representing various characteristic ages $\tau_c = P/(2\dot{P})$ and dashed lines for representative surface magnetic field strengths $B_s = 3.2\times10^{19}\sqrt{P\dot{P}}$\,G.  A histogram indicating the time-spans of our observations is shown in Figure~\ref{fg:tspan}.  The mean and median time spans are 18.5 and 18.3\,yr respectively.  For the recycled pulsars, the unweighted rms of the residuals after subtracting a cubic polynomial ranges from 8\,$\mu$s for PSR~J1744$-$1134 to $460$\,$\mu$s for PSR~B1913$+$16\footnote{Our sample includes 31 recycled pulsars defined with spin-periods $P < 0.1$\,s and spin-down rates $\dot{P} < 10^{-17}$. We note that lower rms values have been published for some of our pulsars.  For instance,  Splaver et al. (2005) obtained TOA timing precisions $<700$\,ns and as small as $\sim 200$\,ns for PSR~J1713$+$0747 (compared with $\sim 10\mu s$ for our data). In contrast to the high precision timing experiments that often use coherent de-dispersion systems and long observing durations the data presented here have been obtained with a stable observing system over many years and the large sample has necessitated short observations and hence poorer TOA precision.}.  For the ordinary pulsars the rms ranges up to 948\,ms for PSR~B1706$-$16, a pulsar with a 653\,ms spin-period\footnote{For a large number of pulsars the timing residuals deviate by more than one pulse period.  In these cases phase jumps of at least one period need to be added to keep track of the pulsar spin-down.}.

\input{table1.tex}

In Figure~\ref{fg:residuals} we plot the timing residuals for all the pulsars in our sample relative to a simple slow-down model including only the pulse-frequency and its first derivative.  The horizontal scale of each panel represents 36 years of observations and the vertical axis is scaled for each separate panel.  The upper value gives the range from the minimum to the maximum residual in milliseconds. The lower values give the same range, but in units of the pulsars' rotational periods. 
%For about 40\% of the pulsars, the variations in the residuals intrinsic to the pulsar rotation are dominated by the measurement errors caused by thermal or cosmic random noise in the observing system (e.g. PSRs~B0011$+$47, B0045$+$33 and J0134$-$2937).

\begin{figure*}
\includegraphics[width=18cm]{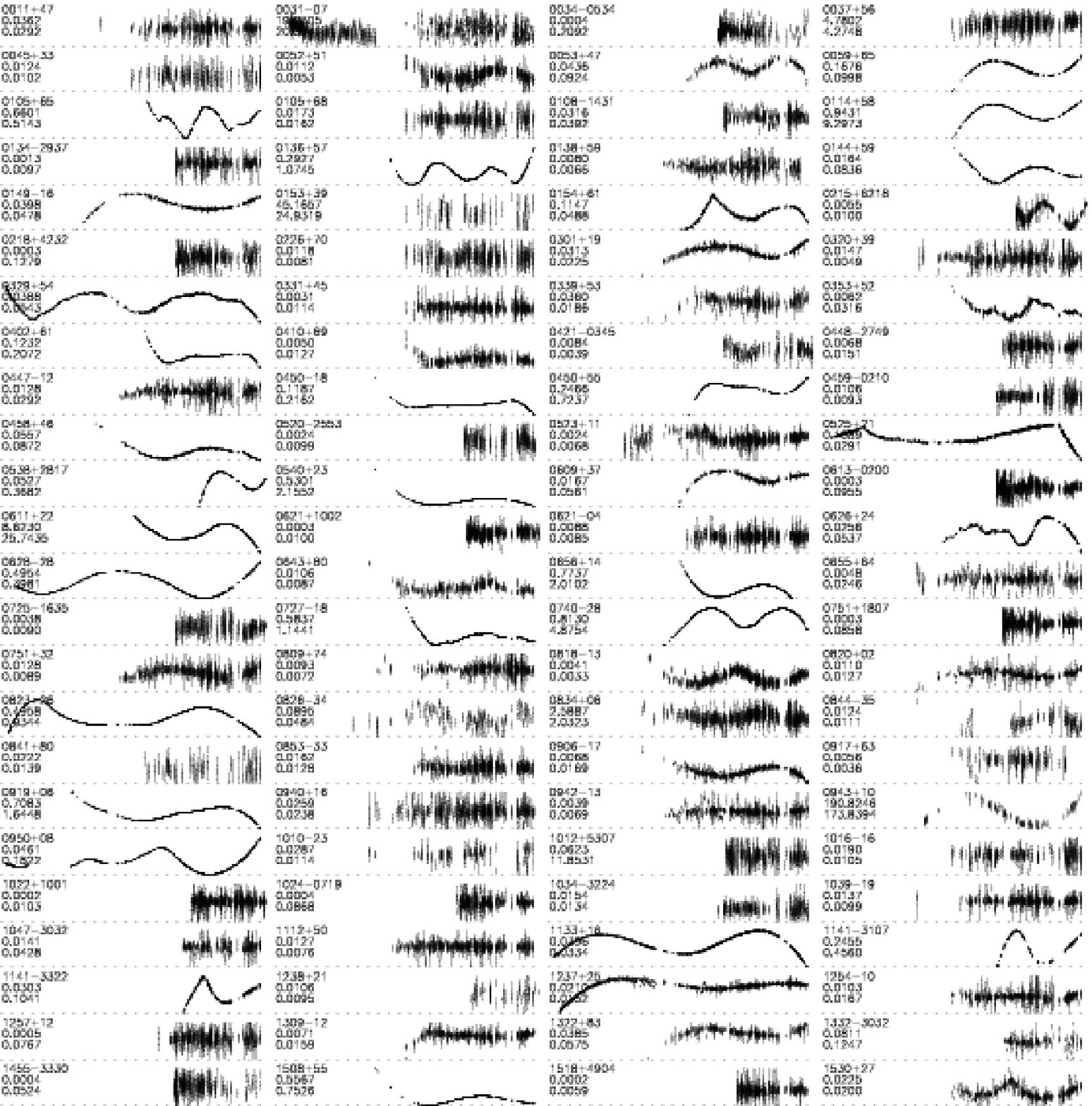}
\caption{Timing residuals after subtraction of the pulsar's spin frequency and its first derivative over the 38 years from MJD 39856 (year 1968) to 53736 (year 2006).   The three labels on the left provide the pulsar name, the range from the minimum to maximum residuals (s) and the same range scales by the pulsar's rotational period.}\label{fg:residuals}
\end{figure*}

\addtocounter{figure}{-1}
\begin{figure*}
\includegraphics[width=18cm]{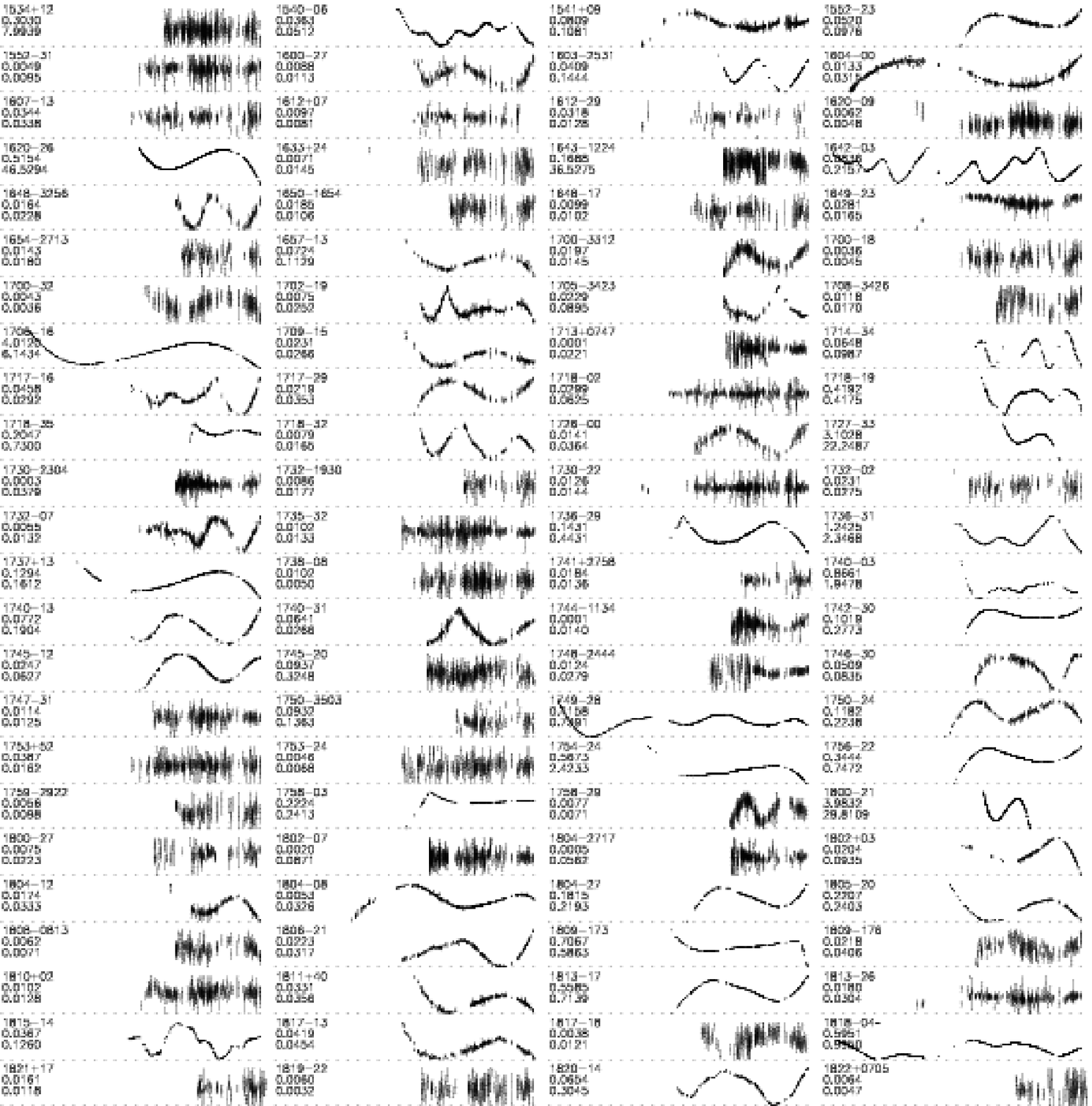}
\caption{\ldots continued}
\end{figure*}

\addtocounter{figure}{-1}
\begin{figure*}
\includegraphics[width=18cm]{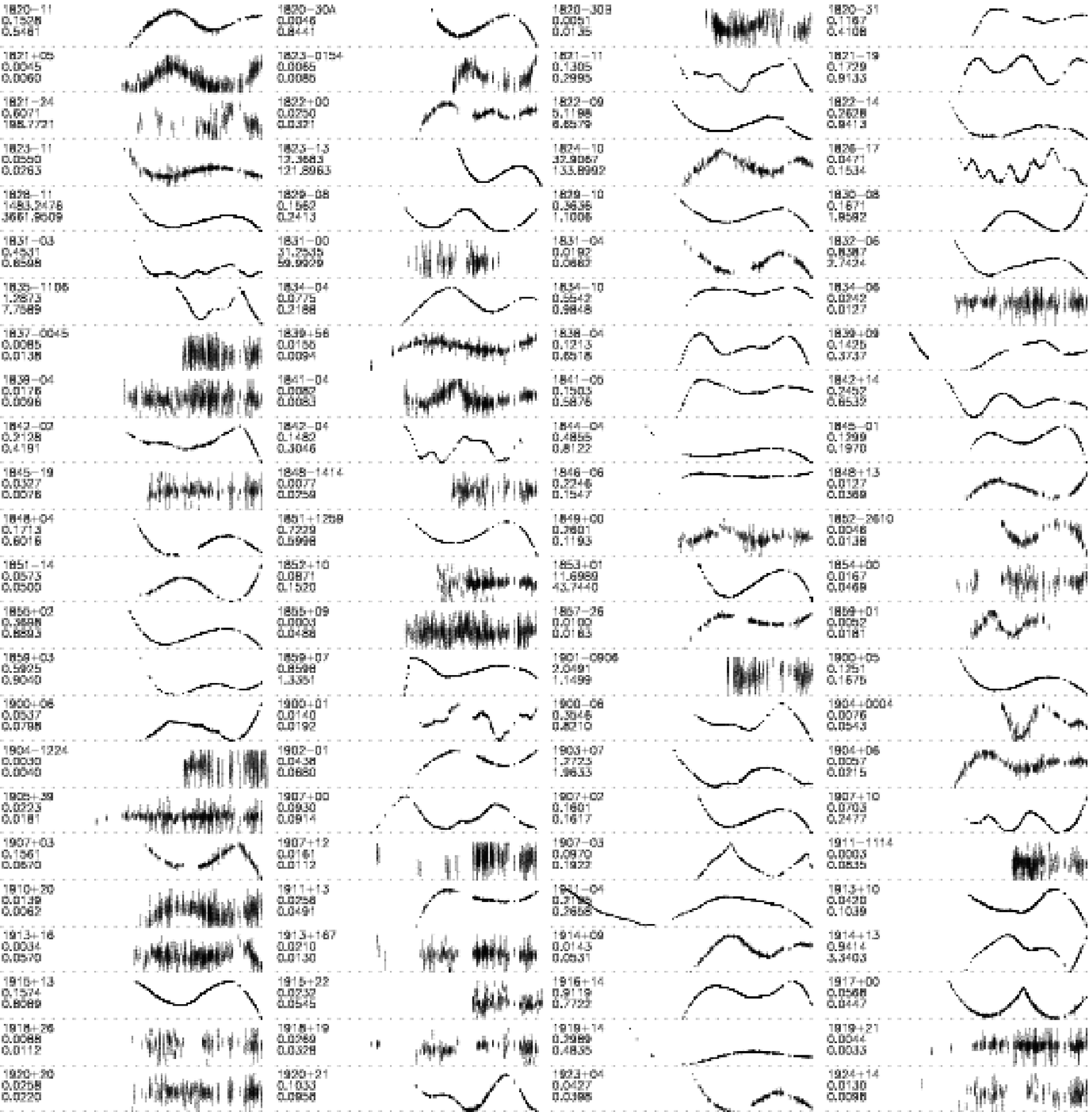}
\caption{\ldots continued}
\end{figure*}

\addtocounter{figure}{-1}
\begin{figure*}
\includegraphics[width=18cm]{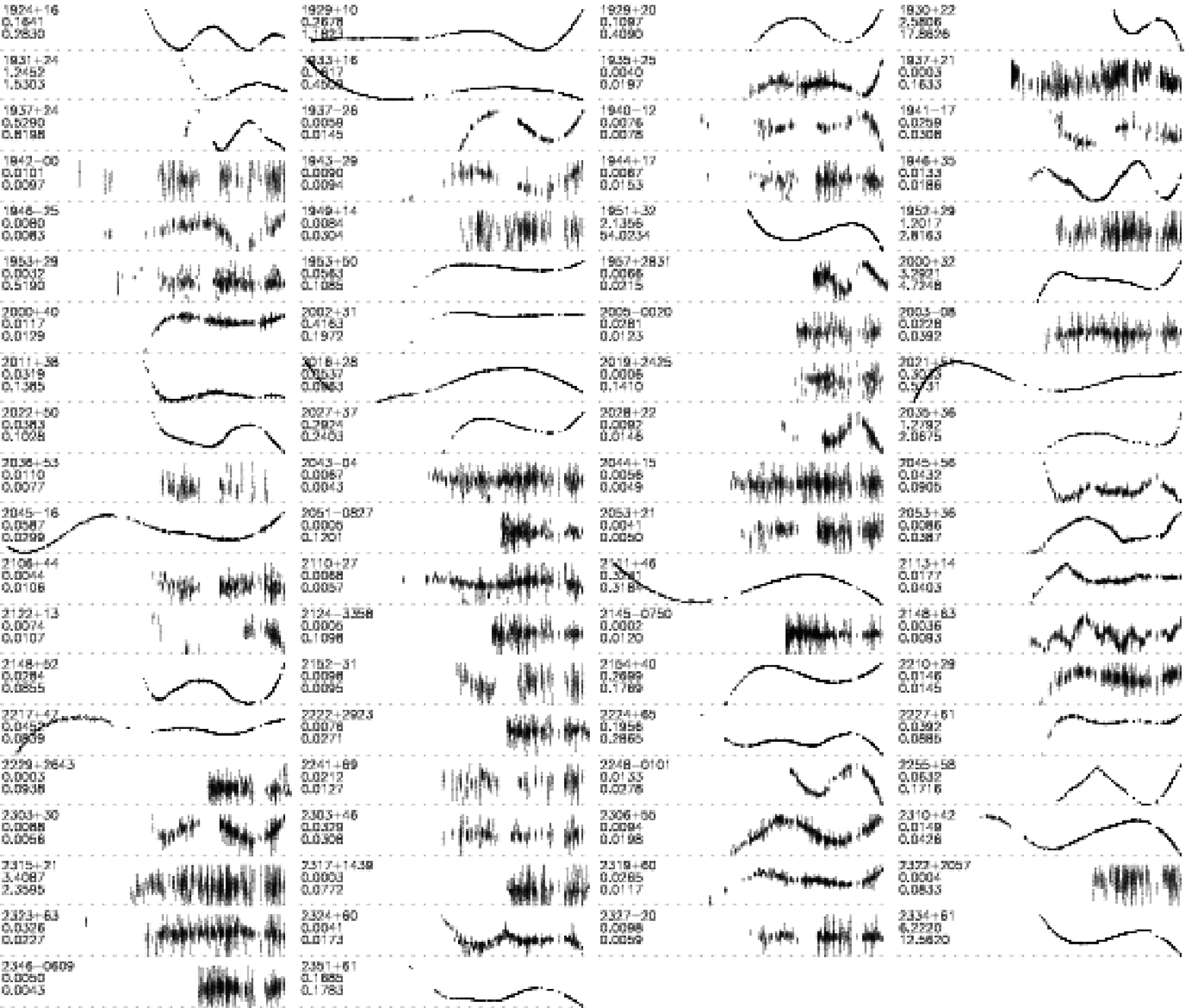}
\caption{\ldots continued}
\end{figure*}

\section{Discussion}\label{sec:discussion}

\begin{figure*}
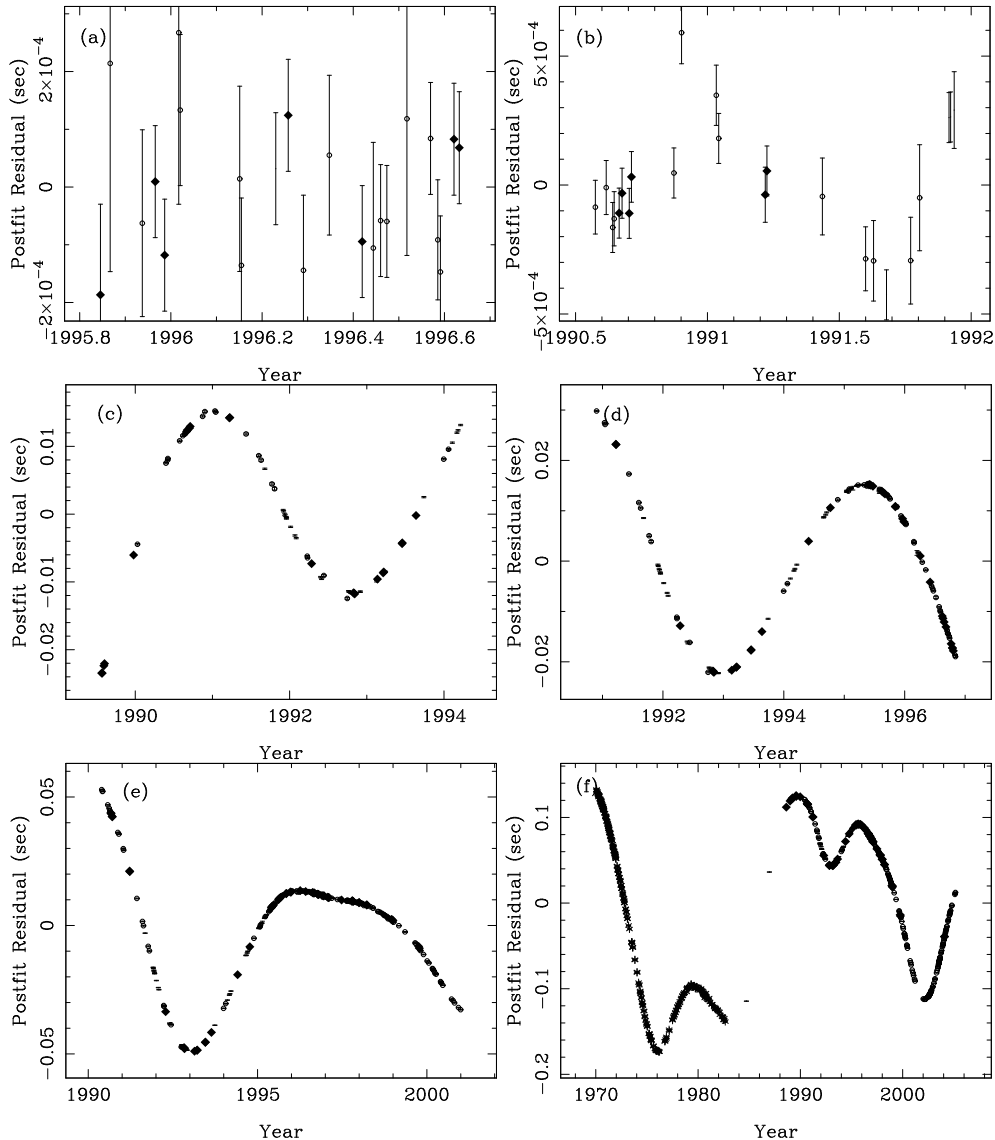

\includegraphics[width=5cm,angle=-90]{oneY_1.ps}
\includegraphics[width=5cm,angle=-90]{1.5_yr.ps}
\includegraphics[width=5cm,angle=-90]{fiveY.ps}
\includegraphics[width=5cm,angle=-90]{sixY.ps}
\includegraphics[width=5cm,angle=-90]{elevenY.ps}
\includegraphics[width=5cm,angle=-90]{full.ps}
\caption{The timing residuals for PSR~B1818$-$04 obtained from different section of the entire data-span available.  In each section the pulsar's spin-frequency and its first derivative have been fitted.  The data spans are approximately a) 1\,yr, b) 1.5\,yr, c) 5\,yr, d) 6\,yr, e) 11\,yr and f) the full 35\,yr.}\label{fg:0950}
\end{figure*}

It is tempting to categorise the pulsars by the structures seen in their timing residuals.  For instance, for approximately 37\% of our sample the variations in the residuals are dominated by the measurement errors caused by thermal or cosmic random noise in the observing system (e.g. PSRs~B0011$+$47, B0045$+$33 and J0134$-$2937). 20\% have residuals which can be modelled by a significant positive $\ddot{\nu}$ value (such residuals take the form of a cubic polynomial), 16\% for a cubic polynomial corresponding to $\ddot{\nu} < 0$ and 27\% which show structures that are more complicated.  However, this simple categorisation has two problems:
\begin{itemize}
\item{The TOAs for some pulsars are measured more precisely than for others.  For instance, in Figure~\ref{fg:overlay} we plot the detailed structure observed in the timing residuals for PSR~B1900$+$01. Any similar structure in the residuals of PSR~B1745$-$20 (that have been overplotted in Figure~\ref{fg:overlay}) would not be observable because of the much larger TOA uncertainties.}

\item{The classification of any particular pulsar is likely to change as more data are obtained.  In Figure~\ref{fg:0950}, we plot the timing residuals that would have been obtained for PSR~B1818$-$04 if we had access to data sets that spanned shorter time intervals of between 1 and 35-years.  Clearly any simple classification scheme would change depending on the length of the data span.}
\end{itemize}

Glitch events lead to clear signatures in the timing residuals.  As the fitting procedure used for Figure~\ref{fg:residuals} removes a quadratic-term, any glitch event will exhibit a cusp (corresponding to a local maximum) in the timing residuals (good examples being PSRs~B0154$+$61, B0525$+$21 and B2255$+$58).  A large number of unpublished glitch events can be identified in these data sets and will be described in a forthcoming publication.  We discuss the phenomenon known as ``slow glitches'' in \S\ref{sec:sglitch}.

%As will be emphasised throughout this paper, if the timing noise has a pure, steep red-noise power spectra then the subtraction of a quadratic polynomial corresponding to the pulsar spin-down will produce timing residuals dominated by a single cubic polynomial.   However, as our data-spans increase we see fewer cubic features. 

\subsection{Ruling out some models of timing noise}\label{sec:ruleOut}

Some pulsars in our sample are also observed as part of long-term timing programs at other observatories.  In all such cases we see the same large-scale features and hence timing noise is not caused by observing systems or the data processing being carried out at a particular observatory.  For instance, Shabanova, Lyne \& Urama (2001)\nocite{slu01} combined data from the Jodrell Bank, Hartebeesthoek and Pushchino observatories for PSR~B1642$-$03 and saw no observatory-dependent artefacts.  

The timing noise features are also not caused by the off-line processing.  We have used the \textsc{Tempo2} timing package which has been designed for high precision timing applications and is accurate for known physical effects at the 1\,ns level of precision (Hobbs, Edwards \& Manchester 2006).  However, we have compared timing residuals using three pulsar timing packages, \textsc{psrtime}, \textsc{tempo} and \textsc{tempo2} for many of our pulsars and see the same features in all cases.  The timing procedure relies on corrections being applied to correct the observatory clocks to the terrestrial time scale.  Inaccuracies in this time transfer and in the terrestrial time scale are expected to be at the 100\,ns-1\,$\mu$s level (see, e.g. Rodin 2008\nocite{rod08}) which will not affect the features of the timing noise described here.  Finally, \textsc{tempo2} converts the observed TOAs to barycentric arrival times using a planetary ephemeris.  Changing between the DE200, DE405 and DE414 ephemerides (Standish 2004\nocite{sta04b}) only changes the rms timing residuals by $\sim 1$\,$\mu$s and does not affect the large-scale timing noise.

\begin{figure}
\includegraphics[angle=-90,width=8cm]{1900_1745.ps}
\caption{The timing residuals for PSR B1900$+$01 (filled cirlces) overlaid on the timing residuals for PSR~B1745$-$20 (cross symbols).}\label{fg:overlay}
\end{figure}

For high-precision millisecond-pulsar timing, small features in the timing residuals can occur from incomplete calibration of the polarisation of the pulse profiles and are likely to be a small fraction of the pulse width.  However, the timing residuals described in this paper are much larger than a typical pulse width.  In addition, changes in pulse shape are likely to produce small step-changes in the timing residuals.  The timing noise phenomena described here produce much larger effects and have different characteristic shapes in the residuals.

\begin{figure}
\includegraphics[width=6cm,angle=-90]{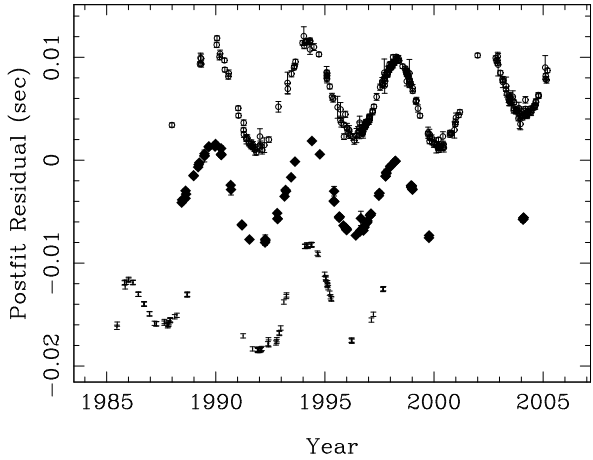}
\caption{The timing residuals for PSR~B1540$-$06 with an arbitrary offset applied between observations at different observing frequencies. The bottom set of residuals contains observations close to 400\,MHz, the middle residuals at 600\,MHz and the top at 1400\,MHz.}\label{fg:diffFreq}
\end{figure}

It has been postulated that the effects of interstellar or interplanetary dispersion measure (DM) variations may lead to timing noise.   However, any effects caused by the interstellar medium (ISM) will be dependent upon the observing frequency.  As shown in Figure~\ref{fg:diffFreq}, the timing noise seen in our data is identical at all observing frequencies and so the large-scale low frequency noise is not caused by the ISM.  It is, however, possible that some smaller scale variations are due to the ISM. A recent analysis of millisecond pulsars (You et al. 2007a)\nocite{yhc+07a} shows typical timing variations caused by variations in the pulsar's DM of up to a few microseconds (at an observing frequency $\sim 1.4$\,GHz).  Interplanetary DM variations, mainly due to the solar wind (You et al. 2007b)\nocite{yhc+07b}, also induce frequency-dependent residuals, but these are modelled with sufficient precision within \textsc{tempo2}.  
 
\subsection{Properties of the timing noise}\label{sec:properties}

\subsubsection{The ``amount'' of timing noise}\label{sec:stab}
 
\begin{figure}
\includegraphics[width=6cm,angle=-90]{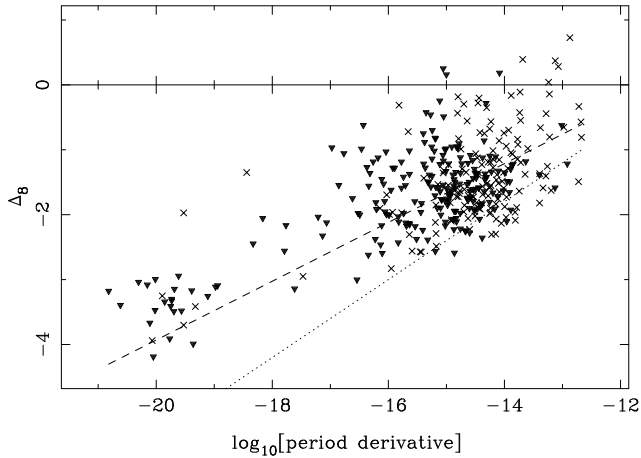}
\caption{The $\Delta_8$ parameter versus the pulse period derivative.  Upper limits are displayed as downward pointing triangles. The dotted line gives the best fit to the data of Arzoumanian et al. (1994). The dashed line is the best fit straight line to our observations.}\label{fg:d8}
\end{figure}

In order to characterise the stability of each pulsar and to determine a measure of the ``amount of timing noise'' we first calculate the $\Delta_8$ value introduced by Arzoumanian et al. (1994),
\begin{equation}
\Delta_8 = \log_{10}\left(\frac{1}{6\nu}|\ddot{\nu}|t^3\right)
\end{equation}
where the spin-frequency, $\nu$, and its second derivative, $\ddot{\nu}$, are measured over a $t = 10^8$\,s interval\footnote{This interval of $\sim 3.16$\,yr has no physical meaning and was the typical data span for the Arzoumanian et al. (1994) data sets.}. As our pulsar data-sets contain at least 10\,yr of data, we compare the average $\Delta_8$ value (listed in Table~\ref{tb:params}) to its variance obtained by fitting for $\nu$ and $\ddot{\nu}$ in unique 3\,yr segments.  Upper bounds are provided for pulsars in which the measured $\ddot{\nu}$ is not significant at the 2$\sigma$ level. In Figure~\ref{fg:d8} we plot $\Delta_8$ versus the pulse period derivative.  The dotted line corresponds to $\Delta_8 = 6.6 + 0.6 \log \dot{P}$ as obtained by Arzoumanian et al. (1994)\nocite{antt94} by eye.  Our data suggest that this is slightly low; a linear least-squares fit to our data for non-recycled pulsars and where the $\Delta_8$ measurement is not an upper-limit (shown as a dashed line) gives that
\begin{equation}
\Delta_8 = 5.1 + 0.5 \log \dot{P}.
\end{equation}
We therefore confirm the correlation between timing noise and spin-down rate (and the implication that younger pulsars with large spin-down rates exhibit more timing noise than older pulsars) that has been described in earlier work.

\begin{figure}
\includegraphics[width=6.5cm,angle=-90]{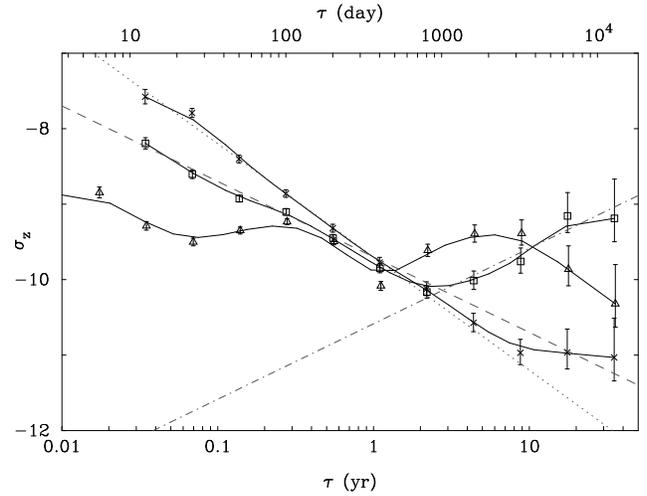}
\caption{$\sigma_z(\tau)$ for PSRs~B0031$-$07 (cross symbols), B0626$-$28 (square) and B1642$-$03 (triangle). The dotted line indicate timing residuals with spectral exponents of $\alpha = 0$, the dashed line for $\alpha = -1$ and the dot-dashed line for $\alpha=-5$.  The solid lines are a cubic spline fit through the data points.}\label{fg:sigmaz}
\end{figure}

The $\Delta_8$ parameter provides a measure of the pulsar noise at a single time scale.  Full information about the spectral components of a data set are needed to characterise the timing noise in detail.  Unfortunately, both the irregular data sampling and the low frequency noise process make power spectral determination difficult (however see \S\ref{sec:pspec}).  Matsakis, Taylor \& Eubanks (1997)\nocite{mte97} generalised the Allan variance (traditionally used for measuring clock stability) to obtain a statistic that provides a measure of pulsar stability at various time scales.  Their parameter,
\begin{equation}
\sigma_z(\tau) = \frac{\tau^2}{2\sqrt{5}}\langle c^2 \rangle^{1/2},
\end{equation}
where the magnitude, $c$, of cubic terms fitted to short sections of the dataspan (of length $\tau$) are averaged.   For timing residuals whose power spectral density can be modelled as
\begin{equation}
S(f) \propto f^{\alpha}
\end{equation}
then, for $\alpha < +1$, $\sigma_z(\tau)$  will also follow a power-law with
\begin{equation}
\sigma_z(\tau) \propto \tau^\beta 
\end{equation}
where $\beta = -(\alpha+3)/2$. Models of timing noise (e.g. D'Alessandro et al. 1995)\nocite{dmh+95} which assume that the pulsar spin is affected by a random walk in the pulse phase, frequency or slow-down rate produce a power-law for the power spectral density with $\alpha = -1$, $-3$ and $-5$ respectively (corresponding to $\beta = -1$, $0$ and $+1$).  Timing residuals not affected by timing noise will have a flat spectrum ($\alpha = 0$) and hence $\beta = -3/2$. In Figure~\ref{fg:sigmaz} we show representative $\sigma_z(\tau)$ plots. As expected for PSR~B0031$-$07, whose residuals are dominated by TOA uncertainties, $\sigma_z(\tau)$ has a gradient of $\beta = -3/2$ (dotted line) with the possibility of a turn-over corresponding to small-amplitude timing noise at large time scales ($> 10$\,yr).  The timing residuals for PSR~B0628$-$28 were reported in D'Alessandro et al. (1995) as being consistent with phase noise ($\beta = -1$).  However, our results show that $\sigma_z(\tau)$ decreases with $\beta = -1$ until a time-scale around 2\,yr before increasing with $\beta = +1$ (or $\alpha = -5$) corresponding, in the idealised model, to a random walk in $\ddot{\nu}$.  $\sigma_z(\tau)$ for PSR~B1642$-$03 has a more complex oscillatory form.  We therefore confirm the conclusion of Cordes \& Downs (1985) that the timing residuals can not be modelled by a simple random walk process.

\begin{figure*}
\includegraphics[width=5cm,angle=-90]{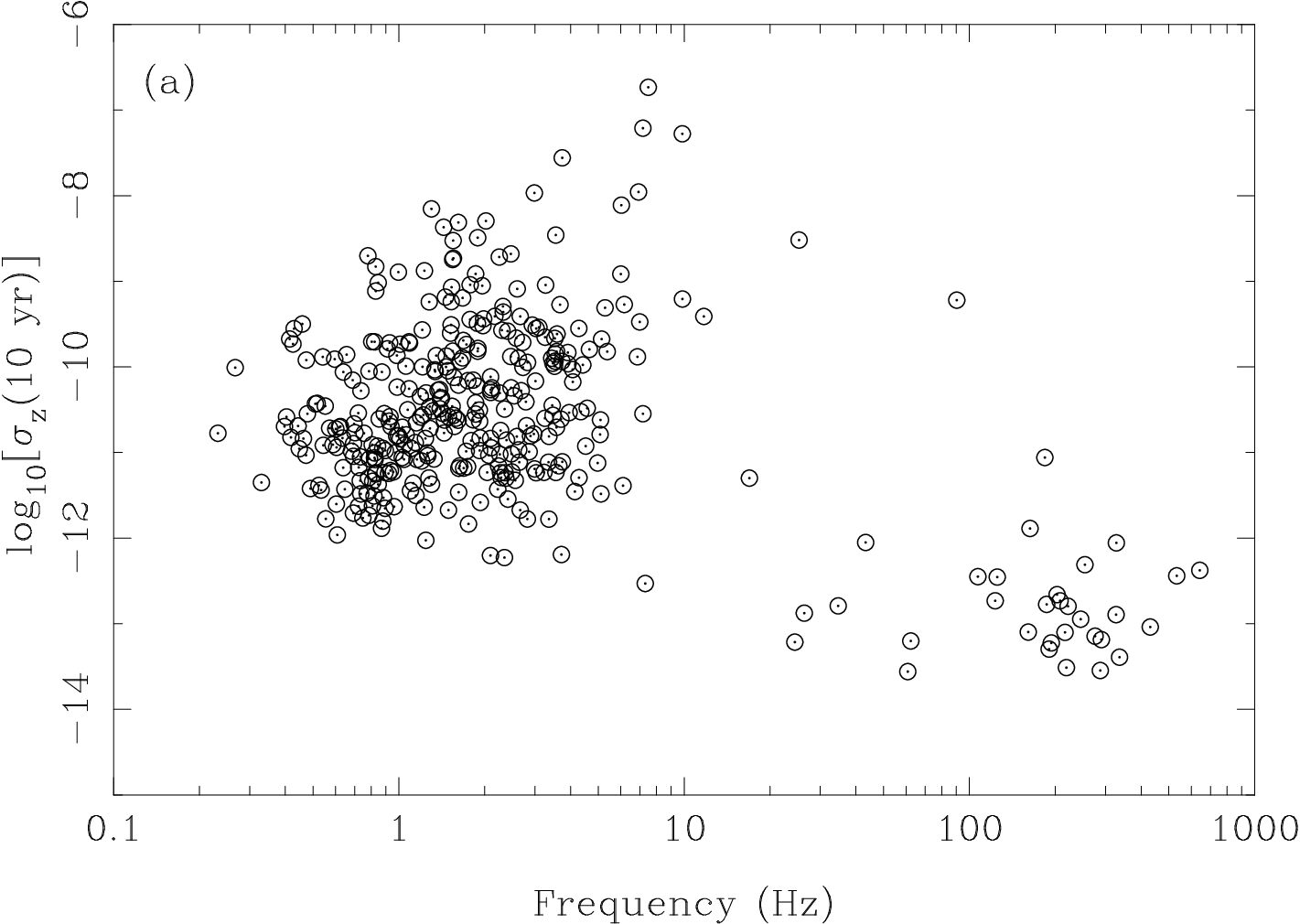}
\includegraphics[width=5cm,angle=-90]{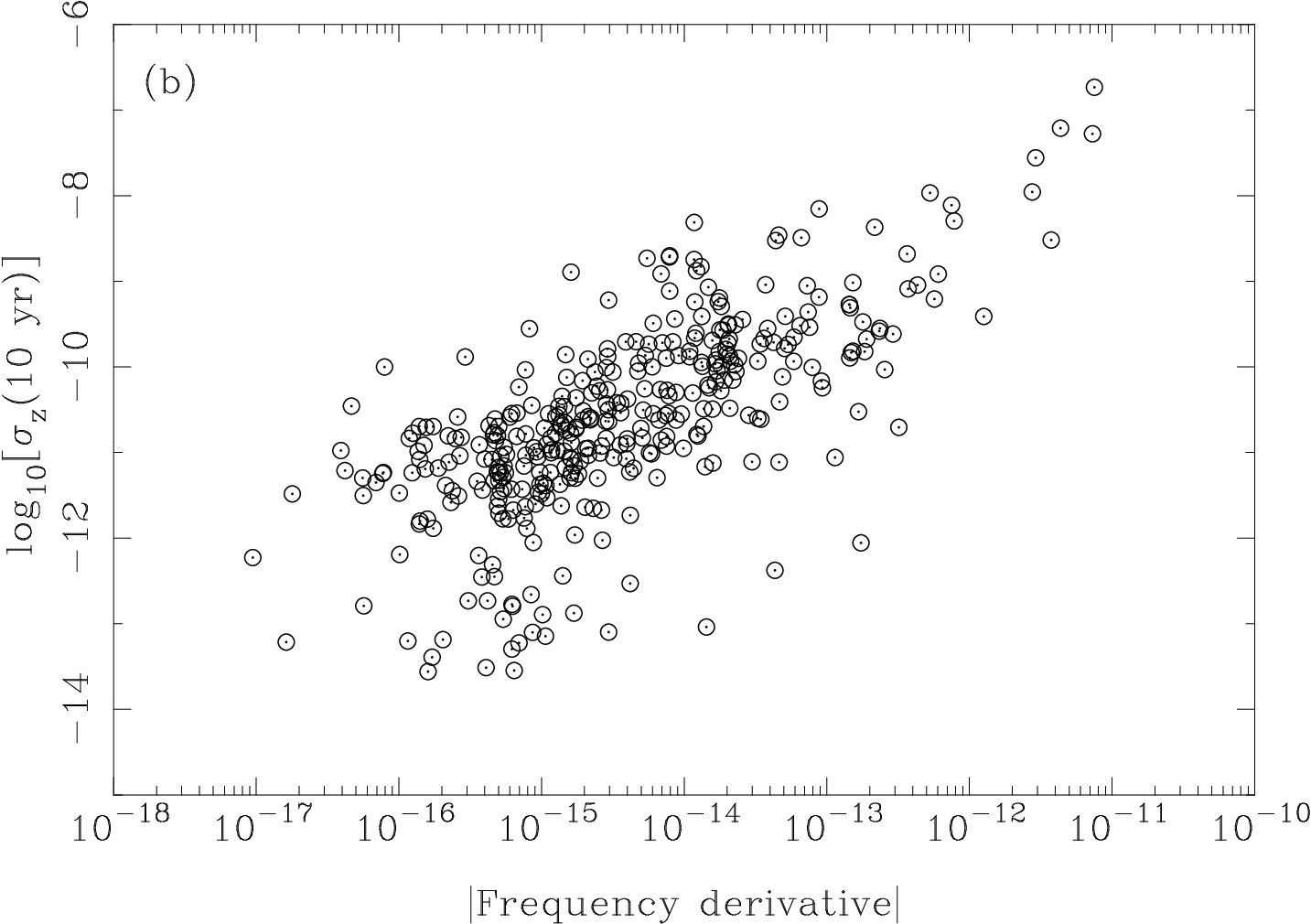}
\includegraphics[width=5cm,angle=-90]{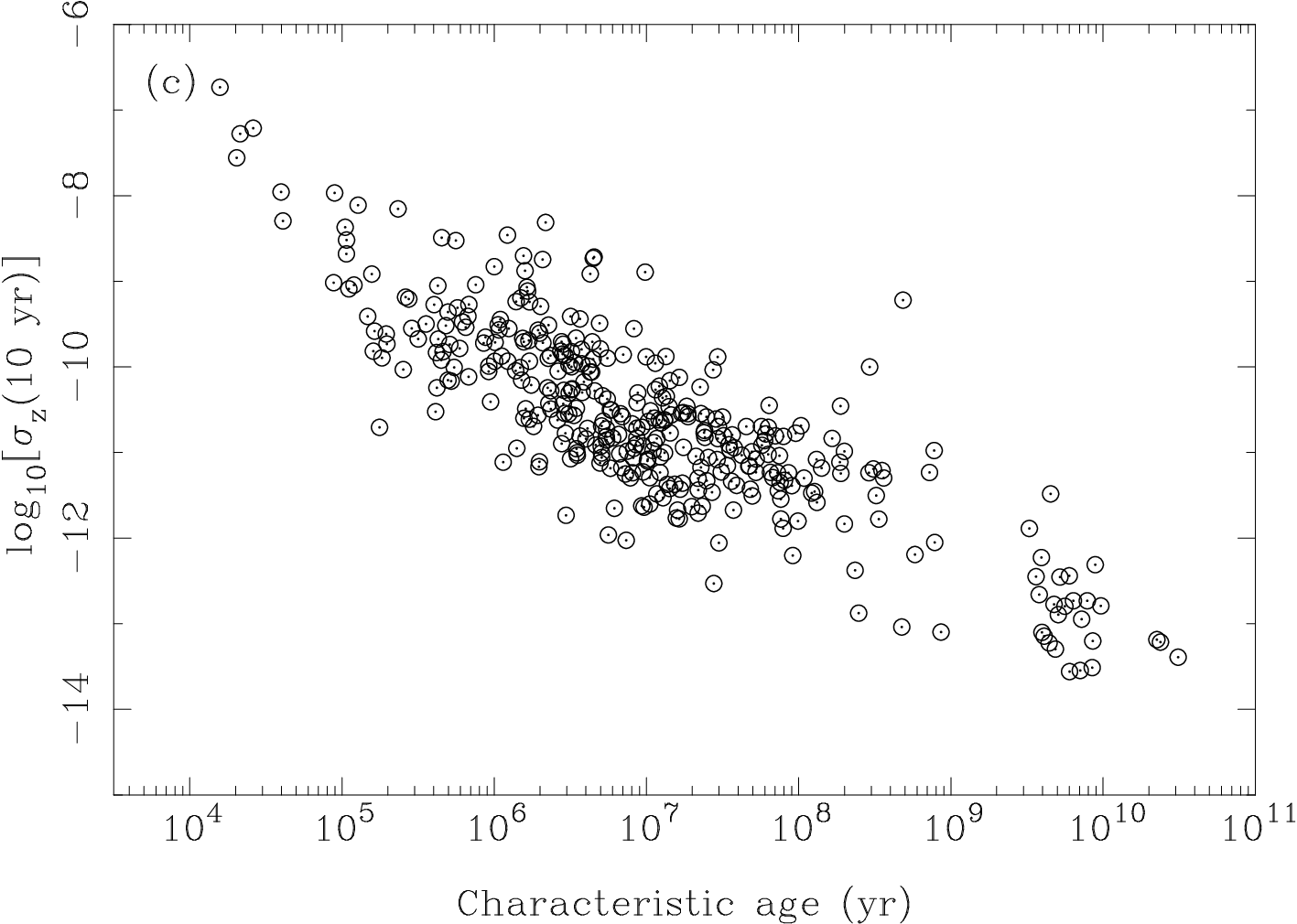}
\includegraphics[width=5cm,angle=-90]{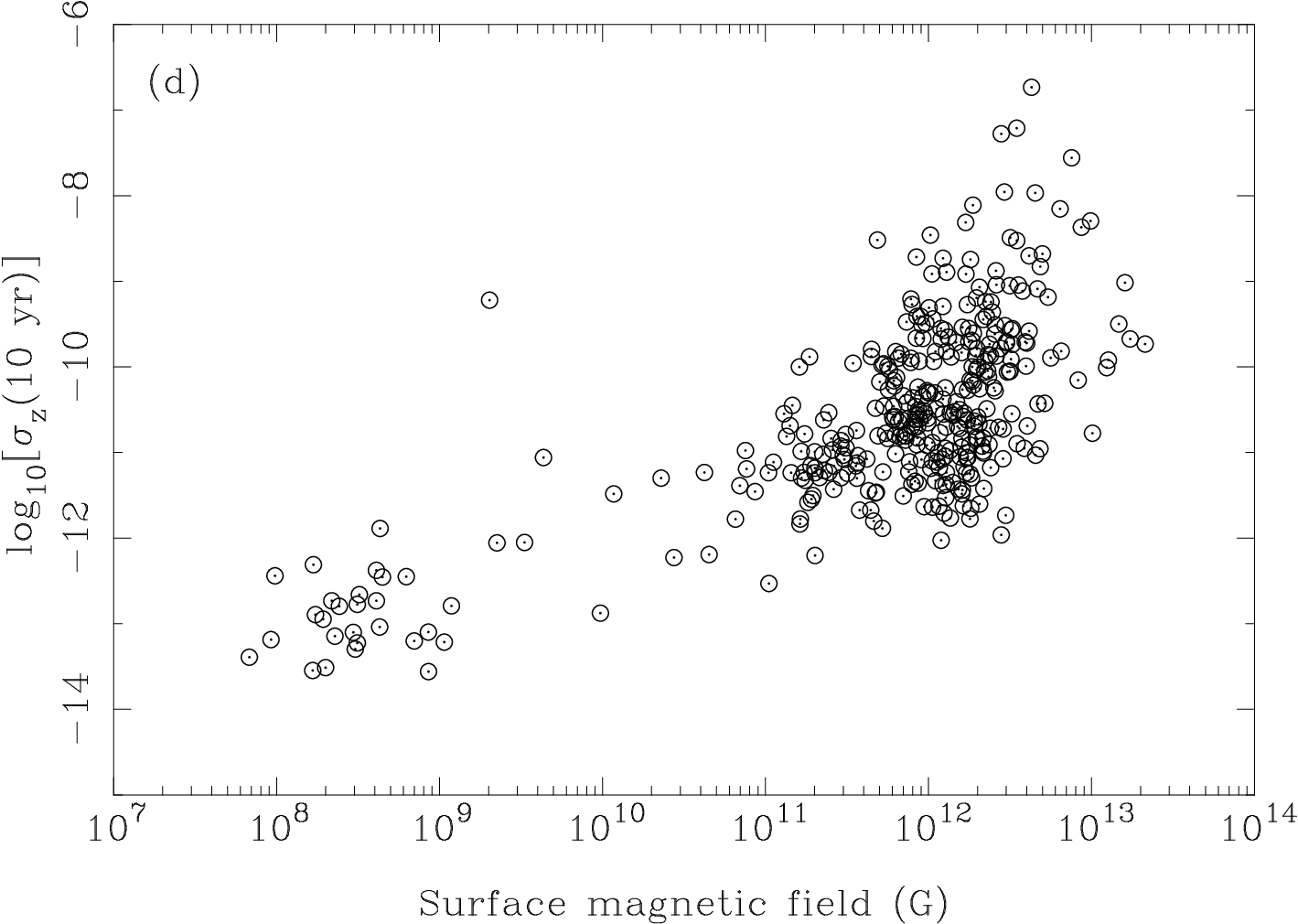}
\includegraphics[width=5cm,angle=-90]{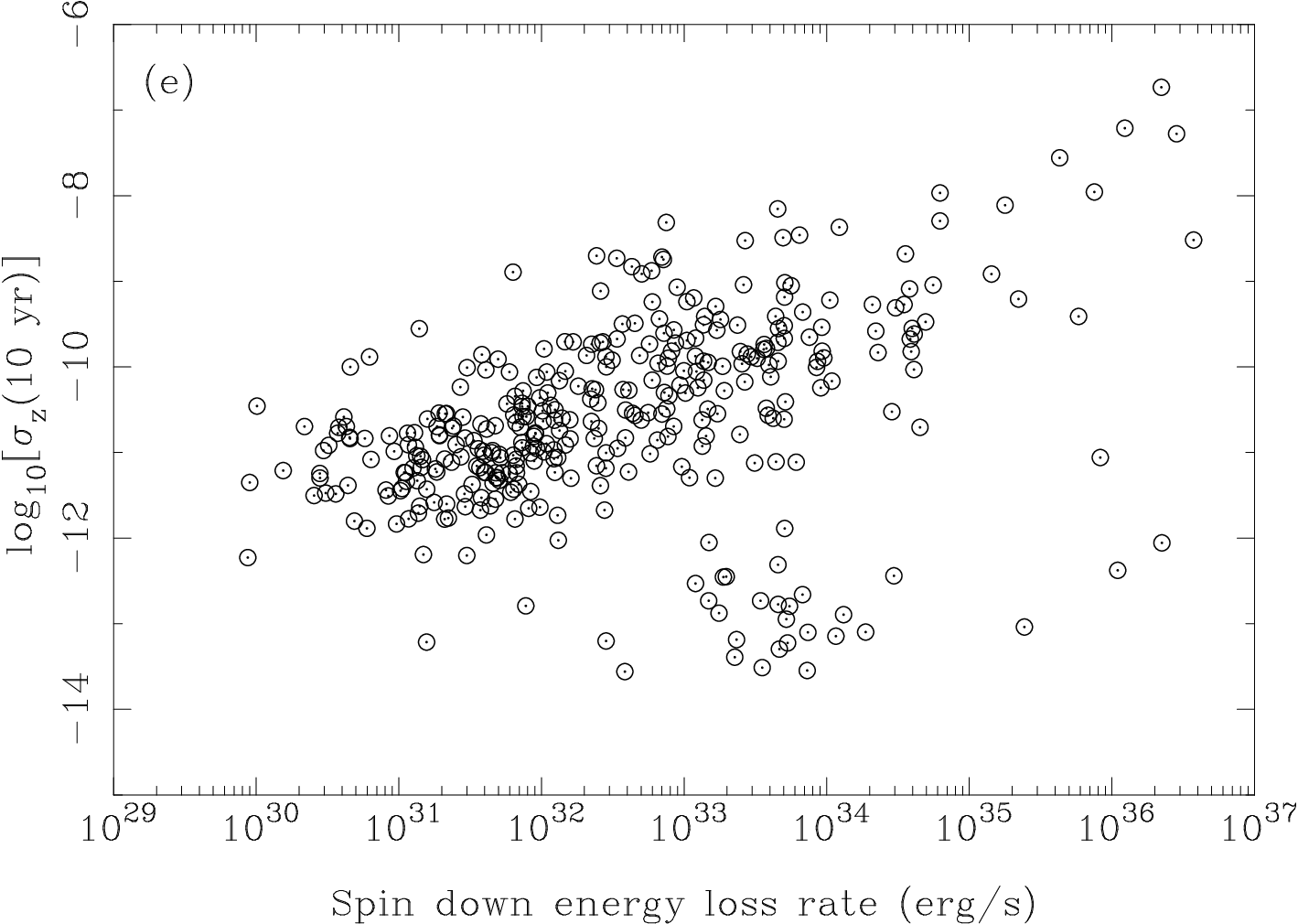}
\includegraphics[width=5cm,angle=-90]{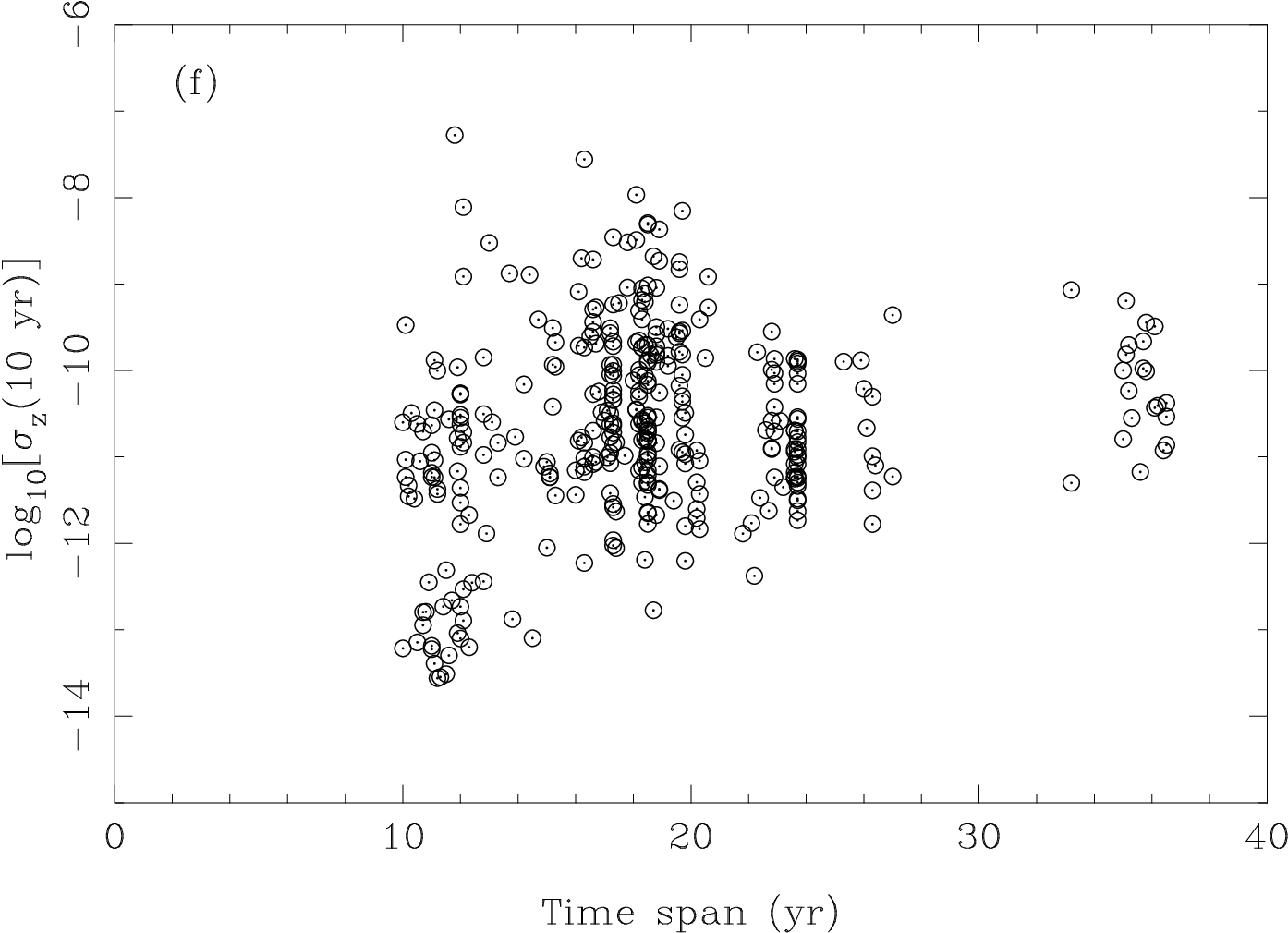}
\caption{$\sigma_z({\rm 10 yr})$ versus (a) pulse frequency, (b) frequency deriative, (c) characteristic age, (d) surface magnetic field strength, (e) energy loss rate and (f) data span.}\label{fg:sz_params}
\end{figure*}

In order to search for correlations with the amount of timing noise and various pulsar parameters we tabulate, in Table~\ref{tb:params}, $\sigma_z(10{\rm yr})$ and plot this parameter in Figure~\ref{fg:sz_params}.  The following correlations are found:

\begin{itemize}
\item{$\sigma_z(10{\rm yr})$ versus $\nu$: the correlation coefficient measured using only the non-recycled pulsars is $\rho = 0.3$.  We therefore confirm the Cordes \& Helfand (1980) conclusion that timing noise is only weakly correlated with pulse period. We also highlight that the recycled pulsars significantly deviate from the trend given by the non-recycled pulsars.}

\item{$\sigma_z(10{\rm yr})$ versus $\dot{\nu}$: In this case $\rho = 0.76$ confirming that timing noise is strongly correlated with the pulse frequency derivative with pulsars having the fastest spin-down rate being more affected by timing noise than pulsars (such as the millisecond pulsars) with slower spin-down rates.}

\item{$\sigma_z(10{\rm yr})$ versus $\tau_c$: The strong anti-correlation with $\dot{\nu}$ and weak correlation with $\nu$ implies that a large anti-correlation exists between timing noise and characteristic age ($\rho = -0.76$);  pulsars with small characteristic ages exhibit more timing noise than older pulsars. We note that the amount of timing noise in millisecond pulsars seems to follow this correlation.}

\item{$\sigma_z(10{\rm yr})$ versus B$_s$: A weak correlation ($\rho = 0.50$) exists with the surface magnetic field strength.  A significant outlying point has $B_s \sim 2\times 10^9$\,G and corresponds to the triple pulsar system PSR~B1620$-$26 in the globular cluster M4.  The high $\sigma_z$ value can be explained as being caused by the acceleration of the pulsar in the cluster's gravitational field.  The timing residuals for this pulsar have been studied by numerous authors (e.g. Thorsett et al., 1999\nocite{tacl99}, Ford et al. 2000\nocite{fjrz00} and Beer, King \& Pringle 2004)\nocite{bkp04}.}

\item{$\sigma_z(10{\rm yr})$ versus $\dot{E}$: A strong anti-correlation ($\rho = -0.71$) exists with the spin-down energy loss rate $\dot{E} \propto \dot{P}/P \propto \dot{\nu}\nu$.}

\item{$\log_{10} [\sigma_z(10{\rm yr})]$ versus time span: No correlation is found between the pulsar stability at 10\,yr and the total observational period ($\rho = -0.03$).}
\end{itemize}

\begin{figure}
\includegraphics[width=6.5cm,angle=-90]{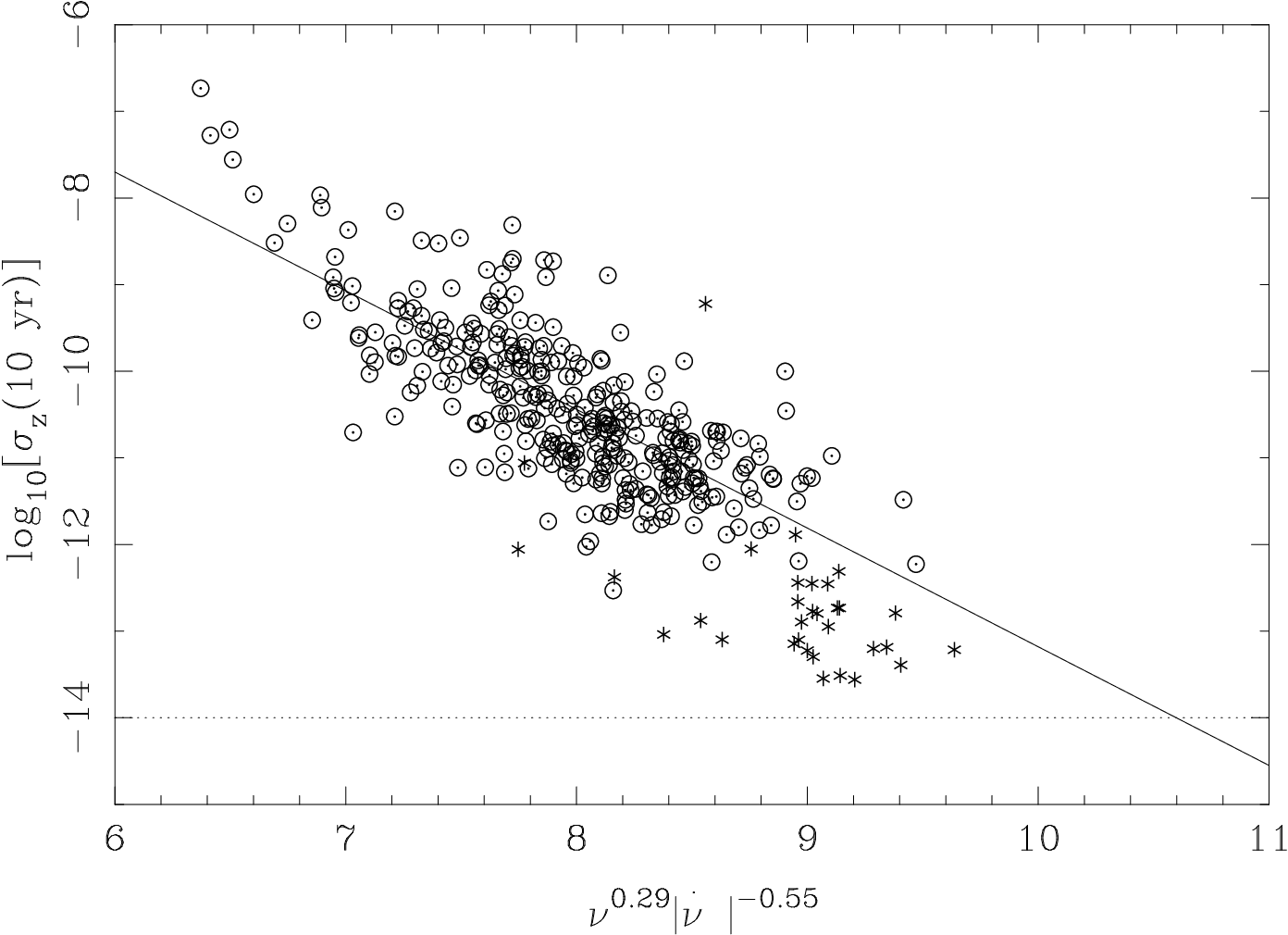}
\caption{$\sigma_z(10{\rm yr})$ is plotted versus $\nu^\alpha |\dot{\nu}|^\beta$ where $\alpha$ and $\beta$ were chosen to produce the highest correlation coefficient for the non-recycled pulsars (circles).  The recycled pulsars are overlaid (star symbols).  The horizontal dotted line is the predicted level of a gravitational wave background.}\label{fg:predGWB}
\end{figure}

In order to predict the amount of timing noise for a given pulsar we searched through multiple combinations of $\nu^\alpha \dot{\nu}^\beta$ and obtained a maximum correlation coefficient of $\rho = 0.77$ for 
\begin{equation}
\log_{10}[\sigma_z(10{\rm yr})] = -1.37 \log_{10}[\nu^{0.29}|\dot{\nu}|^{-0.55}] + 0.52.
\end{equation}
This expression\footnote{Note that $\nu^{0.29}|\dot{\nu}|^{-0.55}$ can be written in terms of the characteristic age and pulse period as $\sim (P\tau_c)^{0.3}$.}, plotted in Figure~\ref{fg:predGWB}, allows the prediction the amount of intrinsic timing noise over a 10\,yr time scale for any pulsar. In order for pulsar timing array experiments to be sensitive to gravitational wave signals it is necessary for $\sigma_z (10\,{\rm yr}) \lesssim 10^{-14}$ (e.g. Hobbs et al. 2009)\nocite{hbb+09}.  On a 10\,yr timescale our results suggest that the residuals for the majority of the millisecond pulsars in our sample will be dominated by timing noise and not by the residuals induced by the gravitational wave background (assuming that such a background exists and that the expression above is valid for millisecond pulsars).  However, there are a few pulsars where the timing noise is at a much lower level. For instance, the timing residuals for PSR~J0437$-$4715 have an rms of $\sim 200$\,ns over a ten-year time scale (Verbiest et al. 2008)\nocite{vbv+08}.  A study of the timing noise for the most stable millisecond pulsars is currently being undertaken and will be published elsewhere.

%\begin{figure}
%\includegraphics[width=6.5cm,angle=-90]{0628-28.ps}
%\caption{$\sigma_z(\tau)$ for PSRs~B0628$-$28.}\label{fg:sigmaz_0628}
%\end{figure}

\subsubsection{Significant $\ddot{\nu}$ values}

The residuals for pulsars whose residuals are dominated by a significant $\ddot{\nu}$ value take the form of a cubic polynomial.  For a pulsar slowing down by magnetic dipole braking then the value of $\ddot{\nu}$ is expected to be 
\begin{equation}
\ddot{\nu} = \frac{n\dot{\nu}^2}{\nu}
\end{equation}
where the braking index, $n=3$.  The measured braking indices for the non-recycled pulsars range from $-$287986 to $+$36246 with a mean of $-$1713 and median of 22.  If we restrict our sample to pulsars where we have data spanning more than 30 years we obtain braking indices ranging from $-1701$ to $+36246$ with a mean of $+3750$ and median of $+29$.  For the simple model of magnetic dipole braking then the $\ddot{\nu}$ value will be positive.  Out of our sample of 366 pulsars, 193 (53\%) have a positive $\ddot{\nu}$ value and the remaining 173 (47\%) have negative $\ddot{\nu}$ (we obtain similar statistics if we restrict our sample to only those pulsars with significantly measured $\ddot{\nu}$ values).  As concluded in H04, the observed $\ddot{\nu}$ values for the majority of pulsars are not caused by magnetic dipole radiation nor any other systematic loss of rotational energy, but are dominated by the amount of timing noise present in the residuals and the data span.  

Johnston \& Galloway (1999, JG99)\nocite{jg99} derived a new method for determining pulsar braking indices which only requires measurements of $\nu$ and $\dot{\nu}$.  Our sample contains 17 pulsars that are in common with the JG99 sample.  In general our braking indices are not consistent with those in JG99. To understand the reason for this inconsistency we consider PSR~B0740$-$28 for which we have a $\sim$21\,yr data-span.  Our measurement for $n$ is inconsistent with JG99 who obtain a positive value $n = 25.6(8)$.  With our entire data-span we obtain $n=-10.0(7)$. If we select a shorter data-set between MJDs 46690 and 51162 we obtain $n = -65.7(7)$ and between MJDs 48892 and 53451 we get $n=+68.5(5)$. If we measure the pulse parameters from both the earliest 3\,yr of data and the most recent then we can use the same methodology as JG99.  Using this method we obtain that $n = 18.5(3)$. A similar analysis using $\sim 200$\,d of data gives $n = -48(2)$. The work of Alpar \& Baykal (2006)\nocite{ab06} was based upon JG99 and concluded that many pulsars with anomalous positive $\ddot{\nu}$ measurements obey a scaling between $\ddot{\nu}$ and glitch parameters.  They also reported that negative $\ddot{\nu}$ values can be understood in terms of glitches that were missed or unresolved.  However, as our results are not consistent with those analysed in the JG99 paper, the Alpar \& Baykal (2006)\nocite{ab06} and JG99 conclusions need to be treated with caution.

\begin{figure}
\includegraphics[angle=-90,width=8cm]{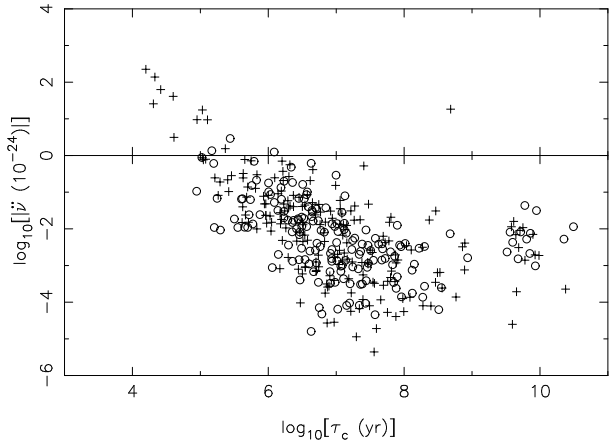}
\caption{The measured $\ddot{\nu}$ values versus characteristic ages. Pulsars with $\ddot{\nu}>0$ are indicated using `plus' signs, with `circles' for those with $\ddot{\nu}<0$.}\label{fg:nuddot_age}
\end{figure}

Plotting the $\ddot{\nu}$ values versus characteristic age (Figure~\ref{fg:nuddot_age} allows us to identify that the six youngest pulsars in our sample all have $\ddot{\nu} > 0$ (as indicated with `plus' symbols on the diagram).  Glitch events prior to the first observation will induce $\ddot{\nu} > 0$ (e.g. Lyne, Shemar \& Graham-Smith 2000\nocite{lsg00}).  Our data are therefore consistent with the timing residuals for all pulsars younger than $\tau_c \sim 10^5$\,yr being dominated by the recovery from a glitch (or from glitch events that occurred during the observation span).  For the very youngest pulsars it seems that glitch activity is small (Shemar \& Lyne 1996) and any timing noise is dominated by magnetic braking.  Hence, for these pulsars braking indices close to 3 are expected and observed. For pulsars with characteristic ages $> 10^5$, 52\% have $\ddot{\nu} > 0$ and the remainder (48\%) have $\ddot{\nu} < 0$. As almost equal numbers of pulsars have each sign of $\ddot{\nu}$ we suggest that the timing noise seen in older pulsars is not caused by glitch recovery nor magnetic braking, but from some other process.

\begin{figure}
\includegraphics[angle=-90,width=8cm]{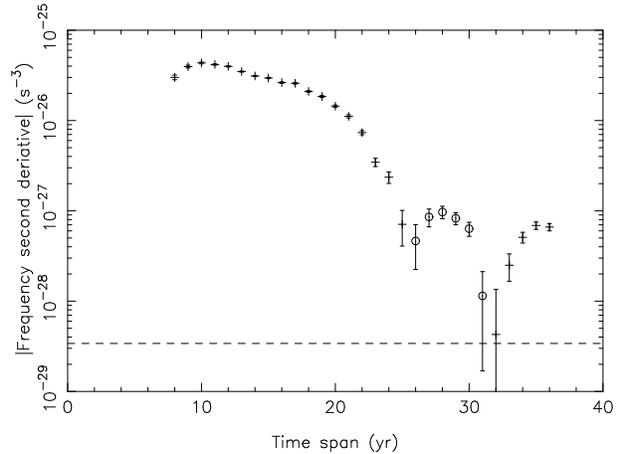}
\caption{Values of $|\ddot{\nu}|$ obtained using differing data spans for PSR~B0329$+$54. The circles indicate $\ddot{\nu} < 0$ and the cross symbols otherwise.   The horizontal dashed line indicates $\ddot{\nu} = 3.4\times10^{-29}$, the value expected if the spin-down is dominated by magnetic dipole braking.}\label{fg:0329_cubic}
\end{figure}

 In the timing residuals that span many years we see many fewer cubic structures (corresponding to large, significant $\ddot{\nu}$ values) than for shorter data spans.  In Figure~\ref{fg:0329_cubic} we plot $|\ddot{\nu}|$ values obtained from the PSR~B0329$+$59 data set with various data spans.  For data spanning $\sim 10$\,yr we measure a large, significant $\ddot{\nu}$ value and, hence, the timing residuals take the form of a cubic polynomial.  For data spanning more than $\sim 25$\,yr we see no significant cubic term.  The braking index measured from the $\ddot{\nu}$ value with the longest data span of $n = 75$ is still significantly greater than that expected from magnetic dipole braking.

For the pulsars dominated by a significant $\ddot{\nu}$ we plot, in Figure~\ref{fg:removeCubic}, the timing residuals after a cubic polynomial has been fitted and removed. In the presence of a steep red-noise spectrum we would expect the resulting residuals to be dominated by a quartic polynomial. However, in most cases we see significantly more quasi-periodic structures (good examples being PSRs~B1540$-$06, B1826$-$17, B1828$-$11 and B1853$+$01).  The recycled pulsars, PSR~B1620$-$26 and B1820$-$30A are unusual in that they are the only recycled pulsars with large frequency derivatives. They both lie close to the cores of globular clusters (M4 and NGC6624 respectively) and the significant $\ddot{\nu}$ values can therefore be explained as being caused by the acceleration of the pulsar in the gravitational field of the cluster as a whole or of neighbouring stars.   

\begin{figure*}
\includegraphics[width=18cm]{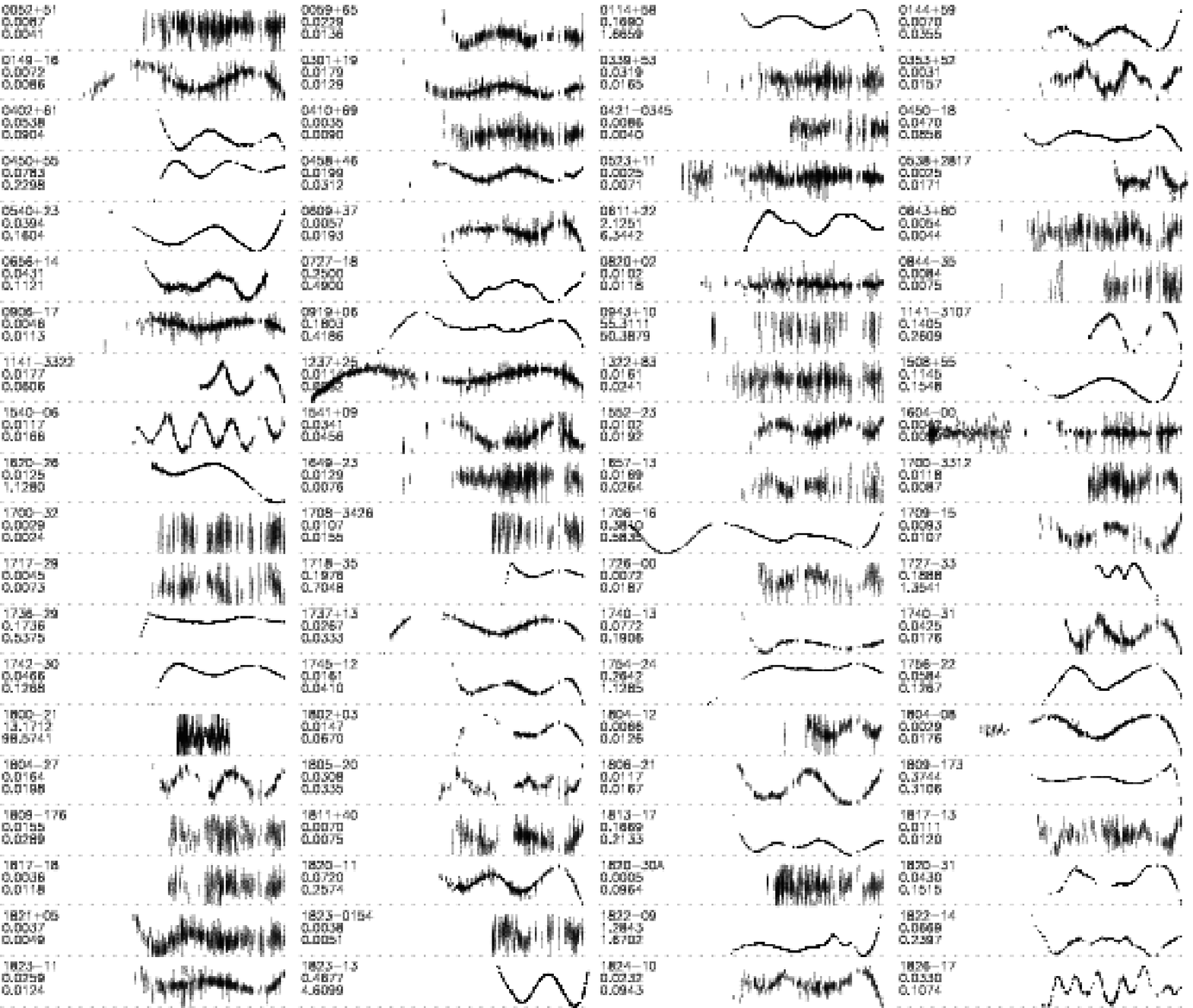}
\caption{Pulsar timing residuals after the removal of a cubic term.}\label{fg:removeCubic}
\end{figure*}

\addtocounter{figure}{-1}
\begin{figure*}
\includegraphics[width=18cm]{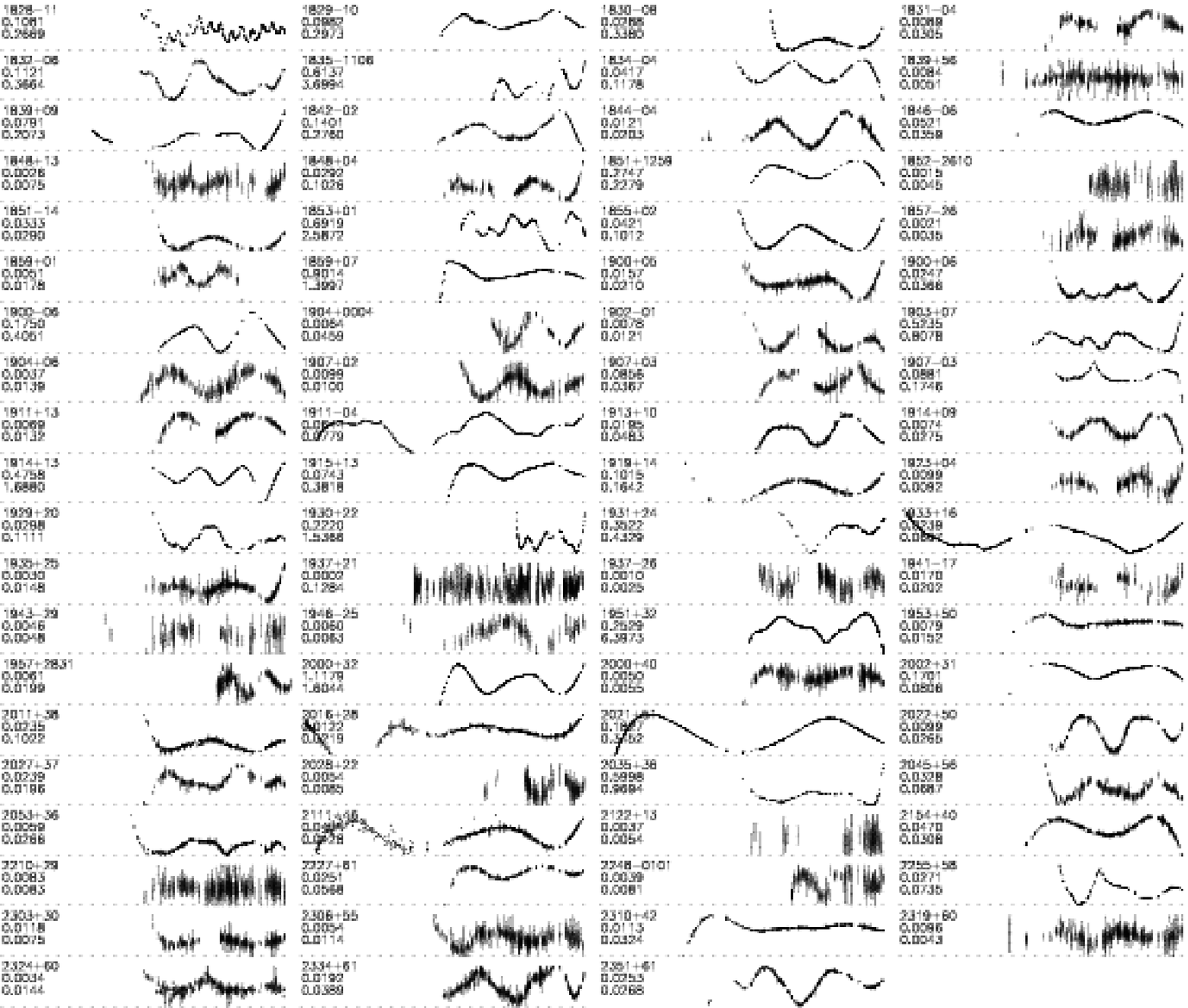}
\caption{\ldots continued}
\end{figure*}

\subsubsection{Periodicities}\label{sec:pspec}

\begin{figure*}
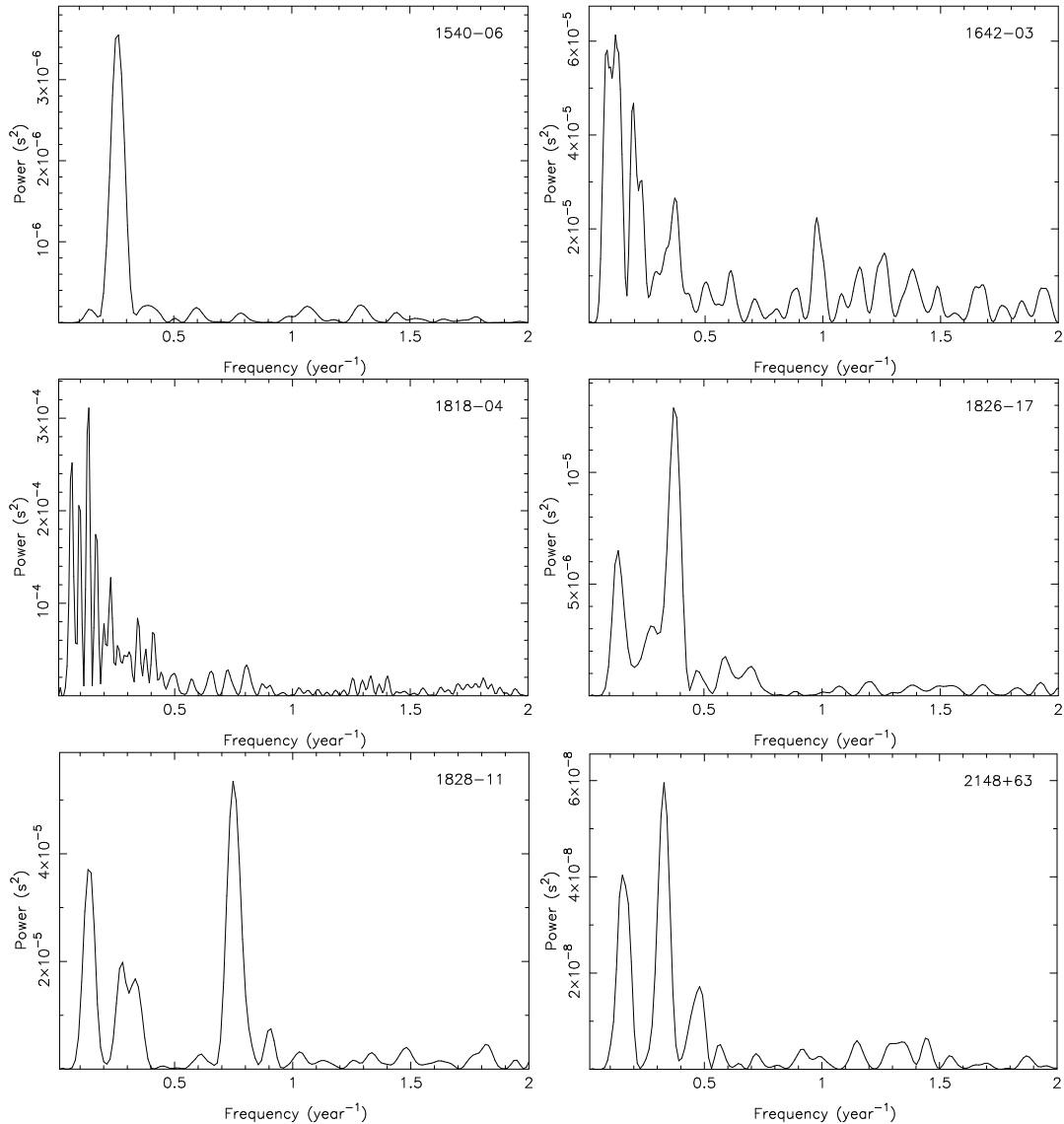

\includegraphics[width=5cm,angle=-90]{1540-06_spec.ps}
\includegraphics[width=5cm,angle=-90]{1642-03_spec.ps}
\includegraphics[width=5cm,angle=-90]{1818-04_spec.ps}
\includegraphics[width=5cm,angle=-90]{1826-17_spec.ps}
\includegraphics[width=5cm,angle=-90]{1828-11_spec.ps}
\includegraphics[width=5cm,angle=-90]{2148+63_spec.ps}
\caption{Power spectra of PSRs~B1540$-$06, B1642$-$03, B1818$-$04, B1826$-$17, B1828$-$11 and B2148$+$63.}\label{fg:pspec}
\end{figure*}

For each of our pulsars we have obtained a power spectrum.  A detailed description of our method for obtaining power spectra in the presence of steep red-noise and irregularly sampled data is beyond the scope of this paper and will be presented elsewhere.  Here we concentrate on those pulsars (PSRs~B1540$-$06, B1642$-$03, B1818$-$04, B1826$-$17, B1828$-$11 and B2148$+$63) which show significant periodicities or quasi-periodic structures.  A Lomb-Scargle power spectrum that has been oversampled by a factor of four (Press et al. 1986)\nocite{pftv86} for each of these pulsars is shown in Figure~\ref{fg:pspec} after whitening the timing residuals by fitting a $\ddot{\nu}$ term. 

\begin{figure*}
\includegraphics[width=5cm,angle=-90]{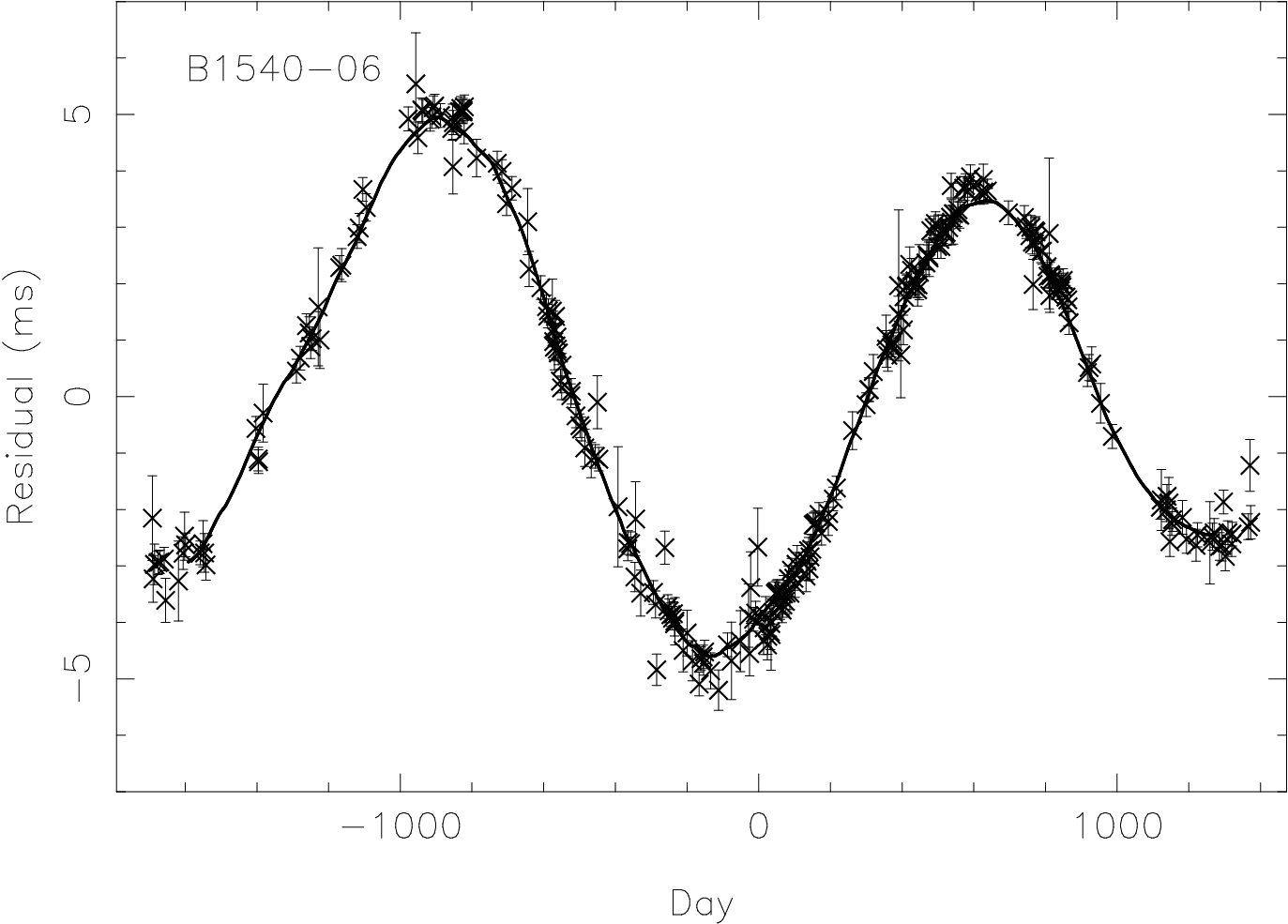}
\includegraphics[width=5cm,angle=-90]{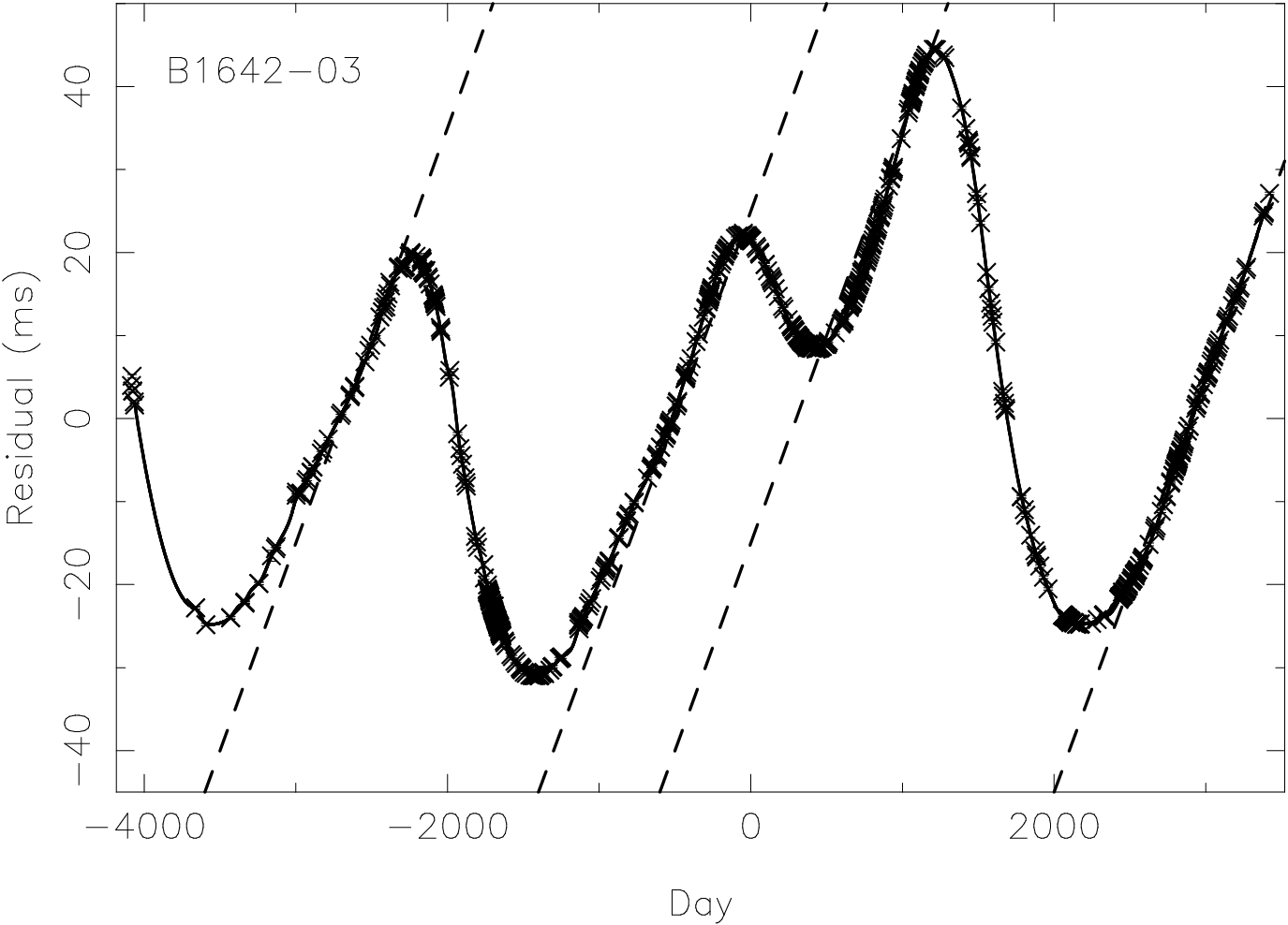}
\includegraphics[width=5cm,angle=-90]{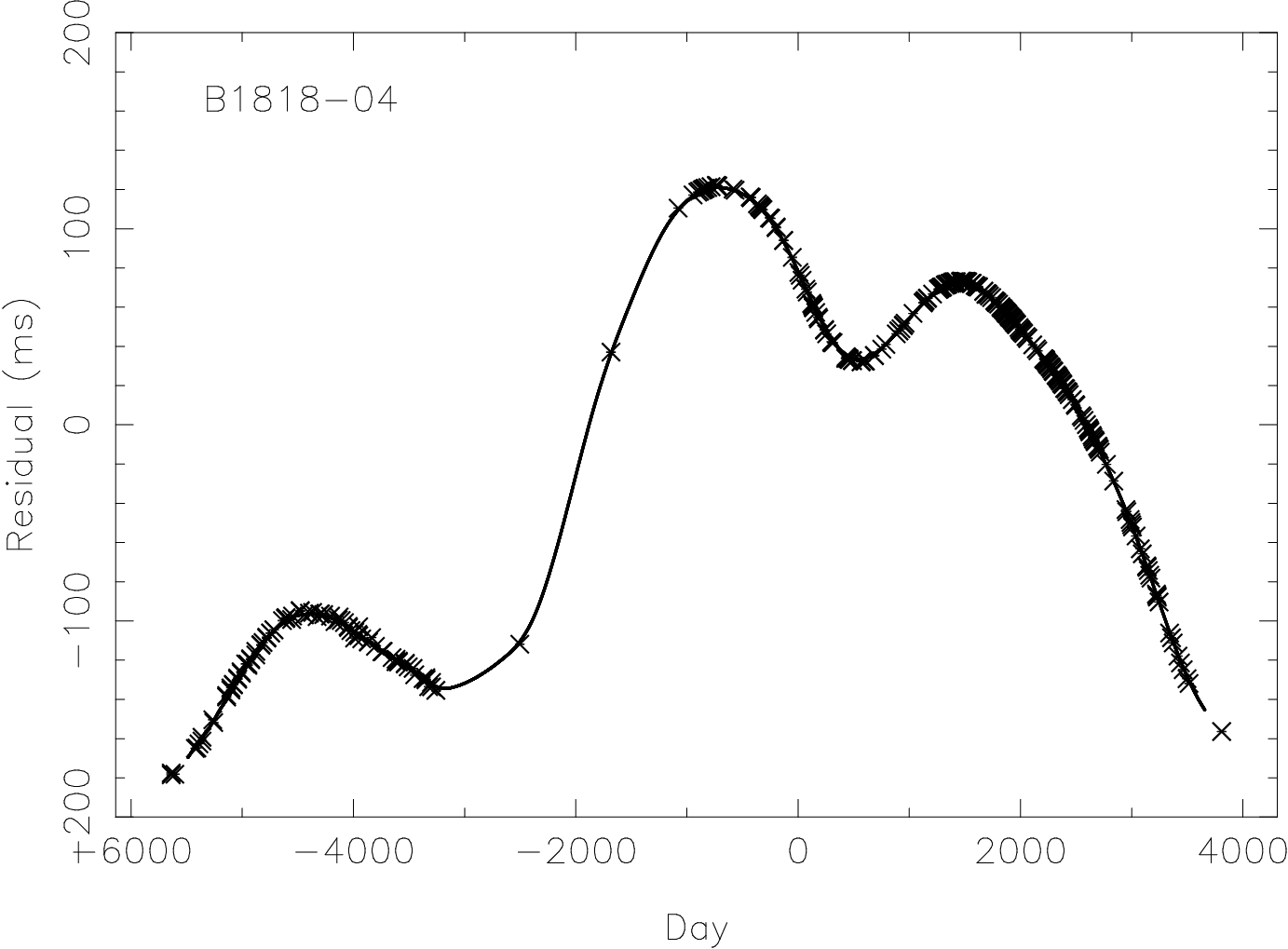}
\includegraphics[width=5cm,angle=-90]{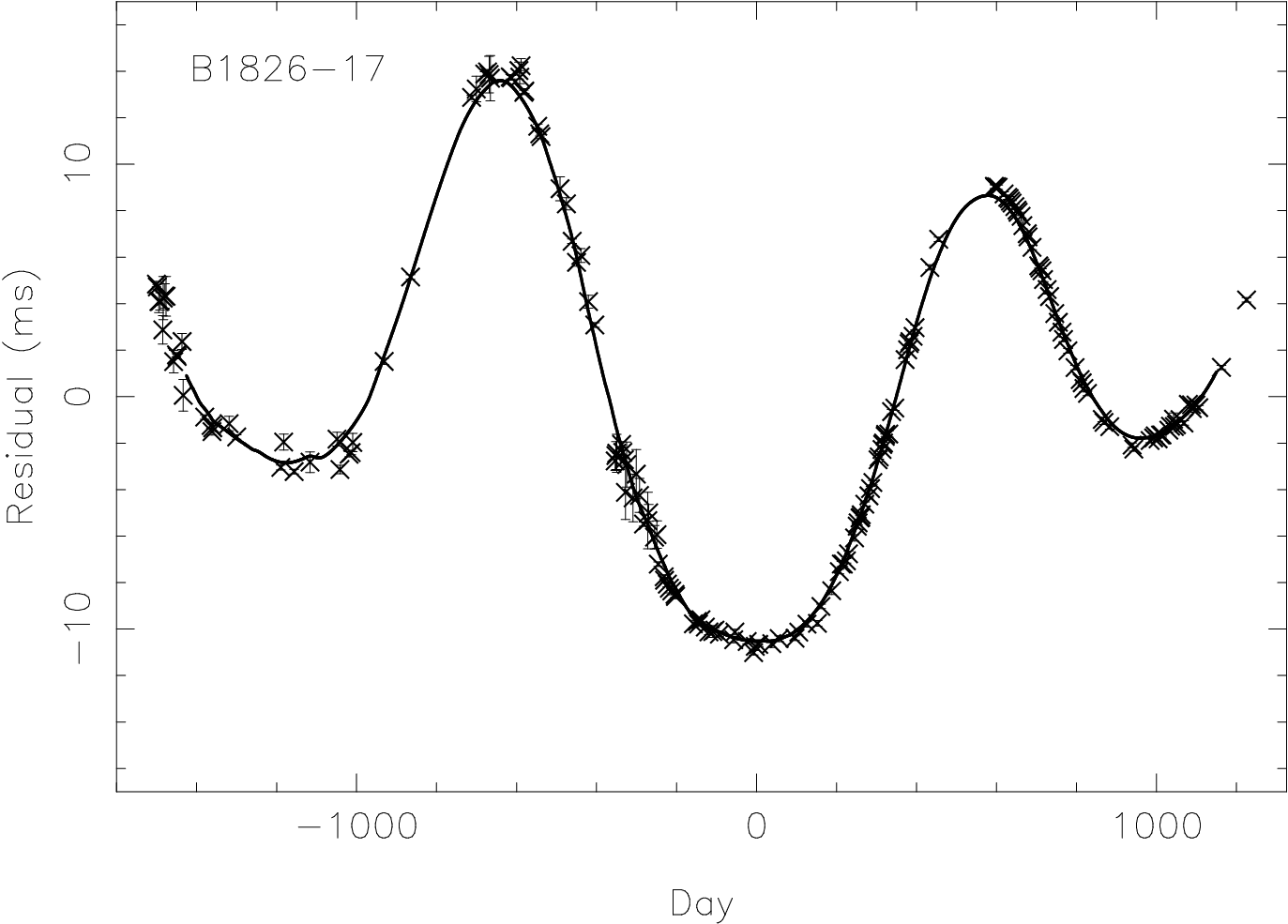}
\includegraphics[width=5cm,angle=-90]{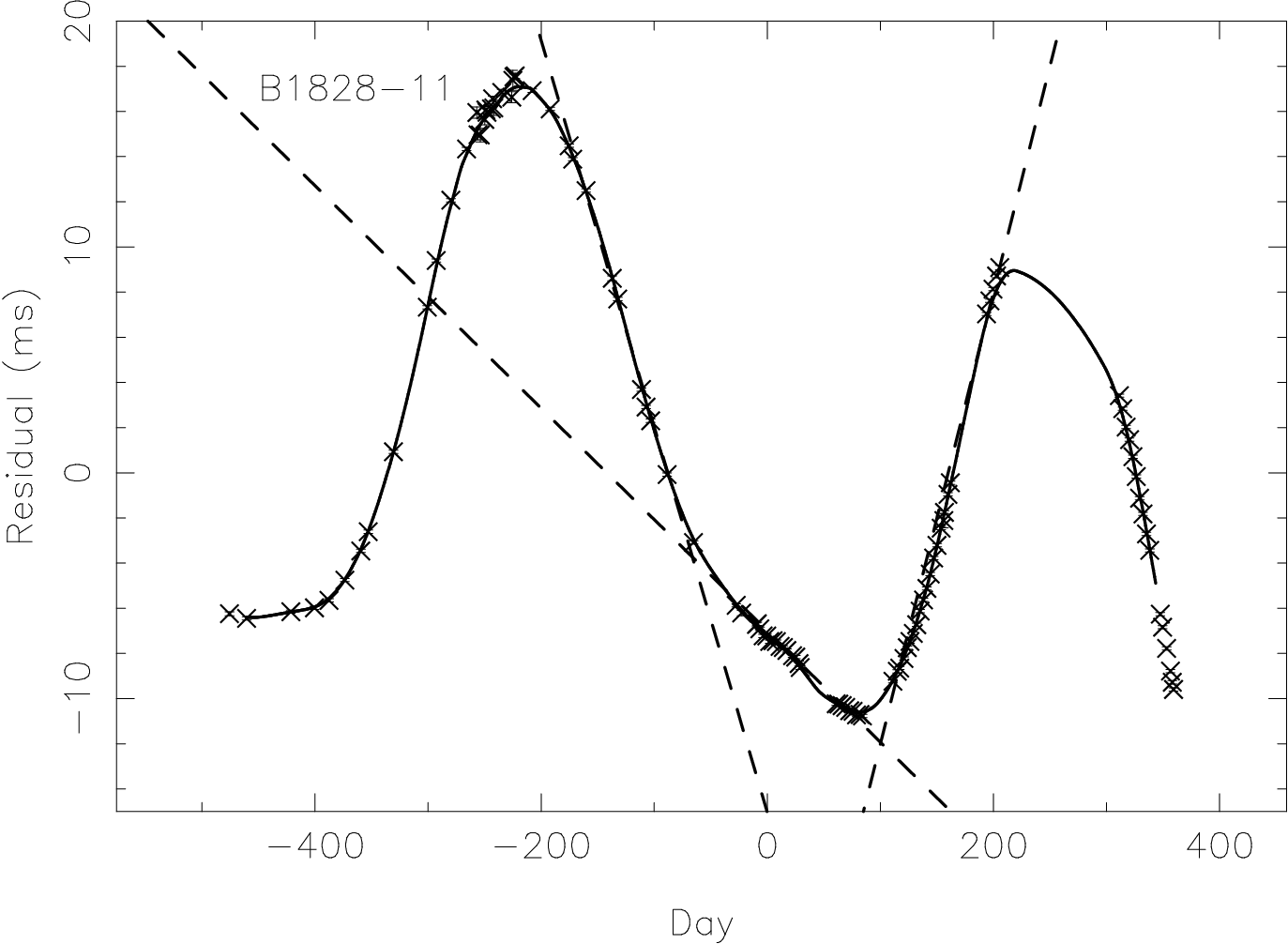}
\includegraphics[width=5cm,angle=-90]{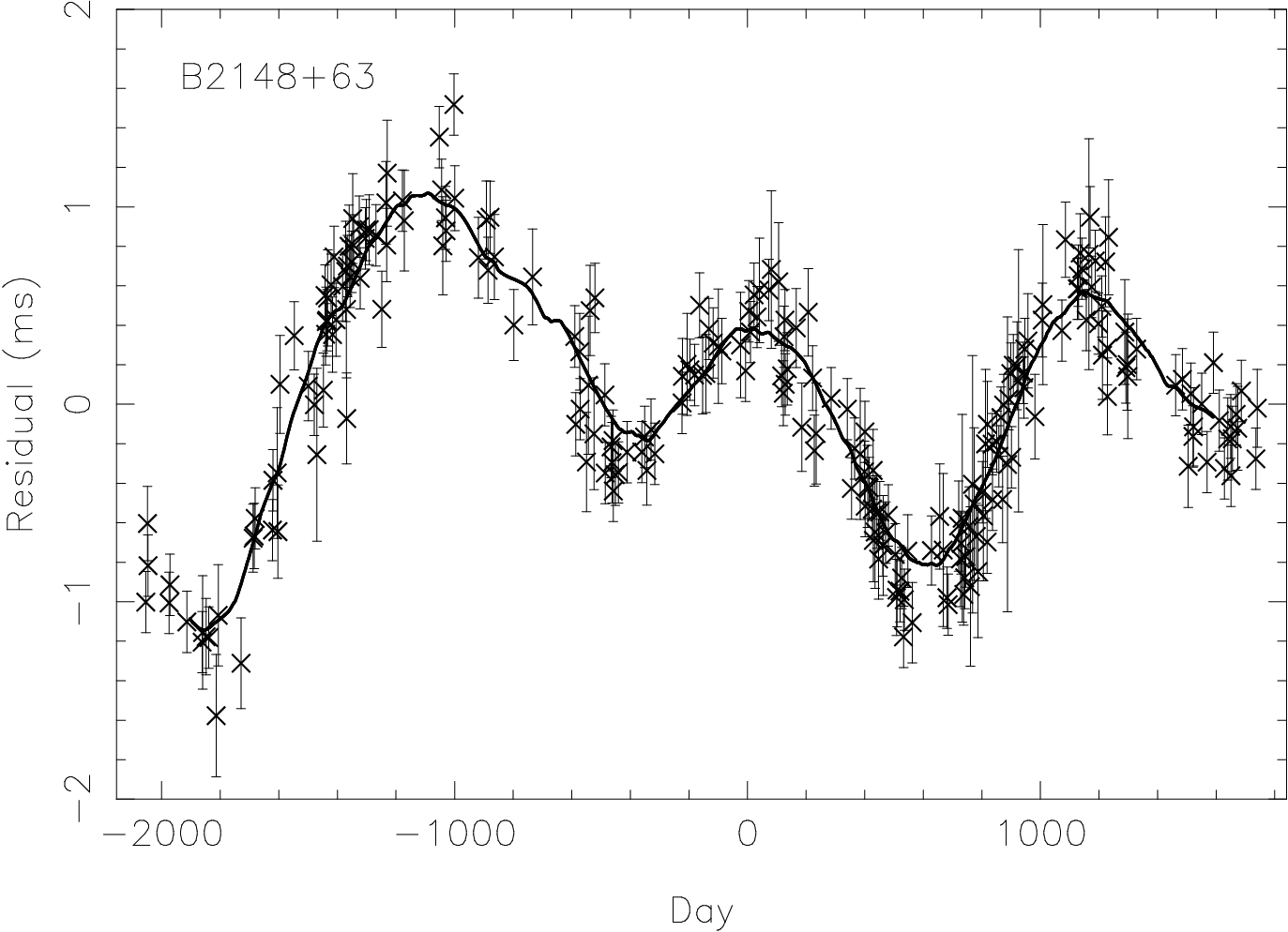}
\caption{Short sections of the timing residuals for PSRs~B1540$-$06, B1642$-$03, B1818$-$04, B1826$-$17, B1828$-$11 and B2148$+$63. A smoothed, constrained cubic spline curve is plotted through the data points.}\label{fg:short}
\end{figure*}

\begin{itemize}
\item{{\bf PSR B1540$-$06:} A significant periodicity is observed with a period of 4.38\,yr.   It is possible that the periodic signal is caused by the orbital motion of an unmodelled Earth-mass planetary companion. Including such a companion in the timing fit indicates an orbital period of $P_b = 1530(3)$\,days, a projected semi-major axis of $a \sin i = 0.00393(7)$ and an epoch of periastron of $T_0 = 48992(5)$.  The rms timing residual decreases from 7.0\,ms without subtracting a cubic term or binary companion, 3.0\,ms after subtracting the cubic term alone and 1.1\,ms after subtracting the cubic and binary parameters. However, significant structure remains in the timing residuals after the removal of these terms and we do not consider this as compelling evidence for a planetary companion. In the top-left panel of Figure~\ref{fg:short} we show one oscillation of the timing residuals. For this pulsar the radii of curvature (and hence the magnitude of the local value of $\dot{\nu}$) at the maxima and minima are similar.}

\item{{\bf PSR B1642$-$03:} The timing residuals for this pulsar show a clear quasi-periodic structure.  The time between successive peaks ranges from 3.4\,yr to 6.6\,yr.  In Figure~\ref{fg:short} we show three local maxima and overplot dashed lines with a gradient of  $+5 \times 10^{-5}$\,s\,day$^{-1}$.  We note that the radius of curvature is smaller at local maxima than minima suggesting that the residuals between a local maxima and the next minima can be modelled simply as two discrete $\dot{\nu}$ values.  This pulsar therefore has three separate slow down rates.  The quasi-periodic nature of the timing residuals implies that the power spectrum (Figure~\ref{fg:pspec}) does not show a single periodicity, but rather multiple low frequency components (which we consider as evidence against the free-precession model of Shabanova, Lyne \& Urama 2001)\nocite{slu01}.}

\item{{\bf PSR B1818$-$04:} The timing residuals shows a clear oscillation where the time between local maxima ranges between $\sim 7$ and $\sim 10$\,years.  The spectrum contains significant low frequency power, but no significant individual periodicities.}

\item{{\bf PSR B1826$-$17: }The power spectrum for this pulsar shows a significant periodicity of 2.9\,yr.  However, the time between successive peaks ranges from 2.4 to 3.2\,yr suggesting a variation in this periodicity of $\sim 10$\%.  It is clear from Figure~\ref{fg:short} that this pulsar is similar to PSR~B1642$-$03 where local maxima have a smaller radii of curvature than local minima.}

\item{{\bf PSR~B1828$-$11: } The timing residuals for PSR~B1828$-$11 are dominated by a significant frequency derivative. However, if this cubic polynomial term is fitted and removed then the timing residuals show the clear periodic structures reported by Stairs, Lyne \& Shemar (2000).  Previously these periodicities, and correlated pulse shape changes, have been explained by free-precession of the neutron star.  However, counter theoretical arguments have been put forward stating that the interior superfluid in the neutron star would damp out any free-precessional effects on a short time scale (Shaham 1977; Sedrakian,Wasserman \& Cordes 1999)\nocite{sha77}\nocite{swc99}.  Our spectral analysis agrees with that of Stairs, Lyne \& Shemar (2000) clearly showing a highly-stable significant periodicity at $\sim 500$\,d. A close-up study of one oscillation (Figure~\ref{fg:short}) shows three components plotted as dashed lines which suggests that the local minima are made up of two components whereas local maxima only have one component.}

\item{{\bf PSR B2148$+$63:} The timing noise is at a much lower level compared with the TOA uncertainties for this pulsar.  However, the periodogram shows clear periodicities with 3.2\,yr ($\sim 1200$\,d), 7.1\,yr and, at a lower level, 2.1\,yr.  In contrast to the residuals for PSR~B1642$-$03 and B1826$-$17 this pulsar exhibits larger radii of curvature at local maxima than at minima.}
\end{itemize}

For completeness we note that it has often been proposed that PSR~B0329$+$59 has planetary companions.  Demia\'nski \& Pr\'oszy\'nski (1979)\nocite{dp79} and Bailes, Lyne \& Shemar (1993)\nocite{bls93} found a three-year periodicity in the timing residuals. However, Shabanova (1995)\nocite{sha95} found no clear evidence for a three-year period, but did present the possibility of a 16.8\,yr periodicity.  Konacki et al. (1999)\nocite{klw+99} argued that there is no evidence for planetary companions. Our data set, which spans 36.5\,yr, also shows no evidence for either a 3\,yr nor a 16.8\,yr periodicity and we believe that the timing noise for PSR~B0329$+$59 has a similar form to the other pulsars in our sample.

\subsubsection{Slow glitches}\label{sec:sglitch}

\begin{figure}
\includegraphics[width=6cm,angle=-90]{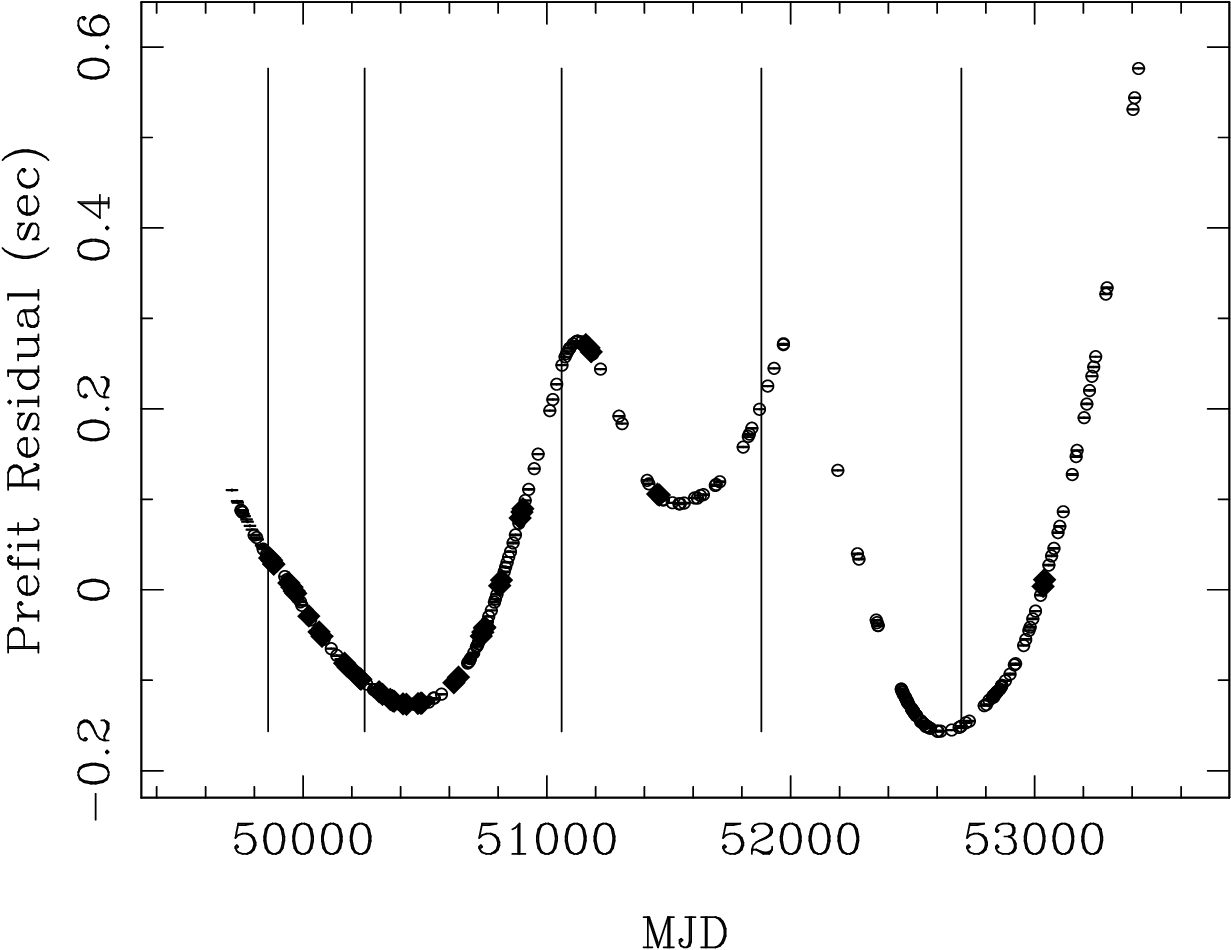}
\caption{The timing residuals for PSR~B1822$-$09 from MJD~49708 to 53426.  The time of ``slow-glitch'' events as reported by Shabanova (2007) are indicated by vertical lines.}\label{fg:1822}
\end{figure}

Zou et al. (2004)\nocite{zww+04} reported a phenomenon known as ``slow glitches'' for PSRs~B1822$-$09 and J1825$-$0935. ``Slow glitches'' are characterised by a permanent increase in frequency, but no significant change in the slow-down rate.  More recently Shabanova (2007) identified a further five slow glitches in the timing of PSR~B1822$-$09. In Figure~\ref{fg:1822} we present our timing residuals for this pulsar between MJD~49708 and 53426.  The epochs of the slow-glitches reported by Shabanova (2007) are indicated by vertical lines.  The observed timing residuals are not dis-similar to those plotted in Figure~\ref{fg:short} and described in \S\ref{sec:pspec}.  We therefore suggest that slow-glitches are not a unique phenomenon, but are caused by the same process as the timing noise seen in our sample of pulsars.

\section{Conclusion}

We have presented the timing residuals of 366 pulsars over the past 36\,yr.  These residuals show that
\begin{itemize}
\item{timing noise is widespread in pulsars (\S2).}
\item{the timing noise described here is not an artefact of the observing systems nor offline processing (\S3.1).}
\item{timing noise is inversely correlated with characteristic age, $\tau_c$ (\S3.2.1).}
\item{the timing noise cannot be explained using a simple random walk model in the pulse phase, frequency or spin-down rate (\S3.2.1).}
\item{the structures seen in the timing noise vary with data span.  As more data is collected more quasi-periodic features are observed.  Shorter data spans generally exhibit a significant $\ddot{\nu}$ term that is not related to the magnetic braking of the neutron star (\S3.2.2).}
\item{the dominant contribution to timing noise for all pulsars with $\tau_c < 10^{5}$ can be explained as being caused by the recovery from previous glitch events (\S3.2.2).}
\item{significant periodicities are seen in the timing residuals of a few pulsars (e.g. PSRs~B1540$-$06, B1826$-$17, B1828$-$11 and B2148$+$63). However, quasi-periodic structures are seen in the timing residuals of many pulsars (\S3.2.3).}
\item{the detailed structure of the timing noise indicates that local maxima usually have different radii of curvature than local minima (\S3.2.3).}
\item{there is no evidence for a planetary companion to PSR~B0329$+$54 (\S3.2.3).}
\item{``slow glitches'' are not likely to be a different phenomenon to that causing timing noise (\S3.2.4).}
\end{itemize}

We emphasise that these results could only have been found by studying the timing residuals over very long data spans and further work is continuing to search for correlated pulse shape changes and to relate the glitch phenomena with timing noise.

\section*{Acknowledgements}

  During the course of this work we made extensive use of NASA's Astrophysics Data System bibliographic database and the astro-ph preprint service. GH is the recipient of an Australian Research Council QEII Fellowship (\#DP0878388).  Many people have been involved in timing pulsars from Jodrell Bank Observatory.  In particular we acknowledge Christine Jordan and the many telescope operators who have overseen the thousands of observations used in this paper.

\bibliography{crossrefs,modrefs,psrrefs}
\bibliographystyle{mn}

\end{document}

%% file: table1.tex
\begin{table*} \caption{Basic parameters for the pulsars in our sample.  $\sigma_1$ is the rms residual after removing a quadratic term, $\sigma_2$, the rms value after removing a cubic term and $\sigma_3$ the value after whitening the data.}\label{tb:params} \begin{tiny}\begin{tabular}{llrrllrlrrrlrr} \hline PSR J & PSR B & $\nu$ & $\dot{\nu}$ & $\ddot{\nu}$ & Epoch & N & $T_s$ & $\sigma_1$ & $\sigma_2$ & $\sigma_3$ & $\Delta_8$ & $\log\sigma_z(10{\rm yr})$\\ & & (s$^{-1}$) & ($10^{-15} s^{-2}$) & ($10^{-24} s^{-3}$) & (MJD) & & (yr) & (ms) & (ms) & (ms) \\ \hline
J0014$+$4746 & B0011$+$47 & 0.806 & $-$0.367 & $0.00035(20)$ & 49285.0 & 365 & 22.8 & 4.26 & 4.23 & 4.23 & <$-$1.14 & $-$10.91 & \\ 
J0034$-$0534 & --- & 532.713 & $-$1.409 & $-0.043(10)$ & 51096.0 & 402 & 12.8 & 0.06 & 0.06 & 0.06 & <$-$3.04 & $-$12.44 & \\ 
J0034$-$0721 & B0031$-$07 & 1.061 & $-$0.459 & $0.00033(3)$ & 47051.0 & 772 & 35.0 & 3.84 & 3.47 & 3.38 & $-$1.44 & $-$10.79 & \\ 
J0040$+$5716 & B0037$+$56 & 0.894 & $-$2.302 & $-0.00005(6)$ & 50091.0 & 426 & 18.5 & 0.62 & 0.62 & 0.62 & <$-$1.79 & $-$11.65 & \\ 
J0048$+$3412 & B0045$+$33 & 0.822 & $-$1.589 & $-0.0003(3)$ & 50083.0 & 201 & 18.4 & 2.17 & 2.15 & 2.15 & <$-$1.21 & $-$11.06 & \\ 
\\
J0055$+$5117 & B0052$+$51 & 0.473 & $-$2.132 & $0.00092(5)$ & 50092.0 & 417 & 18.5 & 1.51 & 1.08 & 1.08 & <$-$1.63 & $-$11.03 & \\ 
J0056$+$4756 & B0053$+$47 & 2.118 & $-$14.938 & $-0.0083(18)$ & 50103.0 & 252 & 18.2 & 6.09 & 5.82 & 2.29 & <$-$1.36 & $-$10.24 & \\ 
J0102$+$6537 & B0059$+$65 & 0.596 & $-$2.113 & $-0.02862(17)$ & 50091.0 & 339 & 18.5 & 65.23 & 2.66 & 1.37 & <$-$1.34 & $-$9.91 & \\ 
J0108$+$6608 & B0105$+$65 & 0.779 & $-$7.918 & $0.107(16)$ & 50482.0 & 427 & 16.2 & 157.09 & 148.26 & 3.51 & $-$0.16 & $-$8.70 & \\ 
J0108$+$6905 & B0105$+$68 & 0.934 & $-$0.042 & $-0.0002(3)$ & 50091.0 & 336 & 18.5 & 2.63 & 2.62 & 2.62 & <$-$1.36 & $-$11.21 & \\ 
\\
J0108$-$1431 & --- & 1.238 & $-$0.118 & $-0.0030(18)$ & 51251.0 & 332 & 12.1 & 4.49 & 4.47 & 4.47 & <$-$1.45 & $-$10.84 & \\ 
J0117$+$5914 & B0114$+$58 & 9.858 & $-$568.596 & $-2.91(3)$ & 50083.0 & 314 & 18.4 & 165.07 & 25.71 & 0.34 & $-$1.55 & $-$9.21 & \\ 
J0134$-$2937 & --- & 7.301 & $-$4.178 & $-0.0004(5)$ & 51251.0 & 298 & 12.1 & 0.19 & 0.19 & 0.19 & <$-$2.59 & $-$12.53 & \\ 
J0139$+$5814 & B0136$+$57 & 3.670 & $-$144.309 & $-0.065(10)$ & 49706.0 & 684 & 20.6 & 56.33 & 52.63 & 0.20 & $-$0.95 & $-$9.27 & \\ 
J0141$+$6009 & B0138$+$59 & 0.818 & $-$0.261 & $0.00029(6)$ & 49495.0 & 451 & 19.4 & 1.09 & 1.06 & 1.06 & <$-$2.24 & $-$11.51 & \\ 
\\
J0147$+$5922 & B0144$+$59 & 5.094 & $-$6.661 & $0.0255(7)$ & 50080.0 & 430 & 18.2 & 2.72 & 1.16 & 0.16 & $-$2.56 & $-$10.62 & \\ 
J0152$-$1637 & B0149$-$16 & 1.201 & $-$1.874 & $-0.00425(5)$ & 48651.0 & 455 & 26.4 & 5.79 & 1.39 & 0.59 & <$-$2.07 & $-$11.10 & \\ 
J0156$+$3949 & B0153$+$39 & 0.552 & $-$0.046 & $-0.0001(13)$ & 50104.0 & 81 & 18.1 & 9.88 & 9.88 & 9.88 & $-$0.31 & $-$10.46 & \\ 
J0157$+$6212 & B0154$+$61 & 0.425 & $-$34.160 & $-0.0093(8)$ & 50126.0 & 463 & 18.3 & 26.74 & 20.09 & 2.18 & $-$1.49 & $-$9.73 & \\ 
J0215$+$6218 & --- & 1.822 & $-$2.198 & $-0.0095(8)$ & 51788.0 & 327 & 10.5 & 1.46 & 0.98 & 0.43 & $-$2.29 & $-$10.62 & \\ 
\\
J0218$+$4232 & --- & 430.461 & $-$14.340 & $-0.007(8)$ & 51268.0 & 387 & 11.9 & 0.05 & 0.05 & 0.05 & <$-$3.26 & $-$13.04 & \\ 
J0231$+$7026 & B0226$+$70 & 0.682 & $-$1.445 & $0.00026(13)$ & 50088.0 & 305 & 18.5 & 1.83 & 1.81 & 1.81 & <$-$1.36 & $-$10.98 & \\ 
J0304$+$1932 & B0301$+$19 & 0.721 & $-$0.673 & $-0.00274(7)$ & 49141.0 & 498 & 23.7 & 4.26 & 1.76 & 1.00 & <$-$1.39 & $-$10.54 & \\ 
J0323$+$3944 & B0320$+$39 & 0.330 & $-$0.069 & $0.00004(4)$ & 49229.0 & 520 & 23.2 & 2.25 & 2.24 & 2.24 & <$-$1.71 & $-$11.35 & \\ 
J0332$+$5434 & B0329$+$54 & 1.400 & $-$4.012 & $0.00092(7)$ & 46775.0 & 1170 & 36.5 & 7.46 & 6.76 & 0.46 & $-$1.86 & $-$10.38 & \\ 
\\
J0335$+$4555 & B0331$+$45 & 3.715 & $-$0.101 & $0.00014(12)$ & 50081.0 & 483 & 18.4 & 0.33 & 0.33 & 0.33 & <$-$2.33 & $-$12.19 & \\ 
J0343$+$5312 & B0339$+$53 & 0.517 & $-$3.587 & $-0.00131(15)$ & 49557.0 & 325 & 22.9 & 4.64 & 4.13 & 4.13 & <$-$1.24 & $-$10.42 & \\ 
J0357$+$5236 & B0353$+$52 & 5.075 & $-$12.277 & $0.0078(4)$ & 50081.0 & 412 & 18.4 & 0.98 & 0.63 & 0.18 & <$-$2.39 & $-$10.79 & \\ 
J0406$+$6138 & B0402$+$61 & 1.682 & $-$15.773 & $0.0476(19)$ & 50428.0 & 406 & 16.7 & 13.77 & 8.43 & 0.43 & $-$1.57 & $-$9.69 & \\ 
J0415$+$6954 & B0410$+$69 & 2.559 & $-$0.502 & $0.00142(11)$ & 50092.0 & 375 & 18.5 & 0.51 & 0.41 & 0.36 & <$-$2.17 & $-$11.32 & \\ 
\\
J0421$-$0345 & --- & 0.463 & $-$0.249 & $0.00110(15)$ & 50847.0 & 226 & 13.3 & 1.27 & 1.13 & 1.13 & <$-$1.71 & $-$10.84 & \\ 
J0448$-$2749 & --- & 2.220 & $-$0.731 & $-0.0029(9)$ & 51427.0 & 332 & 11.2 & 0.88 & 0.85 & 0.85 & <$-$2.03 & $-$11.43 & \\ 
J0450$-$1248 & B0447$-$12 & 2.283 & $-$0.535 & $-0.0010(3)$ & 49751.0 & 527 & 20.2 & 1.68 & 1.64 & 1.64 & <$-$2.00 & $-$11.29 & \\ 
J0452$-$1759 & B0450$-$18 & 1.822 & $-$19.092 & $0.0137(5)$ & 49295.0 & 576 & 22.9 & 8.21 & 5.15 & 0.29 & $-$1.97 & $-$10.16 & \\ 
J0454$+$5543 & B0450$+$55 & 2.935 & $-$20.441 & $-0.184(4)$ & 50336.0 & 335 & 17.2 & 29.42 & 10.56 & 0.33 & $-$1.70 & $-$9.51 & \\ 
\\
J0459$-$0210 & --- & 0.883 & $-$1.089 & $-0.0004(5)$ & 51269.0 & 238 & 12.0 & 1.34 & 1.34 & 1.34 & <$-$1.50 & $-$11.53 & \\ 
J0502$+$4654 & B0458$+$46 & 1.566 & $-$13.690 & $0.01057(16)$ & 49129.0 & 497 & 23.6 & 7.35 & 2.19 & 0.54 & $-$1.98 & $-$10.69 & \\ 
J0520$-$2553 & --- & 4.138 & $-$0.515 & $0.0026(12)$ & 51657.0 & 181 & 10.2 & 0.42 & 0.41 & 0.41 & <$-$1.98 & $-$11.46 & \\ 
J0525$+$1115 & B0523$+$11 & 2.821 & $-$0.586 & $-0.000238(20)$ & 48663.0 & 643 & 26.3 & 0.32 & 0.28 & 0.26 & <$-$2.46 & $-$11.78 & \\ 
J0528$+$2200 & B0525$+$21 & 0.267 & $-$2.854 & $0.00082(4)$ & 46909.0 & 954 & 35.8 & 20.34 & 15.25 & 1.39 & <$-$1.38 & $-$10.01 & \\ 
\\
J0538$+$2817 & --- & 6.985 & $-$179.088 & $-0.6951(12)$ & 51821.0 & 353 & 10.1 & 21.28 & 0.34 & 0.18 & $-$1.91 & $-$9.48 & \\ 
J0543$+$2329 & B0540$+$23 & 4.065 & $-$254.898 & $0.2173(15)$ & 49294.0 & 691 & 22.9 & 46.67 & 7.89 & 0.16 & $-$2.09 & $-$10.03 & \\ 
J0612$+$3721 & B0609$+$37 & 3.356 & $-$0.670 & $-0.0125(3)$ & 50092.0 & 456 & 18.5 & 2.07 & 0.80 & 0.35 & <$-$2.38 & $-$10.81 & \\ 
J0613$-$0200 & --- & 326.601 & $-$1.019 & $0.009(4)$ & 51240.0 & 656 & 12.1 & 0.04 & 0.04 & 0.04 & <$-$3.47 & $-$12.89 & \\ 
J0614$+$2229 & B0611$+$22 & 2.985 & $-$529.522 & $9.51(8)$ & 50164.0 & 973 & 18.1 & 1735.68 & 514.15 & 1.82 & $-$0.14 & $-$7.97 & \\ 
\\
J0621$+$1002 & --- & 34.657 & $-$0.057 & $0.0019(3)$ & 51145.0 & 685 & 10.8 & 0.03 & 0.03 & 0.03 & $-$3.41 & $-$12.79 & \\ 
J0624$-$0424 & B0621$-$04 & 0.962 & $-$0.769 & $0.00001(10)$ & 50302.0 & 459 & 17.4 & 0.87 & 0.87 & 0.87 & <$-$1.77 & $-$11.63 & \\ 
J0629$+$2415 & B0626$+$24 & 2.098 & $-$8.785 & $0.0049(8)$ & 49846.0 & 541 & 19.7 & 5.75 & 5.53 & 0.41 & $-$1.96 & $-$10.30 & \\ 
J0630$-$2834 & B0628$-$28 & 0.804 & $-$4.601 & $-0.0175(7)$ & 47013.0 & 807 & 35.2 & 131.89 & 91.99 & 1.52 & $-$1.84 & $-$9.71 & \\ 
J0653$+$8051 & B0643$+$80 & 0.823 & $-$2.576 & $0.00069(4)$ & 49141.0 & 283 & 23.7 & 1.23 & 0.83 & 0.70 & <$-$1.84 & $-$11.01 & \\ 
\\
J0659$+$1414 & B0656$+$14 & 2.598 & $-$371.288 & $0.7636(19)$ & 49721.0 & 772 & 16.1 & 120.42 & 7.27 & 1.49 & $-$1.26 & $-$9.09 & \\ 
J0700$+$6418 & B0655$+$64 & 5.111 & $-$0.018 & $0.00019(13)$ & 49138.0 & 402 & 23.7 & 0.64 & 0.64 & 0.64 & <$-$2.06 & $-$11.48 & \\ 
J0725$-$1635 & --- & 2.357 & $-$0.515 & $-0.0018(4)$ & 50884.0 & 277 & 13.3 & 0.57 & 0.54 & 0.54 & <$-$2.00 & $-$11.24 & \\ 
J0729$-$1836 & B0727$-$18 & 1.960 & $-$72.832 & $0.245(7)$ & 50126.0 & 635 & 18.3 & 76.04 & 40.15 & 0.55 & $-$0.55 & $-$9.05 & \\ 
J0742$-$2822 & B0740$-$28 & 5.997 & $-$604.876 & $-0.61(5)$ & 49696.0 & 1009 & 20.6 & 165.35 & 145.16 & 0.40 & $-$0.84 & $-$8.92 & \\ 
\\
J0751$+$1807 & --- & 287.458 & $-$0.643 & $-0.002(4)$ & 51389.0 & 531 & 11.3 & 0.04 & 0.04 & 0.04 & <$-$3.67 & $-$13.55 & \\ 
J0754$+$3231 & B0751$+$32 & 0.693 & $-$0.519 & $-0.00081(6)$ & 49763.0 & 492 & 20.3 & 1.54 & 1.28 & 1.21 & <$-$1.79 & $-$11.04 & \\ 
J0814$+$7429 & B0809$+$74 & 0.774 & $-$0.101 & $0.00014(5)$ & 49369.0 & 474 & 22.4 & 1.07 & 1.06 & 1.06 & <$-$1.59 & $-$11.47 & \\ 
J0820$-$1350 & B0818$-$13 & 0.808 & $-$1.373 & $0.000028(19)$ & 49299.0 & 631 & 22.7 & 0.55 & 0.55 & 0.26 & <$-$2.24 & $-$11.62 & \\ 
J0823$+$0159 & B0820$+$02 & 1.156 & $-$0.140 & $-0.00075(5)$ & 49144.0 & 573 & 23.7 & 1.10 & 0.88 & 0.84 & $-$2.00 & $-$11.08 & \\ 
\\
J0826$+$2637 & B0823$+$26 & 1.884 & $-$6.069 & $0.0240(12)$ & 46857.0 & 979 & 36.1 & 110.89 & 90.82 & 0.29 & $-$1.63 & $-$9.49 & \\ 
J0828$-$3417 & B0826$-$34 & 0.541 & $-$0.292 & $-0.0035(5)$ & 48508.0 & 146 & 25.9 & 19.16 & 16.03 & 14.04 & <0.16 & $-$9.88 & \\ 
J0837$+$0610 & B0834$+$06 & 0.785 & $-$4.191 & $0.000096(11)$ & 49138.0 & 604 & 23.7 & 0.36 & 0.34 & 0.34 & <$-$1.91 & $-$11.73 & \\ 
J0846$-$3533 & B0844$-$35 & 0.896 & $-$1.285 & $0.00107(13)$ & 49145.0 & 119 & 23.7 & 1.95 & 1.52 & 1.52 & <$-$0.95 & $-$10.98 & \\ 
J0849$+$8028 & B0841$+$80 & 0.624 & $-$0.174 & $0.00005(74)$ & 50419.0 & 90 & 16.6 & 4.80 & 4.79 & 4.79 & <$-$0.42 & $-$10.70 & \\ 
\\
J0855$-$3331 & B0853$-$33 & 0.789 & $-$3.934 & $0.00059(15)$ & 49735.5 & 435 & 17.2 & 1.63 & 1.60 & 1.60 & <$-$1.66 & $-$11.07 & \\ 
J0908$-$1739 & B0906$-$17 & 2.490 & $-$4.151 & $0.00165(5)$ & 49140.0 & 639 & 23.7 & 0.75 & 0.43 & 0.29 & $-$2.49 & $-$11.23 & \\ 
J0921$+$6254 & B0917$+$63 & 0.638 & $-$1.468 & $0.00018(12)$ & 49687.0 & 156 & 16.3 & 0.86 & 0.85 & 0.85 & <$-$1.58 & $-$11.18 & \\ 
J0922$+$0638 & B0919$+$06 & 2.322 & $-$73.982 & $0.1385(12)$ & 48518.0 & 766 & 27.0 & 99.33 & 22.52 & 0.35 & $-$1.07 & $-$9.36 & \\ 
J0943$+$1631 & B0940$+$16 & 0.920 & $-$0.077 & $0.00008(17)$ & 49143.0 & 443 & 23.7 & 4.04 & 4.04 & 4.04 & <$-$1.03 & $-$11.24 & \\ 
\\
J0944$-$1354 & B0942$-$13 & 1.754 & $-$0.139 & $-0.00027(5)$ & 49764.0 & 533 & 20.3 & 0.41 & 0.40 & 0.40 & <$-$2.62 & $-$11.83 & \\ 
J0946$+$0951 & B0943$+$10 & 0.911 & $-$2.902 & $-0.0377(8)$ & 49352.0 & 117 & 22.3 & 47.86 & 10.29 & 10.29 & $-$0.28 & $-$9.79 & \\ 
J0953$+$0755 & B0950$+$08 & 3.952 & $-$3.588 & $-0.0048(3)$ & 46777.0 & 953 & 36.5 & 11.22 & 9.53 & 0.12 & $-$2.31 & $-$10.53 & \\ 
J1012$+$5307 & --- & 190.268 & $-$0.620 & $0.0031(12)$ & 51331.0 & 521 & 11.6 & 0.02 & 0.02 & 0.02 & <$-$3.91 & $-$13.30 & \\ 
J1012$-$2337 & B1010$-$23 & 0.397 & $-$0.139 & $0.00012(12)$ & 49127.0 & 137 & 23.6 & 4.66 & 4.56 & 4.56 & <0.25 & $-$10.70 & \\ 
\\
J1018$-$1642 & B1016$-$16 & 0.554 & $-$0.535 & $0.00006(18)$ & 50093.0 & 243 & 18.5 & 2.79 & 2.78 & 2.78 & <$-$1.12 & $-$11.78 & \\ 
J1022$+$1001 & --- & 60.779 & $-$0.160 & $0.0014(4)$ & 51638.0 & 625 & 11.2 & 0.02 & 0.02 & 0.02 & <$-$3.99 & $-$13.56 & \\ 
J1024$-$0719 & --- & 193.716 & $-$0.695 & $0.007(6)$ & 51422.0 & 366 & 11.0 & 0.06 & 0.06 & 0.06 & <$-$3.32 & $-$13.22 & \\ 
J1034$-$3224 & --- & 0.869 & $-$0.174 & $-0.0005(6)$ & 51123.0 & 225 & 12.9 & 1.83 & 1.82 & 1.82 & <$-$1.37 & $-$11.89 & \\ 
J1041$-$1942 & B1039$-$19 & 0.721 & $-$0.492 & $-0.0000009(500)$ & 49143.0 & 477 & 23.7 & 1.31 & 1.30 & 1.30 & <$-$1.78 & $-$11.63 & \\ 
\\
J1047$-$3032 & --- & 3.027 & $-$0.559 & $-0.004(3)$ & 51445.0 & 264 & 11.1 & 1.84 & 1.84 & 1.84 & <$-$1.81 & $-$11.24 & \\ 
J1115$+$5030 & B1112$+$50 & 0.604 & $-$0.909 & $-0.00006(5)$ & 49753.0 & 491 & 20.2 & 1.28 & 1.28 & 1.28 & <$-$1.90 & $-$11.60 & \\ 
J1136$+$1551 & B1133$+$16 & 0.842 & $-$2.646 & $0.00125(6)$ & 46407.0 & 900 & 36.4 & 15.22 & 13.76 & 0.24 & <$-$2.07 & $-$10.92 & \\ 
J1141$-$3107 & --- & 1.857 & $-$6.880 & $-0.61(3)$ & 51271.0 & 179 & 12.1 & 71.87 & 37.79 & 2.94 & $-$0.30 & $-$8.91 & \\ 
J1141$-$3322 & --- & 6.862 & $-$10.921 & $-0.291(14)$ & 51446.0 & 328 & 11.1 & 6.64 & 4.22 & 0.51 & <$-$2.15 & $-$9.88 & \\ 
\end{tabular}\end{tiny}\end{table*}\addtocounter{table}{-1}\begin{table*}\caption{\ldots continued}\begin{tiny}\begin{tabular}{llrrllrlrrrlrr} \hline PSR J & PSR B & $\nu$ & $\dot{\nu}$ & $\ddot{\nu}$ & Epoch & N & $T_s$ & $\sigma_1$ & $\sigma_2$ & $\sigma_3$  & $\Delta_8$ & $\log\sigma_z(10{\rm yr})$\\ & & (s$^{-1}$) & ($10^{-15} s^{-2}$) & ($10^{-24} s^{-3}$) & (MJD) & & (yr) & (ms) & (ms) & (ms) \\ \hline
 J1238$+$21 & B1238$+$21 & 0.894 & $-$1.155 & $-0.0010(16)$ & 51822.0 & 82 & 10.1 & 1.81 & 1.80 & 1.80 & <$-$1.35 & $-$11.23 & \\ 
J1239$+$2453 & B1237$+$25 & 0.723 & $-$0.502 & $-0.000305(9)$ & 46942.0 & 1026 & 35.6 & 3.15 & 2.12 & 0.57 & <$-$2.11 & $-$11.17 & \\ 
J1257$-$1027 & B1254$-$10 & 1.620 & $-$0.952 & $-0.00009(16)$ & 49348.2 & 475 & 18.4 & 1.12 & 1.12 & 1.09 & <$-$1.96 & $-$11.46 & \\ 
J1300$+$1240 & B1257$+$12 & 160.810 & $-$2.957 & $-0.002(3)$ & 50788.0 & 360 & 14.5 & 0.08 & 0.07 & 0.07 & <$-$3.09 & $-$13.10 & \\ 
J1311$-$1228 & B1309$-$12 & 2.235 & $-$0.753 & $-0.00141(12)$ & 50094.0 & 433 & 18.5 & 0.65 & 0.56 & 0.49 & <$-$2.11 & $-$11.16 & \\ 
\\
J1321$+$8323 & B1322$+$83 & 1.492 & $-$1.262 & $-0.0056(3)$ & 49282.0 & 289 & 22.8 & 3.97 & 2.30 & 2.30 & <$-$1.41 & $-$10.58 & \\ 
J1332$-$3032 & --- & 1.537 & $-$1.322 & $0.017(9)$ & 51447.0 & 192 & 11.1 & 9.62 & 9.52 & 9.52 & <$-$0.96 & $-$10.46 & \\ 
J1455$-$3330 & --- & 125.200 & $-$0.381 & $-0.008(4)$ & 51190.0 & 269 & 12.4 & 0.07 & 0.07 & 0.07 & <$-$2.94 & $-$12.46 & \\ 
J1509$+$5531 & B1508$+$55 & 1.352 & $-$9.144 & $0.0819(15)$ & 49294.0 & 471 & 22.9 & 63.60 & 22.20 & 0.52 & $-$1.52 & $-$9.87 & \\ 
J1518$+$4904 & --- & 24.429 & $-$0.016 & $0.0002(5)$ & 51619.0 & 309 & 10.0 & 0.03 & 0.03 & 0.03 & <$-$3.48 & $-$13.21 & \\ 
\\
J1532$+$2745 & B1530$+$27 & 0.889 & $-$0.616 & $-0.0024(3)$ & 50096.0 & 414 & 18.5 & 3.37 & 3.03 & 1.62 & <$-$1.64 & $-$10.55 & \\ 
J1537$+$1155 & B1534$+$12 & 26.382 & $-$1.686 & $0.00008(30)$ & 50300.0 & 400 & 13.8 & 0.04 & 0.04 & 0.04 & <$-$3.14 & $-$12.88 & \\ 
J1543$+$0929 & B1541$+$09 & 1.336 & $-$0.772 & $-0.0093(4)$ & 49141.0 & 452 & 23.7 & 10.61 & 6.46 & 3.27 & <$-$1.28 & $-$10.03 & \\ 
J1543$-$0620 & B1540$-$06 & 1.410 & $-$1.749 & $0.0130(3)$ & 49839.0 & 503 & 19.7 & 7.06 & 3.02 & 0.36 & $-$1.36 & $-$10.36 & \\ 
J1555$-$2341 & B1552$-$23 & 1.878 & $-$2.447 & $-0.0280(4)$ & 50301.0 & 346 & 17.3 & 6.42 & 1.40 & 0.88 & <$-$1.90 & $-$10.23 & \\ 
\\
J1555$-$3134 & B1552$-$31 & 1.930 & $-$0.232 & $-0.00018(19)$ & 50299.0 & 289 & 17.3 & 0.69 & 0.69 & 0.69 & <$-$1.92 & $-$11.58 & \\ 
J1603$-$2531 & --- & 3.533 & $-$19.875 & $-0.068(12)$ & 51121.0 & 192 & 12.8 & 9.21 & 8.47 & 0.15 & $-$1.72 & $-$9.85 & \\ 
J1603$-$2712 & B1600$-$27 & 1.285 & $-$4.968 & $-0.0025(3)$ & 50302.0 & 199 & 17.3 & 1.61 & 1.36 & 0.67 & <$-$1.91 & $-$10.72 & \\ 
J1607$-$0032 & B1604$-$00 & 2.371 & $-$1.720 & $-0.002015(10)$ & 47386.0 & 729 & 33.2 & 3.53 & 0.47 & 0.44 & <$-$2.38 & $-$11.30 & \\ 
J1610$-$1322 & B1607$-$13 & 0.982 & $-$0.222 & $-0.0003(6)$ & 50094.0 & 213 & 18.5 & 4.77 & 4.75 & 4.75 & <$-$1.02 & $-$10.81 & \\ 
\\
J1614$+$0737 & B1612$+$07 & 0.829 & $-$1.620 & $-0.0007(3)$ & 49897.0 & 160 & 15.1 & 1.19 & 1.12 & 1.12 & <$-$1.78 & $-$11.24 & \\ 
J1615$-$2940 & B1612$-$29 & 0.404 & $-$0.258 & $0.00012(17)$ & 49026.0 & 99 & 23.1 & 4.91 & 4.89 & 4.89 & $-$0.18 & $-$10.58 & \\ 
J1623$-$0908 & B1620$-$09 & 0.783 & $-$1.584 & $0.00027(3)$ & 49142.0 & 394 & 23.7 & 0.85 & 0.75 & 0.75 & <$-$1.08 & $-$11.30 & \\ 
J1623$-$2631 & B1620$-$26 & 90.287 & $-$2.953 & $18.346(11)$ & 50292.0 & 768 & 17.5 & 103.55 & 3.06 & 0.15 & $-$1.35 & $-$9.22 & \\ 
J1635$+$2418 & B1633$+$24 & 2.039 & $-$0.496 & $0.00070(20)$ & 49130.0 & 149 & 23.6 & 1.23 & 1.18 & 1.18 & <$-$1.28 & $-$11.23 & \\ 
\\
J1643$-$1224 & --- & 216.373 & $-$0.865 & $-0.0043(19)$ & 51251.0 & 433 & 12.0 & 0.03 & 0.03 & 0.03 & <$-$3.30 & $-$13.10 & \\ 
J1645$-$0317 & B1642$-$03 & 2.579 & $-$11.846 & $0.0067(6)$ & 46930.0 & 957 & 35.7 & 21.63 & 20.68 & 0.22 & $-$0.99 & $-$9.66 & \\ 
J1648$-$3256 & --- & 1.390 & $-$6.815 & $0.004(3)$ & 51276.0 & 149 & 12.0 & 4.54 & 4.50 & 0.76 & $-$1.39 & $-$10.26 & \\ 
J1650$-$1654 & --- & 0.572 & $-$1.046 & $-0.0027(8)$ & 51272.0 & 172 & 12.1 & 3.07 & 2.95 & 2.95 & $-$1.06 & $-$10.72 & \\ 
J1651$-$1709 & B1648$-$17 & 1.027 & $-$3.205 & $0.0007(4)$ & 50430.0 & 168 & 16.7 & 1.56 & 1.53 & 1.31 & <$-$1.30 & $-$11.06 & \\ 
\\
J1652$-$2404 & B1649$-$23 & 0.587 & $-$1.088 & $-0.00155(5)$ & 49132.0 & 348 & 23.6 & 3.36 & 1.70 & 1.60 & <$-$1.61 & $-$10.91 & \\ 
J1654$-$2713 & --- & 1.263 & $-$0.268 & $0.002(3)$ & 51414.0 & 154 & 11.1 & 2.39 & 2.38 & 2.38 & <$-$1.19 & $-$11.04 & \\ 
J1659$-$1305 & B1657$-$13 & 1.560 & $-$1.503 & $0.0309(7)$ & 50094.0 & 147 & 18.5 & 10.86 & 2.66 & 2.66 & <$-$1.02 & $-$10.12 & \\ 
J1700$-$3312 & --- & 0.736 & $-$2.555 & $-0.0163(6)$ & 51276.0 & 250 & 12.0 & 3.96 & 1.87 & 1.41 & <$-$1.42 & $-$10.28 & \\ 
J1703$-$1846 & B1700$-$18 & 1.243 & $-$2.676 & $0.00003(12)$ & 50314.0 & 209 & 17.3 & 0.56 & 0.56 & 0.56 & <$-$1.98 & $-$12.02 & \\ 
\\
J1703$-$3241 & B1700$-$32 & 0.825 & $-$0.449 & $0.00122(9)$ & 50427.0 & 159 & 16.6 & 0.78 & 0.51 & 0.51 & <$-$1.58 & $-$11.08 & \\ 
J1705$-$1906 & B1702$-$19 & 3.345 & $-$46.284 & $-0.0009(6)$ & 50497.0 & 410 & 16.3 & 1.28 & 1.24 & 0.25 & <$-$2.36 & $-$11.11 & \\ 
J1705$-$3423 & --- & 3.915 & $-$16.484 & $0.085(6)$ & 51256.0 & 152 & 11.9 & 4.72 & 2.85 & 1.01 & $-$1.55 & $-$9.96 & \\ 
J1708$-$3426 & --- & 1.445 & $-$8.775 & $-0.0100(15)$ & 51276.0 & 124 & 12.0 & 2.42 & 2.05 & 2.05 & <$-$1.12 & $-$10.62 & \\ 
J1709$-$1640 & B1706$-$16 & 1.531 & $-$14.783 & $0.3704(12)$ & 47389.0 & 498 & 33.2 & 948.18 & 63.69 & 0.43 & $-$1.14 & $-$9.07 & \\ 
\\
J1711$-$1509 & B1709$-$15 & 1.151 & $-$1.461 & $0.0067(3)$ & 50094.0 & 197 & 18.5 & 3.78 & 1.72 & 0.67 & <$-$1.71 & $-$10.65 & \\ 
J1713$+$0747 & --- & 218.812 & $-$0.409 & $0.0019(13)$ & 51318.0 & 345 & 11.5 & 0.01 & 0.01 & 0.01 & $-$3.94 & $-$13.51 & \\ 
J1717$-$3425 & B1714$-$34 & 1.524 & $-$22.761 & $0.022(8)$ & 50673.0 & 139 & 15.2 & 17.88 & 17.35 & 0.82 & $-$0.74 & $-$9.51 & \\ 
J1720$-$0212 & B1718$-$02 & 2.093 & $-$0.363 & $-0.0001(6)$ & 49853.0 & 320 & 19.8 & 3.44 & 3.44 & 3.44 & <$-$1.53 & $-$12.20 & \\ 
J1720$-$1633 & B1717$-$16 & 0.639 & $-$2.366 & $-0.00002(72)$ & 50084.0 & 214 & 18.4 & 8.95 & 8.95 & 1.57 & $-$1.20 & $-$10.06 & \\ 
\\
J1720$-$2933 & B1717$-$29 & 1.612 & $-$1.938 & $-0.0142(3)$ & 50290.0 & 164 & 17.2 & 3.88 & 0.87 & 0.69 & <$-$1.58 & $-$10.62 & \\ 
J1721$-$1936 & B1718$-$19 & 0.996 & $-$1.609 & $0.153(13)$ & 50840.0 & 288 & 14.4 & 67.75 & 54.68 & 2.99 & $-$0.83 & $-$8.89 & \\ 
J1721$-$3532 & B1718$-$35 & 3.566 & $-$320.260 & $-0.07(6)$ & 51505.0 & 139 & 10.7 & 20.82 & 20.24 & 0.90 & $-$0.98 & $-$10.71 & \\ 
J1722$-$3207 & B1718$-$32 & 2.096 & $-$2.838 & $-0.0010(7)$ & 50495.0 & 180 & 16.3 & 1.72 & 1.70 & 0.16 & $-$2.12 & $-$10.84 & \\ 
J1728$-$0007 & B1726$-$00 & 2.591 & $-$7.535 & $-0.0167(7)$ & 50513.0 & 152 & 16.1 & 2.88 & 1.28 & 0.93 & <$-$1.64 & $-$10.81 & \\ 
\\
J1730$-$2304 & --- & 123.110 & $-$0.306 & $-0.0053(13)$ & 51240.0 & 444 & 12.0 & 0.03 & 0.03 & 0.03 & <$-$3.49 & $-$12.73 & \\ 
J1730$-$3350 & B1727$-$33 & 7.170 & $-$4355.079 & $62.8(4)$ & 51419.0 & 220 & 8.1 & 408.75 & 30.19 & 0.70 & 0.28 & $-$7.21 & \\ 
J1732$-$1930 & --- & 2.067 & $-$0.776 & $-0.004(3)$ & 51644.0 & 100 & 10.1 & 1.40 & 1.38 & 1.38 & <$-$1.42 & $-$11.03 & \\ 
J1733$-$2228 & B1730$-$22 & 1.147 & $-$0.056 & $-0.00006(8)$ & 49140.0 & 395 & 23.7 & 1.40 & 1.40 & 1.40 & <$-$1.27 & $-$11.50 & \\ 
J1734$-$0212 & B1732$-$02 & 1.191 & $-$0.597 & $0.0026(8)$ & 50117.0 & 127 & 18.3 & 3.53 & 3.36 & 3.36 & <$-$0.88 & $-$10.59 & \\ 
\\
J1735$-$0724 & B1732$-$07 & 2.385 & $-$6.908 & $0.0019(4)$ & 50312.0 & 342 & 17.3 & 1.22 & 1.18 & 0.25 & $-$2.08 & $-$10.85 & \\ 
J1738$-$3211 & B1735$-$32 & 1.301 & $-$1.346 & $0.00044(14)$ & 50018.0 & 365 & 18.9 & 1.34 & 1.32 & 1.29 & <$-$1.90 & $-$11.37 & \\ 
J1739$-$2903 & B1736$-$29 & 3.097 & $-$75.578 & $0.127(4)$ & 49872.0 & 394 & 19.7 & 28.59 & 12.75 & 0.85 & $-$1.52 & $-$9.54 & \\ 
J1739$-$3131 & B1736$-$31 & 1.889 & $-$66.269 & $0.78(5)$ & 49884.0 & 244 & 18.1 & 326.94 & 174.40 & 0.90 & $-$0.11 & $-$8.49 & \\ 
J1740$+$1311 & B1737$+$13 & 1.245 & $-$2.250 & $0.01946(14)$ & 48669.0 & 442 & 26.3 & 24.89 & 3.58 & 0.55 & <$-$1.94 & $-$10.30 & \\ 
\\
J1741$+$2758 & --- & 0.735 & $-$0.994 & $-0.0007(15)$ & 51750.0 & 104 & 10.4 & 2.48 & 2.34 & 2.34 & <$-$1.21 & $-$11.48 & \\ 
J1741$-$0840 & B1738$-$08 & 0.489 & $-$0.545 & $-0.00008(10)$ & 50306.0 & 297 & 17.2 & 1.38 & 1.37 & 1.37 & <$-$1.53 & $-$11.42 & \\ 
J1743$-$0339 & B1740$-$03 & 2.249 & $-$7.873 & $0.41(4)$ & 50199.0 & 139 & 16.6 & 111.71 & 76.44 & 1.71 & $-$0.44 & $-$8.72 & \\ 
J1743$-$1351 & B1740$-$13 & 2.467 & $-$2.910 & $-0.080(3)$ & 50095.0 & 184 & 18.5 & 19.27 & 8.52 & 1.13 & <$-$1.49 & $-$9.88 & \\ 
J1743$-$3150 & B1740$-$31 & 0.414 & $-$20.716 & $-0.0185(7)$ & 50674.0 & 284 & 15.3 & 17.60 & 8.73 & 2.76 & <$-$1.22 & $-$9.67 & \\ 
\\
J1744$-$1134 & --- & 245.426 & $-$0.539 & $-0.0076(12)$ & 51490.0 & 282 & 10.7 & 0.01 & 0.01 & 0.01 & <$-$4.19 & $-$12.95 & \\ 
J1745$-$3040 & B1742$-$30 & 2.722 & $-$79.021 & $-0.0629(18)$ & 50311.0 & 393 & 17.3 & 10.69 & 4.96 & 0.18 & $-$2.04 & $-$10.01 & \\ 
J1748$-$1300 & B1745$-$12 & 2.537 & $-$7.808 & $-0.0347(7)$ & 50315.0 & 289 & 17.3 & 5.91 & 1.86 & 0.26 & $-$2.10 & $-$10.34 & \\ 
J1748$-$2021 & B1745$-$20 & 3.465 & $-$4.803 & $0.059(8)$ & 50677.0 & 362 & 15.3 & 14.84 & 13.67 & 13.67 & <$-$0.75 & $-$9.96 & \\ 
J1748$-$2444 & --- & 2.258 & $-$0.568 & $-0.0024(12)$ & 50917.0 & 190 & 14.2 & 1.82 & 1.80 & 1.73 & <$-$1.62 & $-$11.02 & \\ 
\\
J1749$-$3002 & B1746$-$30 & 1.640 & $-$21.163 & $-0.052(2)$ & 50653.0 & 193 & 15.2 & 11.54 & 5.34 & 1.96 & <$-$1.33 & $-$9.93 & \\ 
J1750$-$3157 & B1747$-$31 & 1.098 & $-$0.237 & $0.0004(4)$ & 50674.0 & 213 & 15.3 & 1.37 & 1.36 & 1.36 & <$-$1.55 & $-$11.45 & \\ 
J1750$-$3503 & --- & 1.462 & $-$0.079 & $0.031(16)$ & 51426.0 & 131 & 11.2 & 15.89 & 15.65 & 15.65 & <$-$0.62 & $-$10.00 & \\ 
J1752$-$2806 & B1749$-$28 & 1.778 & $-$25.687 & $0.0087(10)$ & 46889.0 & 655 & 35.8 & 67.62 & 63.41 & 0.44 & $-$1.15 & $-$9.45 & \\ 
J1753$-$2501 & B1750$-$24 & 1.893 & $-$50.565 & $-0.014(5)$ & 49883.0 & 216 & 19.6 & 22.35 & 21.63 & 3.82 & <$-$1.22 & $-$9.78 & \\ 
\\
J1754$+$5201 & B1753$+$52 & 0.418 & $-$0.274 & $-0.0003(3)$ & 50095.0 & 296 & 18.5 & 5.08 & 5.06 & 5.06 & <$-$0.97 & $-$10.82 & \\ 
J1756$-$2435 & B1753$-$24 & 1.491 & $-$0.633 & $-0.00005(12)$ & 49848.0 & 232 & 18.8 & 0.75 & 0.75 & 0.75 & <$-$1.80 & $-$11.67 & \\ 
J1757$-$2421 & B1754$-$24 & 4.272 & $-$235.755 & $0.283(3)$ & 49121.0 & 397 & 22.8 & 92.18 & 29.55 & 0.76 & $-$1.04 & $-$9.55 & \\ 
J1759$-$2205 & B1756$-$22 & 2.169 & $-$51.171 & $-0.213(3)$ & 50126.0 & 388 & 18.3 & 49.92 & 11.50 & 0.31 & $-$1.73 & $-$9.41 & \\ 
J1759$-$2922 & --- & 1.741 & $-$14.028 & $0.0017(9)$ & 51256.0 & 133 & 11.9 & 0.96 & 0.95 & 0.95 & <$-$1.46 & $-$11.17 & \\ 
\\
J1801$-$0357 & B1758$-$03 & 1.085 & $-$3.897 & $-0.032(3)$ & 50084.0 & 201 & 18.4 & 22.82 & 16.73 & 0.67 & <$-$1.36 & $-$9.71 & \\ 
J1801$-$2920 & B1758$-$29 & 0.924 & $-$2.814 & $-0.0041(8)$ & 51417.0 & 229 & 11.0 & 1.58 & 1.49 & 0.59 & $-$1.81 & $-$10.63 & \\ 
J1803$-$2137 & B1800$-$21 & 7.484 & $-$7511.371 & $224.89(3)$ & 49527.0 & 352 & 6.7 & 2375.83 & 1.87 & 0.99 & 0.73 & $-$6.74 & \\ 
J1803$-$2712 & B1800$-$27 & 2.990 & $-$0.153 & $0.0006(10)$ & 50661.0 & 151 & 15.1 & 1.18 & 1.17 & 1.17 & <$-$1.06 & $-$11.19 & \\ 
J1804$-$0735 & B1802$-$07 & 43.288 & $-$0.875 & $0.0041(20)$ & 50709.0 & 433 & 15.0 & 0.28 & 0.28 & 0.28 & <$-$2.45 & $-$12.05 & \\ 
\end{tabular}\end{tiny}\end{table*}\addtocounter{table}{-1}\begin{table*}\caption{\ldots continued}\begin{tiny}\begin{tabular}{llrrllrlrrrlrr} \hline PSR J & PSR B & $\nu$ & $\dot{\nu}$ & $\ddot{\nu}$ & Epoch & N & $T_s$ & $\sigma_1$ & $\sigma_2$ & $\sigma_3$  & $\Delta_8$ & $\log\sigma_z(10{\rm yr})$\\ & & (s$^{-1}$) & ($10^{-15} s^{-2}$) & ($10^{-24} s^{-3}$) & (MJD) & & (yr) & (ms) & (ms) & (ms) \\ \hline
 J1804$-$2717 & --- & 107.032 & $-$0.468 & $-0.008(5)$ & 51440.0 & 261 & 10.9 & 0.07 & 0.07 & 0.07 & <$-$3.17 & $-$12.45 & \\ 
J1805$+$0306 & B1802$+$03 & 4.572 & $-$20.892 & $0.0444(18)$ & 50331.0 & 198 & 16.9 & 3.85 & 1.83 & 0.26 & <$-$1.76 & $-$10.48 & \\ 
J1806$-$1154 & B1804$-$12 & 1.913 & $-$5.157 & $0.0287(9)$ & 51096.0 & 180 & 12.8 & 3.01 & 1.06 & 0.75 & <$-$1.59 & $-$10.50 & \\ 
J1807$-$0847 & B1804$-$08 & 6.108 & $-$1.074 & $-0.00326(10)$ & 48671.0 & 456 & 26.3 & 1.13 & 0.54 & 0.09 & <$-$3.00 & $-$11.39 & \\ 
J1807$-$2715 & B1804$-$27 & 1.208 & $-$17.768 & $-0.0863(7)$ & 50291.0 & 233 & 17.2 & 27.64 & 3.29 & 0.76 & <$-$1.66 & $-$9.57 & \\ 
\\
J1808$-$0813 & --- & 1.141 & $-$1.616 & $0.0026(5)$ & 51271.0 & 154 & 12.0 & 0.99 & 0.90 & 0.90 & <$-$1.63 & $-$10.88 & \\ 
J1808$-$2057 & B1805$-$20 & 1.089 & $-$20.243 & $0.0773(6)$ & 49937.0 & 188 & 18.8 & 48.80 & 4.60 & 0.97 & $-$0.92 & $-$9.72 & \\ 
J1809$-$2109 & B1806$-$21 & 1.424 & $-$7.747 & $-0.0082(5)$ & 49995.0 & 227 & 18.8 & 4.93 & 3.09 & 0.34 & $-$1.91 & $-$10.54 & \\ 
J1812$+$0226 & B1810$+$02 & 1.260 & $-$5.711 & $0.0009(3)$ & 50336.0 & 244 & 17.1 & 1.32 & 1.28 & 1.28 & <$-$1.59 & $-$11.00 & \\ 
J1812$-$1718 & B1809$-$173 & 0.830 & $-$13.128 & $0.082(4)$ & 49887.0 & 234 & 19.6 & 84.15 & 48.55 & 1.03 & $-$1.20 & $-$8.83 & \\ 
\\
J1812$-$1733 & B1809$-$176 & 1.858 & $-$3.390 & $-0.0137(13)$ & 50688.0 & 191 & 15.2 & 3.81 & 2.99 & 2.99 & <$-$1.33 & $-$10.42 & \\ 
J1813$+$4013 & B1811$+$40 & 1.074 & $-$2.939 & $0.01304(20)$ & 50300.0 & 266 & 17.2 & 5.10 & 1.16 & 0.85 & <$-$1.64 & $-$10.50 & \\ 
J1816$-$1729 & B1813$-$17 & 1.278 & $-$11.873 & $-0.199(3)$ & 49887.0 & 238 & 19.6 & 106.40 & 18.59 & 0.63 & $-$1.46 & $-$9.24 & \\ 
J1816$-$2650 & B1813$-$26 & 1.687 & $-$0.189 & $-0.00109(16)$ & 49573.9 & 261 & 23.6 & 2.04 & 1.88 & 1.88 & <$-$1.12 & $-$11.18 & \\ 
J1818$-$1422 & B1815$-$14 & 3.431 & $-$23.995 & $-0.015(4)$ & 49993.0 & 218 & 18.8 & 8.75 & 8.32 & 0.28 & $-$1.61 & $-$9.90 & \\ 
\\
J1820$-$0427 & B1818$-$04 & 1.672 & $-$17.700 & $0.0277(7)$ & 47020.0 & 690 & 35.1 & 93.58 & 45.77 & 0.66 & $-$1.01 & $-$9.19 & \\ 
J1820$-$1346 & B1817$-$13 & 1.085 & $-$5.294 & $0.0129(3)$ & 50018.0 & 218 & 18.9 & 8.06 & 2.03 & 1.18 & <$-$1.64 & $-$10.26 & \\ 
J1820$-$1818 & B1817$-$18 & 3.227 & $-$0.975 & $0.0042(5)$ & 50663.0 & 170 & 15.1 & 0.71 & 0.58 & 0.58 & $-$1.69 & $-$11.23 & \\ 
J1821$+$1715 & B1821$+$17 & 0.732 & $-$0.466 & $0.0006(15)$ & 51819.0 & 112 & 10.2 & 2.41 & 2.39 & 2.39 & <$-$1.08 & $-$11.33 & \\ 
J1822$+$0705 & --- & 0.734 & $-$0.941 & $0.0013(7)$ & 51748.0 & 102 & 10.6 & 1.24 & 1.15 & 1.15 & <$-$1.36 & $-$11.05 & \\ 
\\
J1822$-$1400 & B1820$-$14 & 4.656 & $-$19.666 & $-0.101(6)$ & 50042.0 & 208 & 18.8 & 16.84 & 10.36 & 0.86 & $-$1.32 & $-$9.79 & \\ 
J1822$-$2256 & B1819$-$22 & 0.534 & $-$0.385 & $0.00007(12)$ & 49120.0 & 164 & 16.0 & 1.10 & 1.10 & 1.10 & <$-$1.46 & $-$11.44 & \\ 
J1823$+$0550 & B1821$+$05 & 1.328 & $-$0.400 & $-0.00156(6)$ & 49844.0 & 365 & 19.8 & 1.02 & 0.60 & 0.36 & <$-$2.13 & $-$11.08 & \\ 
J1823$-$0154 & --- & 1.316 & $-$1.960 & $-0.0085(5)$ & 51276.0 & 157 & 12.0 & 1.36 & 0.71 & 0.54 & <$-$1.68 & $-$10.51 & \\ 
J1823$-$1115 & B1820$-$11 & 3.574 & $-$17.612 & $-0.163(4)$ & 49993.0 & 469 & 18.8 & 26.54 & 11.44 & 2.88 & $-$1.63 & $-$9.82 & \\ 
\\
J1823$-$3021A & B1820$-$30A & 183.823 & $-$114.336 & $0.522(4)$ & 50719.0 & 302 & 15.0 & 0.85 & 0.09 & 0.09 & $-$2.95 & $-$11.06 & \\ 
J1823$-$3021B & B1820$-$30B & 2.641 & $-$0.225 & $0.0031(5)$ & 50700.0 & 303 & 14.9 & 0.79 & 0.72 & 0.61 & <$-$2.00 & $-$11.11 & \\ 
J1823$-$3106 & B1820$-$31 & 3.520 & $-$36.277 & $-0.134(5)$ & 50315.0 & 238 & 17.3 & 17.68 & 8.26 & 0.20 & $-$1.73 & $-$9.67 & \\ 
J1824$-$1118 & B1821$-$11 & 2.295 & $-$18.716 & $0.059(5)$ & 49874.0 & 233 & 19.6 & 26.36 & 20.37 & 0.86 & $-$1.35 & $-$9.57 & \\ 
J1824$-$1945 & B1821$-$19 & 5.282 & $-$145.939 & $-0.152(19)$ & 50101.0 & 351 & 18.2 & 37.37 & 33.81 & 0.24 & $-$1.19 & $-$9.31 & \\ 
\\
J1824$-$2452 & B1821$-$24 & 327.406 & $-$173.519 & $0.047(7)$ & 50238.0 & 163 & 17.4 & 0.13 & 0.11 & 0.09 & <$-$2.56 & $-$12.06 & \\ 
J1825$+$0004 & B1822$+$00 & 1.284 & $-$1.445 & $-0.0076(5)$ & 50336.0 & 240 & 17.1 & 3.29 & 2.21 & 0.61 & <$-$1.54 & $-$10.46 & \\ 
J1825$-$0935 & B1822$-$09 & 1.300 & $-$88.369 & $1.541(16)$ & 49831.0 & 506 & 19.7 & 855.97 & 172.69 & 1.21 & $-$0.44 & $-$8.15 & \\ 
J1825$-$1446 & B1822$-$14 & 3.582 & $-$290.947 & $0.190(4)$ & 49862.0 & 234 & 19.5 & 41.52 & 9.05 & 0.77 & $-$1.38 & $-$9.62 & \\ 
J1826$-$1131 & B1823$-$11 & 0.478 & $-$1.121 & $0.00454(13)$ & 49867.0 & 340 & 19.7 & 7.12 & 3.17 & 2.36 & <$-$1.42 & $-$10.55 & \\ 
\\
J1826$-$1334 & B1823$-$13 & 9.856 & $-$7293.991 & $138.2(4)$ & 50639.9 & 591 & 11.8 & 2267.62 & 123.04 & 0.58 & 0.37 & $-$7.28 & \\ 
J1827$-$0958 & B1824$-$10 & 4.069 & $-$16.597 & $-0.0399(15)$ & 49883.0 & 253 & 19.6 & 7.28 & 3.63 & 1.73 & <$-$1.60 & $-$10.18 & \\ 
J1829$-$1751 & B1826$-$17 & 3.256 & $-$58.852 & $0.048(3)$ & 50101.0 & 363 & 18.2 & 10.66 & 7.31 & 0.44 & $-$0.68 & $-$9.65 & \\ 
J1830$-$1059 & B1828$-$11 & 2.469 & $-$365.843 & $0.865(3)$ & 50031.0 & 755 & 18.7 & 217.18 & 16.82 & 0.47 & $-$0.80 & $-$8.68 & \\ 
J1832$-$0827 & B1829$-$08 & 1.545 & $-$152.462 & $-0.011(4)$ & 49868.0 & 359 & 19.7 & 40.85 & 28.31 & 0.41 & $-$1.63 & $-$9.82 & \\ 
\\
J1832$-$1021 & B1829$-$10 & 3.027 & $-$38.493 & $0.247(3)$ & 49883.0 & 372 & 19.6 & 52.02 & 10.37 & 0.52 & $-$1.57 & $-$9.55 & \\ 
J1833$-$0338 & B1831$-$03 & 1.456 & $-$88.139 & $0.104(5)$ & 50123.0 & 499 & 18.3 & 45.86 & 30.84 & 0.50 & $-$0.66 & $-$9.18 & \\ 
J1833$-$0827 & B1830$-$08 & 11.725 & $-$1260.890 & $-1.357(9)$ & 50740.0 & 351 & 14.7 & 37.77 & 3.72 & 0.24 & $-$1.49 & $-$9.41 & \\ 
J1834$-$0010 & B1831$-$00 & 1.920 & $-$0.039 & $0.0008(53)$ & 49123.0 & 148 & 12.8 & 5.45 & 5.40 & 5.40 & <$-$0.97 & $-$10.98 & \\ 
J1834$-$0426 & B1831$-$04 & 3.447 & $-$0.854 & $0.0208(6)$ & 50165.0 & 250 & 18.1 & 3.22 & 1.24 & 0.46 & $-$1.90 & $-$10.45 & \\ 
\\
J1835$-$0643 & B1832$-$06 & 3.270 & $-$432.507 & $0.790(9)$ & 49993.0 & 256 & 18.8 & 164.18 & 27.53 & 1.84 & $-$1.37 & $-$9.04 & \\ 
J1835$-$1106 & --- & 6.027 & $-$748.776 & $9.4(3)$ & 51289.0 & 243 & 12.1 & 357.22 & 263.09 & 0.53 & 0.39 & $-$8.11 & \\ 
J1836$-$0436 & B1834$-$04 & 2.823 & $-$13.239 & $-0.066(3)$ & 49957.0 & 246 & 19.2 & 18.22 & 7.52 & 0.33 & $-$1.67 & $-$9.95 & \\ 
J1836$-$1008 & B1834$-$10 & 1.777 & $-$37.278 & $-0.024(9)$ & 50123.0 & 279 & 17.8 & 37.79 & 36.84 & 0.32 & $-$0.70 & $-$9.04 & \\ 
J1837$-$0045 & --- & 1.621 & $-$4.424 & $-0.0033(15)$ & 51400.0 & 193 & 11.0 & 1.40 & 1.38 & 1.38 & <$-$1.52 & $-$11.18 & \\ 
\\
J1837$-$0653 & B1834$-$06 & 0.525 & $-$0.213 & $0.00002(17)$ & 50014.0 & 242 & 18.9 & 3.10 & 3.10 & 3.10 & <$-$1.22 & $-$11.38 & \\ 
J1840$+$5640 & B1839$+$56 & 0.605 & $-$0.547 & $-0.00092(3)$ & 49129.0 & 385 & 23.6 & 2.13 & 1.14 & 1.07 & $-$1.72 & $-$10.94 & \\ 
J1841$+$0912 & B1839$+$09 & 2.622 & $-$7.496 & $0.0464(13)$ & 48854.0 & 374 & 25.3 & 28.65 & 12.80 & 0.50 & $-$1.23 & $-$9.90 & \\ 
J1841$-$0425 & B1838$-$04 & 5.372 & $-$184.443 & $0.024(12)$ & 49849.0 & 319 & 18.6 & 22.54 & 21.58 & 0.21 & $-$1.28 & $-$9.82 & \\ 
J1842$-$0359 & B1839$-$04 & 0.543 & $-$0.150 & $0.00037(13)$ & 49848.0 & 303 & 19.6 & 2.94 & 2.68 & 2.68 & <$-$1.39 & $-$10.92 & \\ 
\\
J1844$+$1454 & B1842$+$14 & 2.663 & $-$13.281 & $0.041(8)$ & 49766.0 & 388 & 20.3 & 42.12 & 40.40 & 0.40 & $-$1.30 & $-$9.41 & \\ 
J1844$-$0244 & B1842$-$02 & 1.970 & $-$64.936 & $0.073(5)$ & 49884.0 & 243 & 19.2 & 36.09 & 25.99 & 2.18 & $-$1.14 & $-$9.52 & \\ 
J1844$-$0433 & B1841$-$04 & 1.009 & $-$3.986 & $-0.00111(11)$ & 50032.0 & 349 & 18.8 & 1.41 & 1.22 & 0.70 & <$-$1.86 & $-$10.84 & \\ 
J1844$-$0538 & B1841$-$05 & 3.911 & $-$148.445 & $-0.066(7)$ & 49867.0 & 295 & 19.2 & 20.26 & 17.26 & 0.34 & $-$1.59 & $-$9.83 & \\ 
J1845$-$0434 & B1842$-$04 & 6.163 & $-$143.413 & $0.30(4)$ & 49610.0 & 211 & 16.7 & 87.29 & 31.31 & 0.71 & $-$0.74 & $-$9.27 & \\ 
\\
J1847$-$0402 & B1844$-$04 & 1.673 & $-$144.702 & $0.0819(3)$ & 49142.0 & 422 & 23.7 & 62.88 & 2.87 & 0.39 & $-$1.40 & $-$9.90 & \\ 
J1848$-$0123 & B1845$-$01 & 1.516 & $-$12.075 & $0.037(5)$ & 50407.0 & 289 & 16.5 & 20.01 & 17.64 & 0.31 & $-$1.71 & $-$9.61 & \\ 
J1848$-$1414 & --- & 3.358 & $-$0.158 & $0.0006(19)$ & 51269.0 & 142 & 12.0 & 1.12 & 1.12 & 1.12 & <$-$1.55 & $-$11.78 & \\ 
J1848$-$1952 & B1845$-$19 & 0.232 & $-$1.254 & $-0.00040(19)$ & 50515.0 & 180 & 16.2 & 4.19 & 4.12 & 4.12 & <$-$1.18 & $-$10.78 & \\ 
J1849$-$0636 & B1846$-$06 & 0.689 & $-$21.953 & $-0.0109(3)$ & 49142.0 & 344 & 23.7 & 16.65 & 5.64 & 0.50 & $-$1.59 & $-$10.16 & \\ 
\\
J1850$+$1335 & B1848$+$13 & 2.894 & $-$12.497 & $-0.01152(17)$ & 50128.0 & 214 & 18.3 & 1.97 & 0.39 & 0.26 & <$-$2.11 & $-$10.81 & \\ 
J1851$+$0418 & B1848$+$04 & 3.513 & $-$13.431 & $0.2067(20)$ & 50113.0 & 254 & 18.2 & 29.56 & 4.36 & 1.31 & <$-$1.22 & $-$9.99 & \\ 
J1851$+$1259 & --- & 0.830 & $-$7.924 & $0.208(5)$ & 50106.0 & 195 & 18.4 & 135.61 & 41.86 & 1.00 & $-$1.29 & $-$9.11 & \\ 
J1852$+$0031 & B1849$+$00 & 0.459 & $-$20.398 & $-0.0109(14)$ & 49995.0 & 232 & 18.8 & 33.11 & 28.91 & 21.44 & <$-$0.62 & $-$9.50 & \\ 
J1852$-$2610 & --- & 2.973 & $-$0.775 & $0.0151(5)$ & 51249.0 & 125 & 11.9 & 0.90 & 0.29 & 0.29 & <$-$1.95 & $-$10.78 & \\ 
\\
J1854$+$1050 & B1852$+$10 & 1.745 & $-$1.938 & $0.012(5)$ & 50869.0 & 266 & 14.2 & 11.43 & 11.30 & 11.30 & <$-$0.82 & $-$10.16 & \\ 
J1854$-$1421 & B1851$-$14 & 0.872 & $-$3.166 & $-0.0215(7)$ & 50313.0 & 193 & 17.3 & 11.56 & 4.41 & 0.69 & <$-$1.83 & $-$10.06 & \\ 
J1856$+$0113 & B1853$+$01 & 3.739 & $-$2911.552 & $25.74(7)$ & 50523.0 & 344 & 16.3 & 3040.36 & 145.15 & 1.12 & $-$0.57 & $-$7.56 & \\ 
J1857$+$0057 & B1854$+$00 & 2.802 & $-$0.429 & $-0.0040(12)$ & 50097.0 & 136 & 18.5 & 2.56 & 2.44 & 2.44 & <$-$1.18 & $-$10.69 & \\ 
J1857$+$0212 & B1855$+$02 & 2.405 & $-$232.901 & $0.2459(20)$ & 49945.0 & 280 & 18.8 & 72.96 & 8.82 & 0.30 & $-$1.82 & $-$9.58 & \\ 
\\
J1857$+$0943 & B1855$+$09 & 186.494 & $-$0.620 & $-0.0015(7)$ & 50027.0 & 448 & 18.7 & 0.04 & 0.04 & 0.04 & <$-$3.41 & $-$12.77 & \\ 
J1900$-$2600 & B1857$-$26 & 1.633 & $-$0.546 & $-0.00391(8)$ & 50105.0 & 228 & 18.2 & 1.13 & 0.31 & 0.22 & <$-$2.17 & $-$11.16 & \\ 
J1901$+$0156 & B1859$+$01 & 3.470 & $-$28.379 & $-0.0173(20)$ & 49345.0 & 133 & 11.6 & 1.05 & 0.83 & 0.27 & $-$2.06 & $-$10.56 & \\ 
J1901$+$0331 & B1859$+$03 & 1.526 & $-$17.359 & $0.091(14)$ & 50312.0 & 252 & 17.3 & 55.53 & 50.01 & 0.24 & $-$1.38 & $-$9.24 & \\ 
J1901$+$0716 & B1859$+$07 & 1.553 & $-$5.498 & $0.052(16)$ & 50016.0 & 365 & 18.9 & 136.35 & 134.19 & 3.26 & $-$0.56 & $-$8.73 & \\ 
\\
J1901$-$0906 & --- & 1.122 & $-$1.032 & $0.00125(20)$ & 51277.0 & 141 & 12.0 & 0.41 & 0.36 & 0.36 & <$-$1.76 & $-$11.36 & \\ 
J1902$+$0556 & B1900$+$05 & 1.339 & $-$23.098 & $0.0481(4)$ & 50112.0 & 354 & 18.2 & 18.98 & 2.22 & 0.60 & <$-$1.93 & $-$10.05 & \\ 
J1902$+$0615 & B1900$+$06 & 1.485 & $-$16.990 & $-0.0318(7)$ & 50300.0 & 403 & 17.2 & 9.25 & 3.36 & 0.36 & $-$1.76 & $-$10.05 & \\ 
J1903$+$0135 & B1900$+$01 & 1.371 & $-$7.582 & $-0.0069(6)$ & 50430.0 & 243 & 16.6 & 2.95 & 2.21 & 0.15 & $-$1.82 & $-$10.27 & \\ 
J1903$-$0632 & B1900$-$06 & 2.315 & $-$18.223 & $0.267(16)$ & 50430.0 & 216 & 16.6 & 62.06 & 40.39 & 0.43 & $-$1.12 & $-$9.29 & \\ 
\end{tabular}\end{tiny}\end{table*}\addtocounter{table}{-1}\begin{table*}\caption{\ldots continued}\begin{tiny}\begin{tabular}{llrrllrlrrrlrr} \hline PSR J & PSR B & $\nu$ & $\dot{\nu}$ & $\ddot{\nu}$ & Epoch & N & $T_s$ & $\sigma_1$ & $\sigma_2$ & $\sigma_3$  & $\Delta_8$ & $\log\sigma_z(10{\rm yr})$\\ & & (s$^{-1}$) & ($10^{-15} s^{-2}$) & ($10^{-24} s^{-3}$) & (MJD) & & (yr) & (ms) & (ms) & (ms) \\ \hline
 J1904$+$0004 & --- & 7.167 & $-$6.065 & $0.042(5)$ & 51269.0 & 147 & 12.0 & 1.79 & 1.40 & 0.44 & $-$1.96 & $-$10.55 & \\ 
J1904$-$1224 & --- & 2.664 & $-$2.634 & $-0.0003(7)$ & 50695.0 & 147 & 12.3 & 0.62 & 0.62 & 0.62 & <$-$1.83 & $-$11.67 & \\ 
J1905$+$0709 & B1903$+$07 & 1.543 & $-$11.751 & $0.562(10)$ & 49856.0 & 357 & 19.6 & 271.41 & 81.01 & 7.60 & $-$0.35 & $-$8.75 & \\ 
J1905$-$0056 & B1902$-$01 & 1.555 & $-$7.378 & $-0.0193(3)$ & 50127.0 & 220 & 18.3 & 5.93 & 1.19 & 0.25 & <$-$1.94 & $-$10.57 & \\ 
J1906$+$0641 & B1904$+$06 & 3.741 & $-$29.895 & $-0.0036(3)$ & 50016.0 & 247 & 18.9 & 1.02 & 0.78 & 0.30 & <$-$2.26 & $-$11.11 & \\ 
\\
J1907$+$4002 & B1905$+$39 & 0.809 & $-$0.354 & $0.000004(97)$ & 49144.0 & 382 & 23.7 & 2.44 & 2.44 & 2.44 & <$-$1.57 & $-$11.34 & \\ 
J1909$+$0007 & B1907$+$00 & 0.983 & $-$5.336 & $0.0078(10)$ & 49121.0 & 329 & 23.6 & 21.56 & 19.41 & 0.47 & $-$1.59 & $-$9.87 & \\ 
J1909$+$0254 & B1907$+$02 & 1.010 & $-$5.634 & $0.0724(4)$ & 50496.0 & 265 & 16.3 & 27.54 & 2.06 & 0.72 & <$-$1.81 & $-$9.73 & \\ 
J1909$+$1102 & B1907$+$10 & 3.526 & $-$32.811 & $-0.030(7)$ & 50316.0 & 379 & 17.3 & 14.12 & 13.23 & 0.19 & $-$1.62 & $-$9.93 & \\ 
J1910$+$0358 & B1907$+$03 & 0.429 & $-$0.822 & $0.0305(12)$ & 50428.0 & 195 & 16.6 & 32.23 & 15.07 & 4.09 & <$-$0.86 & $-$9.55 & \\ 
\\
J1910$+$1231 & B1907$+$12 & 0.694 & $-$3.959 & $0.00079(15)$ & 49275.0 & 174 & 22.8 & 2.49 & 2.29 & 2.29 & <0.18 & $-$10.90 & \\ 
J1910$-$0309 & B1907$-$03 & 1.982 & $-$8.593 & $-0.086(5)$ & 50430.0 & 187 & 16.6 & 21.94 & 11.99 & 0.66 & $-$1.35 & $-$9.44 & \\ 
J1911$-$1114 & --- & 275.805 & $-$1.065 & $0.016(9)$ & 51513.0 & 254 & 10.5 & 0.05 & 0.05 & 0.05 & <$-$3.34 & $-$13.15 & \\ 
J1912$+$2104 & B1910$+$20 & 0.448 & $-$2.042 & $-0.00146(12)$ & 50299.0 & 253 & 17.2 & 2.18 & 1.72 & 1.72 & <$-$1.35 & $-$10.96 & \\ 
J1913$+$1400 & B1911$+$13 & 1.918 & $-$2.955 & $-0.0139(4)$ & 50314.0 & 256 & 17.3 & 3.16 & 1.18 & 0.24 & <$-$2.00 & $-$10.64 & \\ 
\\
J1913$-$0440 & B1911$-$04 & 1.211 & $-$5.963 & $0.01485(17)$ & 47036.0 & 543 & 35.0 & 53.91 & 13.10 & 0.44 & $-$1.23 & $-$10.00 & \\ 
J1915$+$1009 & B1913$+$10 & 2.472 & $-$93.191 & $0.0398(12)$ & 50389.0 & 455 & 16.8 & 8.22 & 4.06 & 0.29 & $-$1.96 & $-$10.24 & \\ 
J1915$+$1606 & B1913$+$16 & 16.941 & $-$2.476 & $0.0035(10)$ & 46444.0 & 346 & 18.3 & 0.47 & 0.46 & 0.38 & <$-$2.12 & $-$11.30 & \\ 
J1915$+$1647 & B1913$+$167 & 0.619 & $-$0.155 & $0.00047(10)$ & 49290.0 & 258 & 22.9 & 2.77 & 2.63 & 2.63 & <$-$1.22 & $-$10.71 & \\ 
J1916$+$0951 & B1914$+$09 & 3.700 & $-$34.482 & $-0.0201(8)$ & 50312.0 & 291 & 17.3 & 2.65 & 1.43 & 0.24 & $-$2.23 & $-$10.61 & \\ 
\\
J1916$+$1312 & B1914$+$13 & 3.548 & $-$46.005 & $-1.25(5)$ & 50312.0 & 310 & 17.3 & 207.46 & 82.80 & 0.36 & $-$0.20 & $-$8.46 & \\ 
J1917$+$1353 & B1915$+$13 & 5.138 & $-$189.996 & $0.323(5)$ & 50161.0 & 517 & 18.1 & 33.81 & 10.81 & 0.17 & $-$2.07 & $-$9.67 & \\ 
J1917$+$2224 & B1915$+$22 & 2.348 & $-$15.798 & $-0.017(6)$ & 51813.0 & 208 & 10.3 & 3.80 & 3.63 & 3.63 & <$-$1.61 & $-$10.49 & \\ 
J1918$+$1444 & B1916$+$14 & 0.847 & $-$152.252 & $-0.106(8)$ & 50088.0 & 476 & 18.5 & 145.13 & 110.76 & 1.04 & $-$0.81 & $-$9.02 & \\ 
J1919$+$0021 & B1917$+$00 & 0.786 & $-$4.739 & $0.0046(8)$ & 49832.0 & 368 & 19.7 & 13.57 & 12.89 & 0.88 & <$-$1.72 & $-$10.05 & \\ 
\\
J1920$+$2650 & B1918$+$26 & 1.273 & $-$0.056 & $0.0002(4)$ & 50095.0 & 124 & 18.5 & 1.54 & 1.52 & 1.52 & <$-$0.99 & $-$11.30 & \\ 
J1921$+$1419 & B1919$+$14 & 1.618 & $-$14.657 & $0.0308(9)$ & 48698.0 & 305 & 26.0 & 30.88 & 13.52 & 2.09 & <$-$1.10 & $-$10.21 & \\ 
J1921$+$1948 & B1918$+$19 & 1.218 & $-$1.329 & $0.0015(3)$ & 48395.9 & 174 & 23.7 & 3.64 & 3.26 & 3.26 & <$-$0.49 & $-$10.56 & \\ 
J1921$+$2153 & B1919$+$21 & 0.748 & $-$0.754 & $-0.00009(3)$ & 48469.7 & 304 & 22.1 & 0.50 & 0.48 & 0.48 & <$-$2.23 & $-$11.77 & \\ 
J1922$+$2018 & B1920$+$20 & 0.853 & $-$0.472 & $-0.0009(4)$ & 50085.0 & 195 & 18.4 & 3.41 & 3.35 & 3.35 & <$-$0.83 & $-$10.61 & \\ 
\\
J1922$+$2110 & B1920$+$21 & 0.928 & $-$7.038 & $0.042(3)$ & 50305.0 & 344 & 17.3 & 26.56 & 18.98 & 1.10 & $-$1.65 & $-$9.72 & \\ 
J1926$+$0431 & B1923$+$04 & 0.931 & $-$2.130 & $0.0153(3)$ & 50300.0 & 206 & 17.0 & 6.66 & 1.41 & 0.67 & <$-$1.50 & $-$10.58 & \\ 
J1926$+$1434 & B1924$+$14 & 0.755 & $-$0.125 & $0.00013(13)$ & 49142.0 & 137 & 23.7 & 2.18 & 2.16 & 2.16 & $-$0.72 & $-$10.77 & \\ 
J1926$+$1648 & B1924$+$16 & 1.725 & $-$53.525 & $0.015(6)$ & 50122.0 & 436 & 18.3 & 32.14 & 31.48 & 0.81 & $-$1.64 & $-$9.74 & \\ 
J1932$+$1059 & B1929$+$10 & 4.415 & $-$22.560 & $-0.0210(16)$ & 46926.0 & 769 & 35.7 & 49.73 & 44.15 & 0.21 & $-$1.82 & $-$9.98 & \\ 
\\
J1932$+$2020 & B1929$+$20 & 3.728 & $-$58.626 & $-0.179(4)$ & 50302.0 & 224 & 17.2 & 21.32 & 5.38 & 0.26 & $-$1.69 & $-$9.94 & \\ 
J1932$+$2220 & B1930$+$22 & 6.922 & $-$2756.784 & $41.01(16)$ & 51867.0 & 411 & 8.9 & 432.99 & 31.86 & 0.55 & 0.04 & $-$7.96 & \\ 
J1933$+$2421 & B1931$+$24 & 1.229 & $-$12.216 & $0.702(13)$ & 51014.0 & 580 & 13.7 & 111.46 & 43.12 & 1.03 & $-$0.86 & $-$8.88 & \\ 
J1935$+$1616 & B1933$+$16 & 2.788 & $-$46.642 & $0.01430(11)$ & 46827.0 & 936 & 36.2 & 29.13 & 6.00 & 0.19 & $-$2.31 & $-$10.41 & \\ 
J1937$+$2544 & B1935$+$25 & 4.976 & $-$15.917 & $-0.0026(3)$ & 50126.0 & 370 & 18.3 & 0.51 & 0.45 & 0.20 & <$-$2.56 & $-$11.12 & \\ 
\\
J1939$+$2134 & B1937$+$21 & 641.928 & $-$43.314 & $0.0174(8)$ & 49389.0 & 517 & 22.2 & 0.10 & 0.03 & 0.03 & <$-$3.11 & $-$12.38 & \\ 
J1939$+$2449 & B1937$+$24 & 1.550 & $-$43.898 & $0.72(4)$ & 50351.0 & 277 & 13.0 & 122.73 & 76.48 & 1.90 & $-$0.43 & $-$8.52 & \\ 
J1941$-$2602 & B1937$-$26 & 2.482 & $-$5.892 & $-0.00815(9)$ & 50498.0 & 193 & 16.3 & 1.31 & 0.18 & 0.10 & <$-$2.17 & $-$11.02 & \\ 
J1943$-$1237 & B1940$-$12 & 1.028 & $-$1.751 & $0.00039(8)$ & 49145.0 & 139 & 23.7 & 1.24 & 1.12 & 0.45 & <$-$0.90 & $-$10.70 & \\ 
J1944$-$1750 & B1941$-$17 & 1.189 & $-$1.394 & $0.0095(7)$ & 50314.0 & 139 & 17.3 & 4.88 & 3.09 & 1.96 & <$-$0.98 & $-$10.34 & \\ 
\\
J1945$-$0040 & B1942$-$00 & 0.956 & $-$0.489 & $0.00008(10)$ & 48532.0 & 141 & 27.0 & 1.91 & 1.90 & 1.90 & <$-$0.46 & $-$11.23 & \\ 
J1946$+$1805 & B1944$+$17 & 2.270 & $-$0.124 & $0.00035(15)$ & 49291.0 & 241 & 22.9 & 0.98 & 0.97 & 0.97 & <$-$1.75 & $-$11.24 & \\ 
J1946$-$2913 & B1943$-$29 & 1.042 & $-$1.617 & $-0.00137(7)$ & 49123.0 & 118 & 23.6 & 1.87 & 0.84 & 0.72 & <$-$0.96 & $-$11.08 & \\ 
J1948$+$3540 & B1946$+$35 & 1.394 & $-$13.723 & $0.0048(3)$ & 49852.0 & 459 & 19.8 & 3.83 & 3.02 & 0.20 & $-$2.30 & $-$10.49 & \\ 
J1949$-$2524 & B1946$-$25 & 1.044 & $-$3.566 & $-0.00068(9)$ & 48533.0 & 129 & 23.7 & 1.41 & 1.17 & 0.53 & <$-$1.43 & $-$10.91 & \\ 
\\
J1952$+$1410 & B1949$+$14 & 3.636 & $-$1.694 & $0.0023(19)$ & 50096.0 & 164 & 16.0 & 1.75 & 1.74 & 1.74 & <$-$1.28 & $-$11.15 & \\ 
J1952$+$3252 & B1951$+$32 & 25.296 & $-$3739.490 & $17.51(10)$ & 50249.0 & 760 & 17.8 & 452.10 & 56.20 & 0.34 & $-$0.57 & $-$8.52 & \\ 
J1954$+$2923 & B1952$+$29 & 2.344 & $-$0.009 & $0.00002(8)$ & 49137.0 & 242 & 16.3 & 0.19 & 0.19 & 0.19 & <$-$2.16 & $-$12.23 & \\ 
J1955$+$2908 & B1953$+$29 & 163.048 & $-$0.789 & $-0.002(4)$ & 49456.0 & 314 & 21.8 & 0.43 & 0.43 & 0.43 & $-$1.97 & $-$11.89 & \\ 
J1955$+$5059 & B1953$+$50 & 1.927 & $-$5.096 & $-0.00908(6)$ & 49145.0 & 406 & 23.7 & 4.62 & 0.58 & 0.13 & <$-$2.59 & $-$10.83 & \\ 
\\
J1957$+$2831 & --- & 3.250 & $-$32.849 & $0.0180(18)$ & 51873.0 & 277 & 10.0 & 1.34 & 1.04 & 0.54 & $-$2.02 & $-$10.60 & \\ 
J2002$+$3217 & B2000$+$32 & 1.435 & $-$216.619 & $-0.88(3)$ & 50015.0 & 497 & 18.9 & 473.25 & 232.21 & 0.98 & $-$0.65 & $-$8.37 & \\ 
J2002$+$4050 & B2000$+$40 & 1.105 & $-$2.123 & $-0.00214(7)$ & 50094.0 & 388 & 18.5 & 1.11 & 0.58 & 0.39 & <$-$2.22 & $-$10.95 & \\ 
J2004$+$3137 & B2002$+$31 & 0.474 & $-$16.723 & $-0.0111(5)$ & 49142.0 & 425 & 23.7 & 28.67 & 18.41 & 0.46 & <$-$1.59 & $-$9.92 & \\ 
J2005$-$0020 & --- & 0.877 & $-$9.876 & $-0.002(3)$ & 51404.0 & 140 & 11.0 & 3.56 & 3.53 & 3.53 & <$-$1.30 & $-$10.95 & \\ 
\\
J2006$-$0807 & B2003$-$08 & 1.722 & $-$0.136 & $-0.0033(7)$ & 50190.0 & 303 & 17.7 & 2.91 & 2.78 & 2.78 & <$-$1.56 & $-$10.99 & \\ 
J2013$+$3845 & B2011$+$38 & 4.344 & $-$167.035 & $0.0150(11)$ & 50094.0 & 458 & 18.5 & 4.99 & 2.60 & 0.58 & $-$2.13 & $-$10.52 & \\ 
J2018$+$2839 & B2016$+$28 & 1.792 & $-$0.476 & $0.00469(3)$ & 46774.0 & 618 & 36.5 & 12.69 & 1.58 & 0.30 & <$-$2.43 & $-$10.86 & \\ 
J2019$+$2425 & --- & 254.160 & $-$0.454 & $-0.032(16)$ & 51351.0 & 170 & 11.5 & 0.09 & 0.09 & 0.09 & <$-$3.08 & $-$12.31 & \\ 
J2022$+$5154 & B2021$+$51 & 1.890 & $-$10.939 & $-0.0188(7)$ & 47028.0 & 781 & 35.1 & 64.59 & 45.72 & 0.33 & $-$1.95 & $-$9.82 & \\ 
\\
J2023$+$5037 & B2022$+$50 & 2.684 & $-$18.091 & $0.0403(10)$ & 50313.0 & 365 & 17.3 & 7.21 & 3.02 & 0.20 & <$-$2.28 & $-$10.28 & \\ 
J2029$+$3744 & B2027$+$37 & 0.822 & $-$8.323 & $-0.0642(4)$ & 50094.0 & 231 & 18.5 & 42.21 & 3.47 & 0.93 & <$-$1.49 & $-$9.71 & \\ 
J2030$+$2228 & B2028$+$22 & 1.586 & $-$2.225 & $0.0134(7)$ & 51088.0 & 161 & 13.1 & 1.80 & 0.93 & 0.53 & <$-$1.17 & $-$10.60 & \\ 
J2037$+$3621 & B2035$+$36 & 1.616 & $-$11.771 & $-0.48(3)$ & 50093.0 & 163 & 18.5 & 183.77 & 99.25 & 1.61 & $-$0.75 & $-$8.31 & \\ 
J2038$+$5319 & B2036$+$53 & 0.702 & $-$0.465 & $0.0012(5)$ & 50122.0 & 93 & 13.9 & 1.95 & 1.87 & 1.87 & <$-$1.00 & $-$10.77 & \\ 
\\
J2046$+$1540 & B2044$+$15 & 0.879 & $-$0.141 & $0.00006(5)$ & 49850.0 & 342 & 19.8 & 0.73 & 0.73 & 0.73 & <$-$1.98 & $-$11.80 & \\ 
J2046$+$5708 & B2045$+$56 & 2.098 & $-$48.934 & $0.0100(12)$ & 50155.0 & 299 & 18.0 & 5.46 & 4.78 & 1.81 & <$-$1.49 & $-$10.12 & \\ 
J2046$-$0421 & B2043$-$04 & 0.646 & $-$0.615 & $0.00018(5)$ & 49763.0 & 353 & 20.3 & 0.89 & 0.87 & 0.87 & <$-$2.05 & $-$11.43 & \\ 
J2048$-$1616 & B2045$-$16 & 0.510 & $-$2.848 & $-0.00144(4)$ & 46827.0 & 594 & 36.1 & 13.05 & 11.28 & 0.53 & <$-$2.10 & $-$10.43 & \\ 
J2051$-$0827 & --- & 221.796 & $-$0.626 & $0.011(11)$ & 51481.0 & 303 & 10.7 & 0.08 & 0.08 & 0.08 & $-$3.25 & $-$12.80 & \\ 
\\
J2055$+$2209 & B2053$+$21 & 1.227 & $-$2.017 & $-0.00035(8)$ & 50095.0 & 254 & 18.5 & 0.54 & 0.51 & 0.51 & <$-$1.55 & $-$11.64 & \\ 
J2055$+$3630 & B2053$+$36 & 4.515 & $-$7.524 & $-0.0116(3)$ & 49753.0 & 472 & 20.2 & 1.86 & 0.69 & 0.12 & $-$2.58 & $-$10.93 & \\ 
J2108$+$4441 & B2106$+$44 & 2.410 & $-$0.501 & $0.0012(3)$ & 50315.0 & 292 & 17.3 & 0.64 & 0.61 & 0.61 & <$-$1.79 & $-$11.55 & \\ 
J2113$+$2754 & B2110$+$27 & 0.831 & $-$1.813 & $0.00031(4)$ & 49133.0 & 384 & 23.6 & 0.89 & 0.79 & 0.71 & <$-$1.73 & $-$11.25 & \\ 
J2113$+$4644 & B2111$+$46 & 0.986 & $-$0.694 & $0.01772(8)$ & 47007.0 & 601 & 35.2 & 84.39 & 8.20 & 2.56 & $-$1.53 & $-$10.24 & \\ 
\\
J2116$+$1414 & B2113$+$14 & 2.272 & $-$1.494 & $-0.0039(4)$ & 49855.0 & 373 & 19.8 & 2.25 & 1.87 & 0.41 & <$-$2.18 & $-$10.74 & \\ 
J2124$+$1407 & B2122$+$13 & 1.441 & $-$1.594 & $0.0042(3)$ & 50276.0 & 79 & 17.2 & 1.54 & 0.73 & 0.73 & <$-$0.25 & $-$10.77 & \\ 
J2124$-$3358 & --- & 202.794 & $-$0.846 & $0.011(8)$ & 51284.0 & 227 & 11.7 & 0.08 & 0.08 & 0.08 & <$-$3.15 & $-$12.66 & \\ 
J2145$-$0750 & --- & 62.296 & $-$0.115 & $-0.0010(4)$ & 51173.0 & 470 & 12.3 & 0.02 & 0.02 & 0.02 & $-$3.70 & $-$13.20 & \\ 
J2149$+$6329 & B2148$+$63 & 2.631 & $-$1.177 & $-0.00139(11)$ & 49852.0 & 388 & 19.8 & 0.67 & 0.56 & 0.23 & $-$1.97 & $-$10.97 & \\ 
\end{tabular}\end{tiny}\end{table*}\addtocounter{table}{-1}\begin{table*}\caption{\ldots continued}\begin{tiny}\begin{tabular}{llrrllrlrrrlrr} \hline PSR J & PSR B & $\nu$ & $\dot{\nu}$ & $\ddot{\nu}$ & Epoch & N & $T_s$ & $\sigma_1$ & $\sigma_2$ & $\sigma_3$  & $\Delta_8$ & $\log\sigma_z(10{\rm yr})$\\ & & (s$^{-1}$) & ($10^{-15} s^{-2}$) & ($10^{-24} s^{-3}$) & (MJD) & & (yr) & (ms) & (ms) & (ms) \\ \hline
 J2150$+$5247 & B2148$+$52 & 3.010 & $-$91.574 & $-0.0243(11)$ & 50094.0 & 401 & 18.5 & 6.18 & 3.68 & 0.20 & $-$1.88 & $-$10.17 & \\ 
J2155$-$3118 & B2152$-$31 & 0.971 & $-$1.169 & $0.0022(4)$ & 50429.0 & 125 & 16.6 & 2.06 & 1.80 & 1.80 & <$-$0.96 & $-$11.00 & \\ 
J2157$+$4017 & B2154$+$40 & 0.656 & $-$1.477 & $-0.0376(4)$ & 49686.0 & 474 & 20.5 & 46.24 & 8.26 & 0.98 & $-$1.89 & $-$9.86 & \\ 
J2212$+$2933 & B2210$+$29 & 0.995 & $-$0.491 & $-0.00257(14)$ & 50093.0 & 304 & 18.5 & 1.75 & 1.18 & 1.12 & <$-$1.50 & $-$10.80 & \\ 
J2219$+$4754 & B2217$+$47 & 1.857 & $-$9.537 & $-0.00296(5)$ & 46471.0 & 510 & 35.3 & 7.64 & 2.61 & 0.61 & $-$1.71 & $-$10.55 & \\ 
\\
J2222$+$2923 & --- & 3.554 & $-$0.078 & $0.0032(15)$ & 51671.0 & 258 & 11.0 & 0.97 & 0.95 & 0.95 & <$-$2.04 & $-$11.23 & \\ 
J2225$+$6535 & B2224$+$65 & 1.465 & $-$20.734 & $0.0151(11)$ & 49142.0 & 517 & 23.7 & 19.39 & 16.46 & 0.80 & $-$1.35 & $-$9.87 & \\ 
J2229$+$2643 & --- & 335.816 & $-$0.171 & $-0.011(6)$ & 51625.0 & 279 & 11.1 & 0.04 & 0.04 & 0.04 & <$-$3.18 & $-$13.39 & \\ 
J2229$+$6205 & B2227$+$61 & 2.257 & $-$11.489 & $-0.0117(9)$ & 50081.0 & 215 & 18.2 & 3.88 & 2.80 & 0.45 & <$-$1.74 & $-$10.30 & \\ 
J2242$+$6950 & B2241$+$69 & 0.601 & $-$1.741 & $-0.0003(3)$ & 50092.0 & 115 & 18.5 & 3.54 & 3.27 & 3.27 & <$-$0.28 & $-$10.72 & \\ 
\\
J2248$-$0101 & --- & 2.095 & $-$2.894 & $0.0354(8)$ & 51270.0 & 171 & 12.0 & 3.20 & 0.81 & 0.48 & $-$1.85 & $-$10.27 & \\ 
J2257$+$5909 & B2255$+$58 & 2.716 & $-$42.428 & $-0.102(3)$ & 50502.0 & 236 & 16.1 & 16.41 & 5.88 & 0.14 & $-$1.90 & $-$9.71 & \\ 
J2305$+$3100 & B2303$+$30 & 0.635 & $-$1.165 & $-0.00131(11)$ & 50302.0 & 278 & 17.4 & 1.44 & 1.14 & 0.70 & <$-$1.44 & $-$10.83 & \\ 
J2305$+$4707 & B2303$+$46 & 0.938 & $-$0.500 & $-0.0012(7)$ & 50088.0 & 186 & 18.5 & 4.49 & 4.44 & 4.44 & <$-$0.70 & $-$10.70 & \\ 
J2308$+$5547 & B2306$+$55 & 2.105 & $-$0.884 & $-0.00551(14)$ & 49841.0 & 394 & 19.7 & 2.02 & 0.87 & 0.51 & <$-$2.12 & $-$10.94 & \\ 
\\
J2313$+$4253 & B2310$+$42 & 2.862 & $-$0.920 & $0.00629(8)$ & 48202.7 & 499 & 26.3 & 3.62 & 0.94 & 0.12 & $-$2.83 & $-$10.99 & \\ 
J2317$+$1439 & --- & 290.255 & $-$0.204 & $-0.005(7)$ & 51656.0 & 230 & 11.0 & 0.04 & 0.04 & 0.04 & <$-$3.39 & $-$13.19 & \\ 
J2317$+$2149 & B2315$+$21 & 0.692 & $-$0.502 & $-0.00008(3)$ & 49142.0 & 300 & 20.2 & 0.55 & 0.54 & 0.54 & <$-$1.85 & $-$11.71 & \\ 
J2321$+$6024 & B2319$+$60 & 0.443 & $-$1.382 & $-0.00118(4)$ & 49332.0 & 348 & 22.6 & 3.14 & 1.38 & 1.12 & <$-$1.11 & $-$10.69 & \\ 
J2322$+$2057 & --- & 207.968 & $-$0.418 & $0.007(11)$ & 51335.0 & 151 & 11.4 & 0.08 & 0.07 & 0.04 & <$-$2.99 & $-$12.73 & \\ 
\\
J2325$+$6316 & B2323$+$63 & 0.696 & $-$1.370 & $0.00053(11)$ & 48714.0 & 381 & 26.1 & 4.58 & 4.42 & 4.42 & <$-$1.16 & $-$10.67 & \\ 
J2326$+$6113 & B2324$+$60 & 4.280 & $-$6.458 & $0.00202(19)$ & 50069.0 & 395 & 18.5 & 0.52 & 0.46 & 0.27 & $-$2.57 & $-$11.29 & \\ 
J2330$-$2005 & B2327$-$20 & 0.608 & $-$1.714 & $-0.00007(12)$ & 50304.0 & 232 & 17.3 & 1.11 & 1.10 & 1.10 & <$-$1.37 & $-$11.96 & \\ 
J2337$+$6151 & B2334$+$61 & 2.019 & $-$781.495 & $3.1256(8)$ & 50094.0 & 465 & 18.5 & 1064.52 & 4.43 & 1.19 & $-$0.33 & $-$8.29 & \\ 
J2346$-$0609 & --- & 0.846 & $-$0.977 & $-0.0014(3)$ & 51426.0 & 236 & 11.2 & 0.63 & 0.60 & 0.60 & <$-$1.94 & $-$11.37 & \\ 
\\
J2354$+$6155 & B2351$+$61 & 1.058 & $-$18.219 & $0.0233(4)$ & 49276.0 & 375 & 22.8 & 23.91 & 6.66 & 0.42 & $-$1.89 & $-$9.99 & \\ 
\hline\end{tabular}\end{tiny}\end{table*}

%% file: hobbs.bbl
\begin{thebibliography}{{{Lyne}, {Shemar} \& {Graham-Smith} }{2000}}

\bibitem[\protect\citename{{Alpar} \& {Baykal} }{2006}]{ab06}
{Alpar}~M.~A., {Baykal}~A., 2006, MNRAS, 372, 489

\bibitem[\protect\citename{Arzoumanian {\rm et~al. }}{1994}]{antt94}
Arzoumanian~Z., Nice~D.~J., Taylor~J.~H., Thorsett~S.~E., 1994, ApJ, 422, 671

\bibitem[\protect\citename{Bailes, Lyne \& Shemar }{1993}]{bls93}
Bailes~M., Lyne~A.~G., Shemar~S.~L., 1993, in Phillips~J.~A., Thorsett~S.~E.,
  Kulkarni~S.~R., eds, Planets Around Pulsars.
\newblock Astronomical Society of the Pacific Conference Series, p.~19

\bibitem[\protect\citename{{Baykal} {\rm et~al. }}{1999}]{babd99}
{Baykal}~A., {Ali Alpar}~M., {Boynton}~P., {Deeter}~J., 1999, MNRAS, 306, 207

\bibitem[\protect\citename{{Beer}, {King} \& {Pringle} }{2004}]{bkp04}
{Beer}~M.~E., {King}~A.~R., {Pringle}~J.~E., 2004, MNRAS, 355, 1244

\bibitem[\protect\citename{Boynton {\rm et~al. }}{1972}]{bgh+72}
Boynton~P.~E., Groth~E.~J., Hutchinson~D.~P., Nanos~G.~P., Partridge~R.~B.,
  Wilkinson~D.~T., 1972, ApJ, 175, 217

\bibitem[\protect\citename{Cordes \& Downs }{1985}]{cd85}
Cordes~J.~M., Downs~G.~S., 1985, ApJS, 59, 343

\bibitem[\protect\citename{Cordes \& Helfand }{1980}]{ch80}
Cordes~J.~M., Helfand~D.~J., 1980, ApJ, 239, 640

\bibitem[\protect\citename{Cordes }{1993}]{cor93}
Cordes~J.~M., 1993, in Phillips~J.~A., Thorsett~S.~E., Kulkarni~S.~R., eds,
  Planets around Pulsars.
\newblock Astron.\ Soc.\ Pac.\ Conf.\ Ser.\ Vol.\ 36, p.~43

\bibitem[\protect\citename{{D'Alessandro} {\rm et~al. }}{1995}]{dmh+95}
{D'Alessandro}~F., {McCulloch}~P.~M., {Hamilton}~P.~A., {Deshpande}~A.~A.,
  1995, MNRAS, 277, 1033

\bibitem[\protect\citename{Demia\'{n}ski \& Pr\'{o}szy\'{n}ski }{1979}]{dp79}
Demia\'{n}ski~M., Pr\'{o}szy\'{n}ski~M., 1979, Nat, 282, 383

\bibitem[\protect\citename{Downs \& Krause-Polstorff }{1986}]{dk86}
Downs~G.~S., Krause-Polstorff~J., 1986, ApJS, 62, 81

\bibitem[\protect\citename{Downs \& Reichley }{1983}]{dr83}
Downs~G.~S., Reichley~P.~E., 1983, ApJS, 53, 169

\bibitem[\protect\citename{{Edwards}, {Hobbs} \& {Manchester} }{2006}]{ehm06}
{Edwards}~R.~T., {Hobbs}~G.~B., {Manchester}~R.~N., 2006, MNRAS, 372, 1549

\bibitem[\protect\citename{{Ford} {\rm et~al. }}{2000}]{fjrz00}
{Ford}~E.~B., {Joshi}~K.~J., {Rasio}~F.~A., {Zbarsky}~B., 2000, ApJ, 528, 336

\bibitem[\protect\citename{Hobbs }{2005}]{hob05}
Hobbs~G., 2005, Proc.\,Astr.\,Soc.\,Aust., 22, 179

\bibitem[\protect\citename{{Hobbs}, {Edwards} \& {Manchester} }{2006}]{hem06}
{Hobbs}~G.~B., {Edwards}~R.~T., {Manchester}~R.~N., 2006, MNRAS, 369, 655

\bibitem[\protect\citename{Hobbs {\rm et~al. }}{2009}]{hbb+09}
Hobbs~G. {\rm et~al.}, 2009, Proc.\,Astr.\,Soc.\,Aust., 26, 103

\bibitem[\protect\citename{Hobbs {\rm et~al. }}{2002}]{hlj+02}
Hobbs~G. {\rm et~al.}, 2002, MNRAS, 333, L7

\bibitem[\protect\citename{Hobbs {\rm et~al. }}{2004}]{hlk+04}
Hobbs~G., Lyne~A.~G., Kramer~M., Martin~C.~E., Jordan~C., 2004, MNRAS, 353,
  1311

\bibitem[\protect\citename{Hobbs {\rm et~al. }}{2005}]{hllk05}
Hobbs~G., Lorimer~D.~R., Lyne~A.~G., Kramer~M., 2005, MNRAS, 360, 974

\bibitem[\protect\citename{Janssen \& Stappers}{2006}]{js06}
Janssen~G., Stappers~B., 2006, A\&A, 457, 611

\bibitem[\protect\citename{Johnston \& Galloway }{1999}]{jg99}
Johnston~S., Galloway~D., 1999, MNRAS, 306, L50

\bibitem[\protect\citename{Konacki {\rm et~al. }}{1999}]{klw+99}
Konacki~M., Lewandowski~W., Wolszczan~A., Doroshenko~O., Kramer~M., 1999, ApJ,
  519, L81

\bibitem[\protect\citename{{Livingstone} {\rm et~al. }}{2005}]{lkgm05}
{Livingstone}~M.~A., {Kaspi}~V.~M., {Gavriil}~F.~P., {Manchester}~R.~N., 2005,
  ApJ, 619, 1046

\bibitem[\protect\citename{Lorimer \& Kramer }{2005}]{lk05}
Lorimer~D.~R., Kramer~M., 2005, Handbook of Pulsar Astronomy.
\newblock Cambridge University Press

\bibitem[\protect\citename{Lyne \& Smith }{2004}]{ls04}
Lyne~A.~G., Smith~F.~G., 2004, Pulsar Astronomy, 3rd ed.
\newblock Cambridge University Press, Cambridge

\bibitem[\protect\citename{{Lyne} }{1999}]{lyn99}
Arzoumanian~Z., {van der Hooft}~F., {van den Heuvel}~E.~P.~J., eds, Pulsar
  Timing, General Relativity, and the Internal Structure of Neutron Stars,
  North Holland, Amsterdam, 1999

\bibitem[\protect\citename{{Lyne}, {Shemar} \& {Graham-Smith} }{2000}]{lsg00}
{Lyne}~A.~G., {Shemar}~S.~L., {Graham-Smith}~F., 2000, MNRAS, 315, 534

\bibitem[\protect\citename{Manchester \& Taylor }{1977}]{mt77}
Manchester~R.~N., Taylor~J.~H., 1977, Pulsars.
\newblock Freeman, San Francisco

\bibitem[\protect\citename{Matsakis, Taylor \& Eubanks }{1997}]{mte97}
Matsakis~D.~N., Taylor~J.~H., Eubanks~T.~M., 1997, AA, 326, 924

\bibitem[\protect\citename{{Melatos}, {Peralta} \& {Wyithe} }{2008}]{mpw08}
{Melatos}~A., {Peralta}~C., {Wyithe}~J.~S.~B., 2008, ApJ, 672, 1103

\bibitem[\protect\citename{Press {\rm et~al. }}{1986}]{pftv86}
Press~W.~H., Flannery~B.~P., Teukolsky~S.~A., Vetterling~W.~T., 1986, Numerical
  Recipes: {T}he Art of Scientific Computing.
\newblock Cambridge University Press, Cambridge

\bibitem[\protect\citename{{Rodin} }{2008}]{rod08}
{Rodin}~A.~E., 2008, MNRAS, 387, 1583

\bibitem[\protect\citename{Sedrakian, Wasserman \& Cordes }{1999}]{swc99}
Sedrakian~A., Wasserman~I., Cordes~J.~M., 1999, ApJ, 524, 341

\bibitem[\protect\citename{Shabanova }{1995}]{sha95}
Shabanova~T.~V., 1995, ApJ, 453, 779

\bibitem[\protect\citename{{Shabanova}, {Lyne} \& {Urama} }{2001}]{slu01}
{Shabanova}~T.~V., {Lyne}~A.~G., {Urama}~J.~O., 2001, ApJ, 552, 321

\bibitem[\protect\citename{Shaham }{1977}]{sha77}
Shaham~J., 1977, ApJ, 214, 251

\bibitem[\protect\citename{Shemar }{1996}]{she96}
Shemar~S., Lyne~A.~G., 1996, MNRAS, 282, 677 

\bibitem[\protect\citename{Splaver {\rm et~al. }}{2005}]{sns+05}
Splaver~E.~M., Nice~D.~J., Stairs~I.~H., Lommen~A.~N., Backer~D.~C., 2005, ApJ,
  620, 405

\bibitem[\protect\citename{{Stairs}, {Lyne} \& {Shemar} }{2000}]{sls00}
{Stairs}~I.~H., {Lyne}~A.~G., {Shemar}~S., 2000, Nat, 406, 484

\bibitem[\protect\citename{Standish }{2004}]{sta04b}
Standish~E.~M., 2004, AA, 417, 1165

\bibitem[\protect\citename{{Thorsett} {\rm et~al. }}{1999}]{tacl99}
{Thorsett}~S.~E., {Arzoumanian}~Z., {Camilo}~F., {Lyne}~A.~G., 1999, ApJ, 523,
  763

\bibitem[\protect\citename{{Verbiest} {\rm et~al. }}{2008}]{vbv+08}
{Verbiest}~J.~P.~W. {\rm et~al.}, 2008, ArXiv e-prints, 801

\bibitem[\protect\citename{{Warszawski} \& {Melatos} }{2008}]{wm08}
{Warszawski}~L., {Melatos}~A., 2008, MNRAS, 390, 175

\bibitem[\protect\citename{{You} {\rm et~al. }}{2007a}]{yhc+07a}
{You}~X.~P., {Hobbs}~G.~B., {Coles}~W.~A., {Manchester}~R.~N., {Han}~J.~L.,
  2007, ApJ, 671, 907

\bibitem[\protect\citename{{You} {\rm et~al. }}{2007b}]{yhc+07b}
{You}~X.~P., {Hobbs}~G.~B., {Coles}~W.~A., {Manchester}~R.~N., {Han}~J.~L.,
  2007, ApJ, 671, 907

\bibitem[\protect\citename{{Zou} {\rm et~al. }}{2004}]{zww+04}
{Zou}~W.~Z., {Wang}~N., {Wang}~H.~X., {Manchester}~R.~N., {Wu}~X.~J.,
  {Zhang}~J., 2004, MNRAS, 354, 811

\end{thebibliography}
